\documentclass[preprint,prd,aps,showpacs,nofootinbib]{revtex4}
\usepackage{inputenc}
\usepackage{epsfig,float}
\usepackage{amsfonts}
\usepackage[hypertex]{hyperref}
\usepackage{amsmath}
\input{epsf}
\newcommand{\be}{\begin{eqnarray}}
\newcommand{\ee}{\end{eqnarray}}
\newcommand{\ba}{\begin{array}}
\newcommand{\ea}{\end{array}}

\newcommand{\s}[1]{{\rlap/ #1}}

\restylefloat{figure}
\restylefloat{table}

\begin{document}

\title{ $\pi N$ transition distribution amplitudes: Their symmetries and
constraints from chiral dynamics}

\author{B.~Pire$^1$,  K.~Semenov-Tian-Shansky$^{1,2}$, L.~Szymanowski$^{3}$ }
\affiliation{$^1$ CPhT, \'{E}cole Polytechnique, CNRS,  91128, Palaiseau, France  \\
$^2$ LPT,   Universit\'{e} d'Orsay, CNRS, 91404 Orsay, France \\
%$^$ St. Petersburg State University, St.Petersburg,  \\ 198504, Petrodvoretz,   Russia \\
$^3$ Soltan Institute for Nuclear Studies, Warsaw,  Poland.
}
%\email[]{Kirill.Semenov@cpht.polytechnique.fr}

\preprint{CPHT-RR038.0511, LPT-ORSAY 11-39}

\begin{abstract}
Baryon to meson transition distribution amplitudes (TDAs) extend
the concept of generalized parton distributions. Baryon to meson  TDAs appear
as  building blocks in the collinear factorized description of amplitudes for a class of
hard exclusive reactions, prominent examples of which being hard exclusive meson electroproduction off a nucleon
in the backward region and baryon-antibaryon annihilation into a meson and a lepton pair.
We study the general
properties of these objects following from
the underlying symmetries of QCD. In particular, the Lorentz symmetry results in
the polynomiality property of  the Mellin moments in longitudinal momentum
fractions.
We present a detailed account of the isotopic and permutation symmetry properties
of nucleon to pion ($\pi N$) TDAs. This restricts the number of independent leading twist $\pi N$ TDAs to
eight functions, providing description of all isotopic channels.
Using chiral symmetry and the crossing relation between $\pi N$ TDAs and $\pi N$ generalized distribution amplitudes
we establish  soft pion theorems for $\pi N$ TDAs, which determine the magnitude of
$\pi N$ TDAs.  
Finally, we build a simple resonance exchange model for $\pi N$ TDAs considering $N$ and $\Delta(1232)$ exchange
contributions into the isospin-$\frac{1}{2}$ and isospin-$\frac{3}{2}$ $\pi N$ TDAs.

\end{abstract}

\maketitle
\thispagestyle{empty}
\renewcommand{\thesection}{\arabic{section}}

\renewcommand{\thesubsection}{\arabic{subsection}}

\section{%$\pi N$ GDA and TDA: physical domains and crossing
Introducing $\pi N$ TDAs}
\label{Section_Intro}
Hadronic matrix elements of nonlocal light-cone operators are
the conventional non-perturbative objects which arise in the
description of hard exclusive electroproduction reactions
within the collinear factorization approach.
Factorization theorems for hard exclusive backward meson electroproduction argued in
\cite{Frankfurt:1999fp,Frankfurt:2002kz}
and  baryon-antibaryon anihilation into a pion and a high energy dilepton pair \cite{LPS}
lead to the introduction of baryon to meson transition distribution amplitudes (TDAs),
non diagonal matrix elements
of light-cone three-quark operators
\be
&&
\widehat{O}^{\alpha \beta \gamma}_{\rho \tau \chi}(z_1,\,z_2,z_3)
%\nonumber \\ &&
= \left.
\varepsilon_{c_1 c_2 c_3}
\Psi^{c_1 \, \alpha}_\rho(z_1) \Psi^{c_2 \beta}_\tau (z_2) \Psi^{c_3 \, \gamma}_\chi (z_3) \right|_{z_i^2=0}
%\ \ \ \ (z_i^2=0)
\label{Def_operator}
\ee
between a baryon and a meson states. In
(\ref{Def_operator})
$\alpha$, $\beta$, $\gamma$
stand for quark flavor indices;
$\rho$, $\tau$
and
$\chi$
denote the Dirac indices and
$c_{1,2,3}$
are indices of the color group.
Throughout this paper we adopt the light-cone
gauge
$A^+=0$,
so that the gauge link is equal to unity and we do not show it explicitly
in the definition of the operator
(\ref{Def_operator}).

In accordance with the usual logic of the collinear factorization approach, baryon to meson TDAs
have well established renormalization group behavior.
The evolution properties of the three-quark nonlocal operator
(\ref{Def_operator})
on the light-cone
\cite{Radyushkin:1977gp,Efremov:1978rn,Lepage:1980,Chernyak:1983ej}
were extensively studied in the literature
(see {\it e.g.} \cite{Stefanis:1999wy,Braun:1999te})
for the case of matrix elements
between a baryon and the vacuum known as baryon distribution amplitudes (DAs).
The definition of baryon to meson TDAs involves the same light-cone operator.
Consequently, its evolution also determines the factorization scale dependence of TDAs
\cite{Pire:2005ax}.

Because of the nonperturbative nature of TDAs the initial conditions for evolution require modeling
at low factorization scale.
In particular, nucleon to pion ($\pi N$) TDAs at low scale were recently studied within a light-front quark model
\cite{Pasquini:2009ki}.
From the physics point of view $\pi N$ TDAs may be seen as an essential object to probe the pion cloud content  of the nucleon
\cite{Strikman:2009bd}.

In
\cite{Pire:2005ax,Pire:2005mt,Lansberg:2007ec}
a factorized framework was introduced to describe the electroproduction process
\be
\gamma^*(q)+N(p_1) \rightarrow \pi(p_\pi) +N(p_2)
\label{Direct_channel_reaction}
\ee
in the generalized Bjorken limit ($Q^2=-q^2$ -- large ; $Q^2/(2 p_1 \cdot q)$ -- fixed )
in the so-called backward region
$ |u| \equiv |(p_\pi-p_1)^2| \ll Q^2$
in terms of
 $\pi N$
TDAs. It is worth   emphasizing that such kinematical regime essentially differs from the more conventional limit
$-t \equiv -(p_2-p_1)^2 \ll Q^2$
in which the factorization theorem for hard exclusive electroproduction of pion off a nucleon
\cite{Collins:1996fb}
applies for
(\ref{Direct_channel_reaction}).
In this later case the description of
(\ref{Direct_channel_reaction})
involves standard nucleon generalized parton distributions (GPDs).

We introduce the standard  Mandelstam variables for the reaction
(\ref{Direct_channel_reaction}):
\be
&&
s=(p_1+q)^2\,; \ \    u=(p_\pi - p_1)^2\,; \ \   t=(p_2-p_1)^2\,.
\nonumber \\ &&
\ee
Therefore, the $t$-channel of
(\ref{Direct_channel_reaction})
corresponds to an exchange with quantum numbers of a meson while in the $u$-channel
an intermediate state with baryon quantum numbers is involved.

Throughout this paper we adopt a reference frame in which the three-momenta
$\vec{q}$
and
$\vec{p_1}$
have only a third component. We define the
light-cone vectors $p$ and $n$ such that $2 p \cdot n=1$ and introduce
standard kinematical quantities: average momentum  $P=\frac{1}{2}(p_1+p_\pi)$,
momentum transfer
$\Delta=p_\pi-p_1$
and its transverse component $\Delta_T$. The skewness parameter $\xi$ is defined with respect to the
$u$-channel momentum transfer in the
usual way $\xi=- \frac{\Delta \cdot n}{2 P \cdot n}$.
The detailed description of kinematics of (\ref{Direct_channel_reaction}) in the backward regime
is presented in \cite{Lansberg:2007ec}.

The definition of the leading twist-$3$ $\pi N$ TDA involved in the description of
the reaction
(\ref{Direct_channel_reaction})
in the backward regime can be symbolically written as
\be
&&
4(P \cdot n)^3 \int   \left[ \prod_{j=1}^3 \frac{d \lambda_j}{2 \pi}   \right]
e^{i \sum_{k=1}^3 x_k \lambda_k (P \cdot n)}
\langle \pi_a(p_\pi)| \widehat{O}_{\rho \, \tau \, \chi}^{\alpha \beta \gamma}( \lambda_1 n, \,\lambda_2 n, \, \lambda_3 n )| N_\iota(p_1)  \rangle
\nonumber \\ &&
=\delta(x_1+x_2+x_3-2 \xi) \; \sum_{s.f.} (f_a)^{\alpha \beta \gamma}_\iota s_{\rho \, \tau, \, \chi} H^{(\pi N)}_{s.f.}(x_1,x_2,x_3,\xi, \Delta^2)\,.
\label{Formal_definition_TDA}
\ee
The spin-flavor ($s.f.$) sum in (\ref{Formal_definition_TDA}) stands over all relevant independent
flavor structures $(f_a)^{\alpha \beta \gamma}_\iota$
and the Dirac structures
$s_{\rho \, \tau, \, \chi}$.  The detailed account
of the Dirac and flavor structure occurring in
(\ref{Formal_definition_TDA})
is given in Sec.~{\ref{Section_polynomiality}} and Sec.~\ref{Section_isospin_TDAs}.

Nucleon to pion TDAs are conceptually much related to pion-nucleon generalized distribution amplitudes (GDAs)
\cite{BLP1,BLP2}
which are defined through the cross-conjugated matrix element of the same three-quark operator
(\ref{Def_operator}). Indeed, a similar correspondance was established
between pion GPD and $2 \pi$ GDA \cite{Polyakov:1998ze,Polyakov:1999gs}.

Therefore, it is natural  to simultaneously consider the cross-conjugated
($p_\pi' \leftrightarrow -p_\pi$, $q' \leftrightarrow -q$)
reaction:
\be
\pi(p_\pi')+ N(p_1)   \rightarrow \gamma^*(q')+N(p_2)\,.
\label{Cross_channel_reaction}
\ee
The formal definition of $\pi N$ GDA that arises in the description of
(\ref{Cross_channel_reaction})
reads
\be
&&
4(P' \cdot n)^3 \int   \left[ \prod_{j=1}^3 \frac{d \lambda_j}{2 \pi}   \right]
e^{i \sum_{k=1}^3 y_k \lambda_k (P' \cdot n)}
\langle 0| \widehat{O}_{\rho \, \tau \, \chi}^{\alpha \beta \gamma}( \lambda_1 n, \,\lambda_2 n, \, \lambda_3 n )| N_\iota(p_1) \pi_a(p_\pi') \rangle
\nonumber \\ &&
=\delta(y_1+y_2+y_3-1) \;\sum_{s.f.}  (f_a)^{\alpha \beta \gamma}_\iota s'_{\rho \, \tau, \, \chi} \Phi^{(\pi N)}_{s.f.}(y_1,y_2,y_3,\zeta, {P'}^2)\,,
\label{Formal_definition_GDA}
\ee
where
we introduce %the usual GDA notation
$P'=p_1+p_\pi'$ for the total momentum of
$\pi N$
state and $\Delta'=p_1-p_\pi'$. The variable $\zeta= \frac{p_1 \cdot n}{P' \cdot n}$
characterizes the distribution of the plus momenta of the
$\pi N$
system.
We choose the Dirac structures in
(\ref{Formal_definition_GDA})
$s'_{\rho \, \tau, \, \chi}$
as being given by crossing of $s_{\rho \, \tau, \, \chi}$ in (\ref{Formal_definition_TDA}).
$\pi N$ TDA and GDA are interrelated
by a crossing transformation
\be
P' \leftrightarrow -\Delta\,; \ \ \ \Delta' \leftrightarrow 2P
\ee
and analytic continuation in the appropriate kinematical variables:
\be
P'^2 \leftrightarrow \Delta^2\,; \ \ \ 2\zeta+1 \leftrightarrow \frac{1}{\xi}\,; \ \ \  y_i \leftrightarrow \frac{x_i}{2 \xi}\,, \ \ i=\{1,\,2,\,3\}\,.
\ee

The physical domain in $(\Delta^2,\,\xi)$-plane for both the direct channel
(\ref{Direct_channel_reaction})
and cross-conjugated
(\ref{Cross_channel_reaction})
reactions is determined by the requirement that the transverse momentum transfer $\Delta_T=-P'_T$ should be spacelike:
\be
\Delta_T^2= \frac{1-\xi}{1+\xi}\left( \Delta^2-2 \xi \left[ \frac{M^2}{1+\xi} - \frac{m^2}{1-\xi} \right] \right) \le 0\,.
\label{Delta_t2}
\ee
Here $M$ and $m$ stand for nucleon and pion masses respectively.

\begin{figure}[H]
 \begin{center}
 \epsfig{figure=  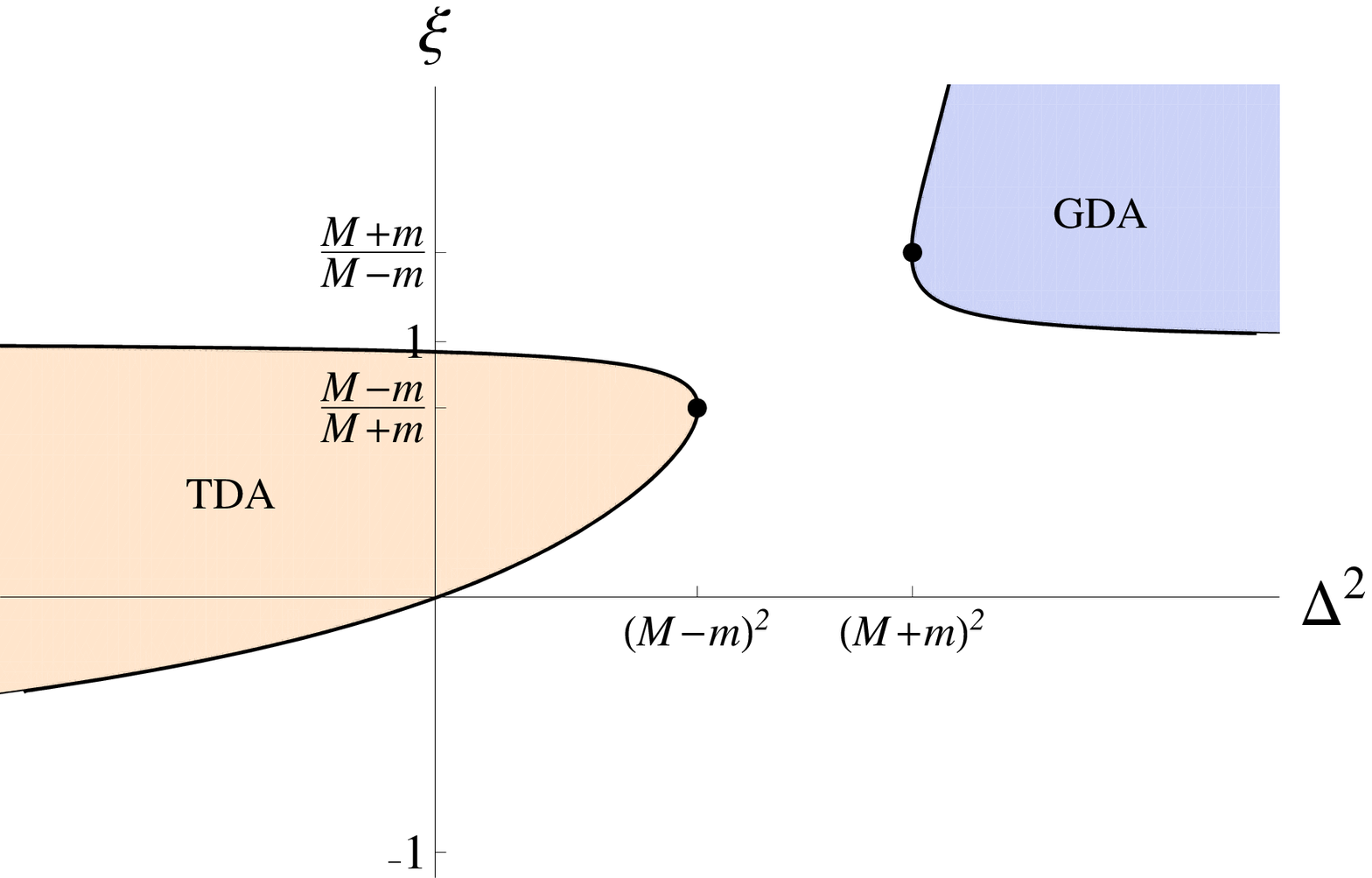 , height=4.5cm}
  \epsfig{figure=  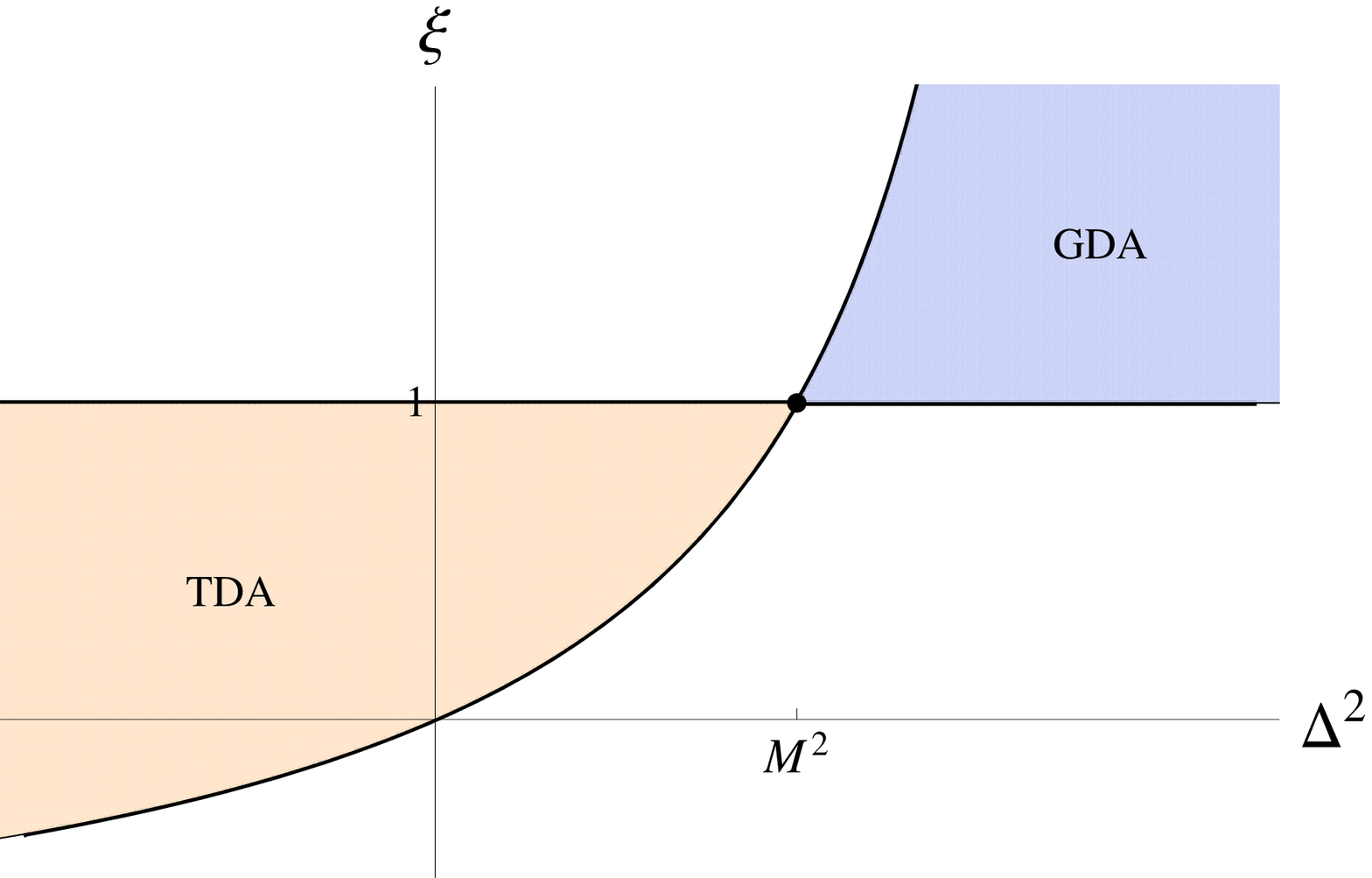 , height=4.5cm}
  \caption{Physical domains (bounded by the condition $\Delta_T^2 \le 0$) in $(\Delta^2, \xi)$ plane for $\pi N$ TDAs for the case  $m \ne 0$   (left panel)
  and in the chiral limit $m=0$ (right panel). }
\label{Fig1}
\end{center}
\end{figure}

On the left panel of Fig.~\ref{Fig1} we show physical domains for the reactions (\ref{Direct_channel_reaction})
and (\ref{Cross_channel_reaction}) for physical pion mass.
One may distinguish  two regimes: the direct channel regime with its threshold at
$\Delta^2 = (M-m)^2$ and the cross-channel one with its threshold at
$\Delta^2 = (M+m)^2$.
The upper (lower) branch of the curve bordering the physical domain in the direct channel regime
tends to $\xi=1$ ($\xi=-1$) when $\Delta^2 \rightarrow -\infty$.
Note that the physical domain of the direct channel of
(\ref{Direct_channel_reaction})
includes both negative and positive values of $\Delta^2$.
Moreover, in the chiral limit ($m=0$) the two thresholds stick together (see the right panel of Fig.~\ref{Fig1}).
We exploit this fact later in Sec.~\ref{Section_Soft_pion} in order to work out the normalization for $\pi N$ TDAs.

\begin{figure}[H]
 \begin{center}
 \epsfig{figure=  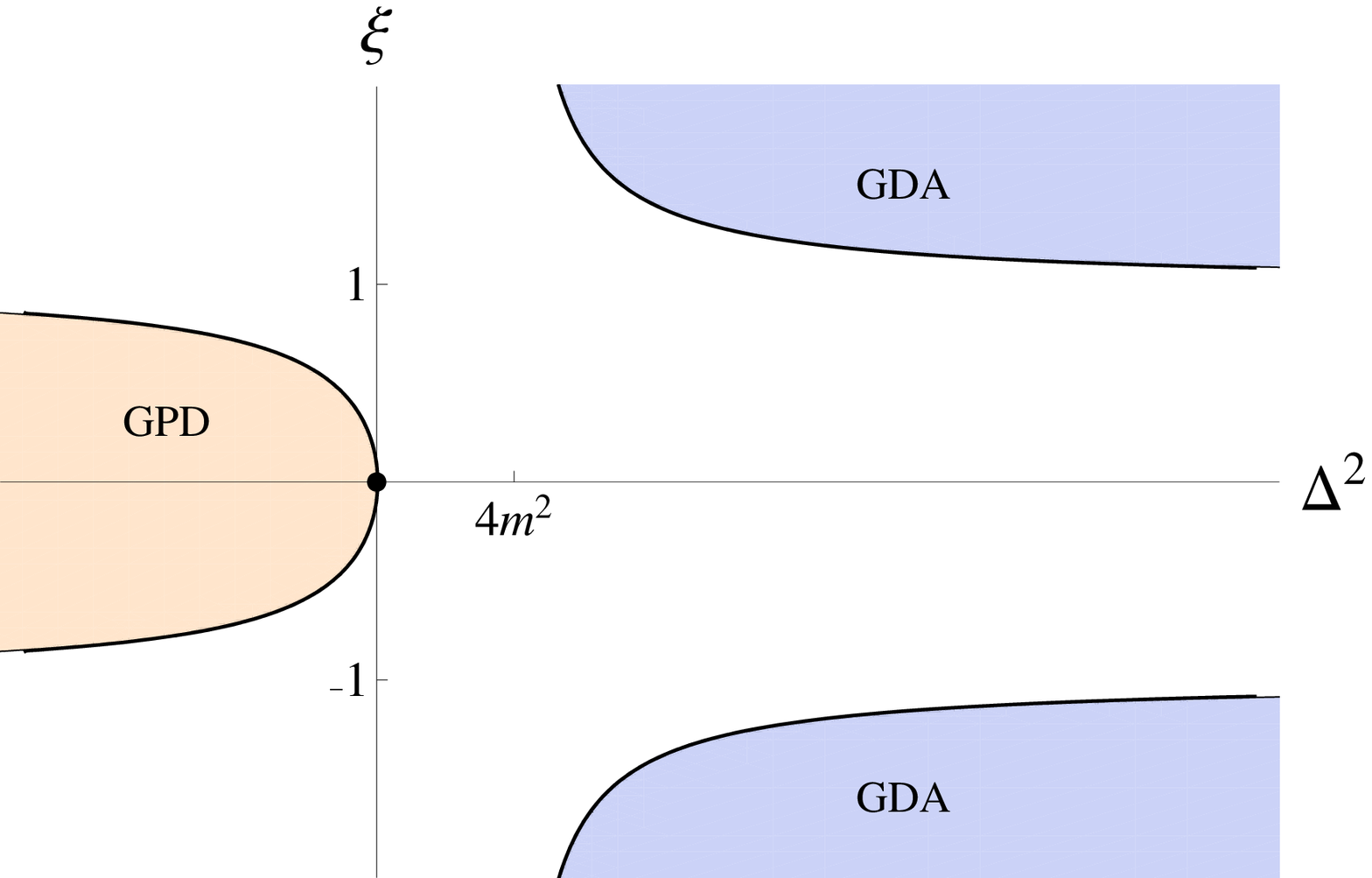 , height=4.5cm}
  \epsfig{figure=  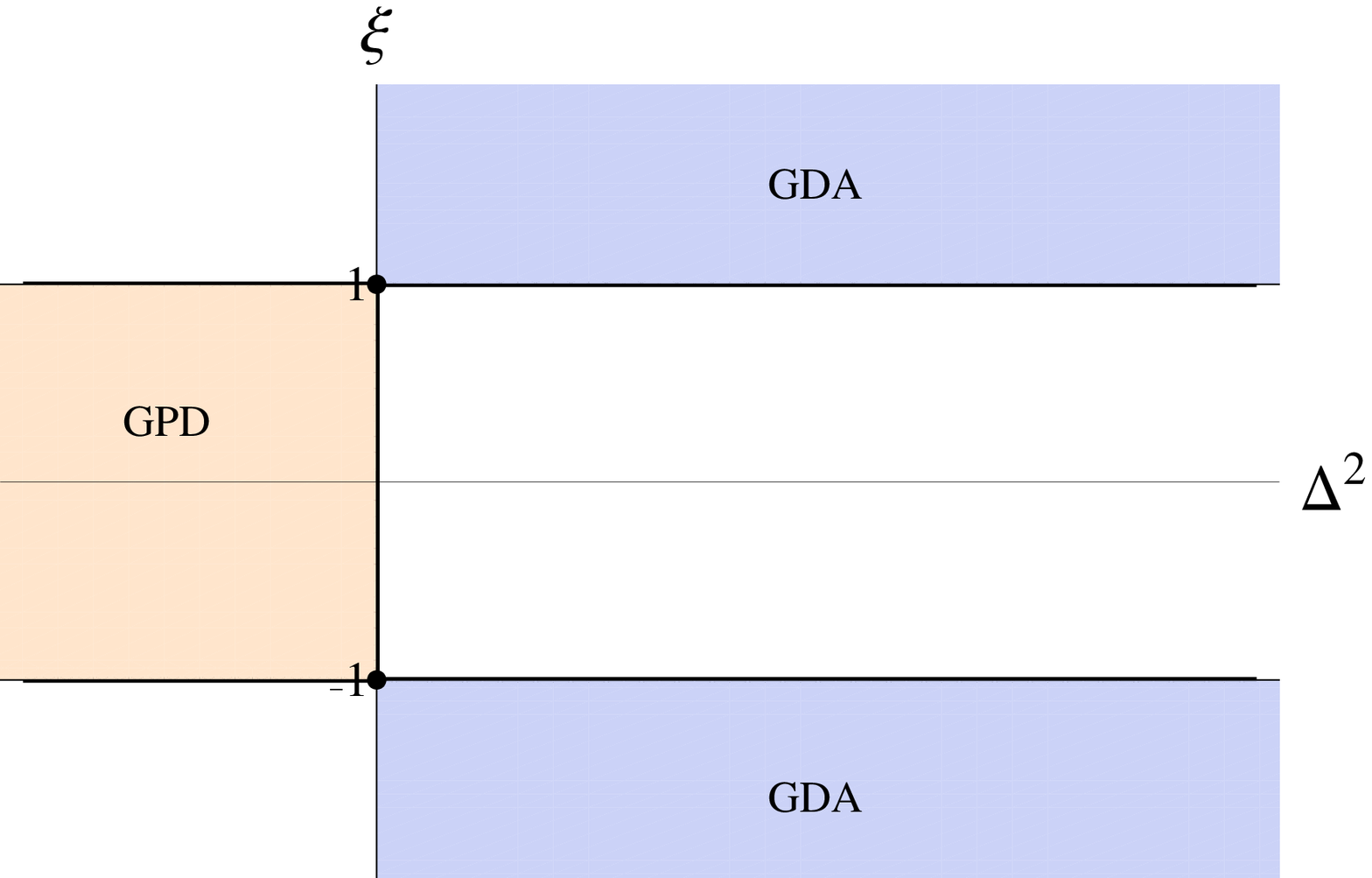 , height=4.5cm}
  \caption{Physical domains in $(\Delta^2, \xi)$ plane for pion GPD and $2 \pi$ GDA for the case  $m \ne 0$ (left panel) and in the
  chiral limit  $m = 0$ (right panel). }
\label{Fig2}
\end{center}
\end{figure}

It is interesting to compare Fig.~\ref{Fig1} to that in the case of equal masses
of particles in
$|{\it in} \rangle$
and
$\langle {\it out}|$
states.
On Fig.~\ref{Fig2} we show the  physical domains in
$(\Delta^2,\,\xi)$-plane  for pion GPD  and $2 \pi$ GDA occurring in the description of
$\gamma^* \pi \rightarrow \gamma \pi$ and $\gamma^* \gamma \rightarrow \pi \pi$
(here
$\Delta$
refers to the momentum transfer between
initial and final pions in
$\gamma^* \pi \rightarrow \gamma \pi$
and the usual definition of
$\xi$
is assumed).
In particular, the physical domains for  pion GPD  and
$2 \pi$
GDA are symmetric under the reflection of
$\xi$. The difference between Fig.~\ref{Fig1} and Fig.~\ref{Fig2} has purely kinematical origin and is not related
to the nature of  QCD operator in the matrix element
in question.

The basic issue of the approach based on $\pi N$ TDAs is that, contrary to the  GPD case, $\pi N$ TDAs
lack an intelligible forward limit $\xi\rightarrow0$.
However, the opposite limit
$\xi \rightarrow 1$
turns out to be very illuminating. For simplicity, let us consider the pion to be massless.
The point
$\xi=1$, $\Delta^2=M^2$
(corresponding to both the direct and cross-channel threshold)
belongs both to the physical regions for $\pi N$ GDAs and TDAs.
Moreover, it is for this very point that the   soft pion theorem
\cite{Pobylitsa:2001cz}
applies for
$\pi N$
GDAs. As argued in
\cite{BLP1,BLP2},
this allows to constrain
$\pi N$
GDAs at the threshold in terms of the nucleon DA.
In the chiral limit the   soft pion theorem for GDAs also constrains $\pi N$ TDAs
exactly  as the  soft pion theorem \cite{Polyakov:1998ze} for $2 \pi$ GDA in the chiral limit links the
isovector pion GPD at $\xi=1$, $\Delta^2=0$ to the pion DA.
Thus, in the chiral limit the soft pion theorem provides us with the desired reference point for
$\pi N$ TDAs. This valuable information may be used as input for realistic modeling of $\pi N$ TDAs based on the
spectral representation in terms of quadruple distributions
\cite{Pire:2010if}.
In
\cite{Cross-sec-paper}
we will argue that a possible approach  consists in evolving from the $\xi=1$ limit for $\pi N$ TDAs through
a procedure analogous to the one used for GPDs;
in this latter case one  employs the forward limit $\xi=0$ to constrain GPDs through the successful Radyushkin's factorized Ansatz
\cite{RDDA4}.

It is worth to mention that the simpler case of $\pi  \gamma$ TDAs has already been discussed in details
\cite{pigamTDA1,pigamTDA2,pigamTDA3,pigamTDA4,pigamTDA5}.
These TDAs share many features with $\pi N$ TDAs, and are also subject to
chiral symmetry constraints. Since the operator in this case is the same as in usual GPDs,
polynomiality properties and isospin relations are straightforwardly extended from
one case to the other.

In this paper, we analyze  the constraints on $\pi N$ TDAs imposed by the symmetries of QCD.
First, we argue that the
underlying Lorentz symmetry results in the polynomiality conditions which restrict the
skewness parameter dependence of TDAs in a way similar to the well known GPD case.
Second, we analyze in details the isospin decomposition of
$\pi N$ TDAs and establish the consequences of the isotopic and permutation symmetries.
Third, we exploit the chiral symmetry of QCD to
calculate  $\pi N$ GDAs and TDAs in the soft pion limit. Finally, we show how a model based
on nucleon and $\Delta(1232)$ exchanges satisfies the revealed polynomiality and isospin
constrains. The paper is organized as follows:
\begin{itemize}
\item In Sec.~\ref{Section_polynomiality} we introduce the new parametrization for $\pi N$ TDAs and show that,
within this parametrization,  $\pi N$ TDAs satisfy the polynomiality conditions.
\item In Sec.~\ref{Section_Isopin_par_DA} we consider the isospin structure of the three-quark operator
and describe the general isospin parametrization of the leading twist baryon
DAs. Next, we consider the consequences of isotopic and permutation symmetries of baryon DAs.
We rederive the familiar isospin identities for the leading twist baryon DAs.
\item In Sec.~\ref{Section_isospin_TDAs} we apply the isospin formalism to the case of $\pi N$ TDAs and GDAs and derive the set of symmetry
relations for $\pi N$ TDAs and GDAs.
\item In  Sec.~\ref{Section_Soft_pion} we derive the   soft pion theorem for  $\pi N$ GDAs and discuss its
consequences for   $\pi N$ TDAs.
\item Sec.~\ref{Section_u-channel_resonance} contains the calculation of $u$-channel nucleon and $\Delta(1232)$ exchange contributions
into $\pi N$ TDAs.
\item Our conclusions are presented in Sec.~\ref{Section_conclusions}.
\end{itemize}

Let us stress that the main goal of the present paper is to  provide the basic formalism for a consistent modelling of TDAs.
The phenomenological applications of this formalism will be addressed in forthcoming  publications.
Measuring pion electroproduction at large angle is a challenging experimental problem. Some preliminary data
are already available from J-Lab \cite{J-lab} and more data are expected from J-Lab at $12$ GeV.
A detailed proposal for measuring the reaction $\bar{p} p \to \gamma^* \pi N$
exists in the PANDA Physics program \cite{PANDA}.

\section{Polynomiality  property of $\pi N$ TDAs}
\label{Section_polynomiality}

In this section our goal is to show that, analogously to
GPDs, nucleon to meson TDAs satisfy the polynomiality property
{\it i.e.} their Mellin moments in longitudinal momentum
fractions $x_i$ are polynomials of variable
$\xi$
of definite power. For definiteness we are going to consider the
case of nucleon to pion TDAs. In this section we omit
flavor indices in the
operator
$\widehat{O}_{\rho \, \tau \, \chi}$
(\ref{Def_operator})
since  flavor symmetry is irrelevant for
the present problem.

It turns out necessary to change the parametrization of $\pi N$ TDA earlier proposed in
Refs.~\cite{Lansberg:2007ec,Pire:2010if}.
The important drawback of our initial parametrization
is that it involves the set of the Dirac structures which leads to spoiling of the polynomiality property of TDAs
by the  kinematical factors $\frac{1}{1+\xi}$.

In order to get rid of these kinematical singularities
we suggest the following parametrization of the leading twist-$3$ $\pi N$ TDAs%
\footnote{Throughout the text we employ Dirac's ``hat'' notation:
$\hat{a} \equiv \gamma_{\mu} a^{\mu}$\,.
The following conventions are adopted:
$\sigma^{\mu \nu}= \frac{1}{2} [\gamma^\mu, \, \gamma^\nu]$;
$\sigma^{v \nu} \equiv v_\mu \sigma^{\mu \nu}$,
 where $v_\mu$ is an arbitrary $4$-vector.}:
\be
&&
4 \mathcal{F} \langle \pi (p_\pi) | \widehat{O}_{\rho \, \tau \, \chi}( \lambda_1 n, \,\lambda_2 n, \, \lambda_3 n )| N (p_1) \rangle=
\delta(x_1+x_2+x_3-2 \xi)
\nonumber \\ &&
 \times i \frac{f_N}{f_\pi M}     \big[ V_1^{ \pi N}(x_1,x_2,x_3,\xi,\Delta^2)  (\hat{P}C)_{\rho \, \tau } (\hat{P} U)_\chi+
   A_1^{\pi N} (x_1,x_2,x_3,\xi,\Delta^2) (\hat{P} \gamma^5 C)_{\rho \, \tau } (\gamma^5 \hat{P} U)_\chi \nonumber \\ &&
+
  T_1^{\pi N} (x_1,x_2,x_3,\xi,\Delta^2) (\sigma_{P \mu} C)_{\rho \, \tau } (\gamma^\mu \hat{P} U)_\chi
\nonumber \\ &&
+    V_2^{\pi N} (x_1,x_2,x_3,\xi,\Delta^2) (\hat{P}C)_{\rho \, \tau }  (\hat{\Delta}  U )_\chi
+
  A_2^{\pi N} (x_1,x_2,x_3,\xi,\Delta^2)(\hat{P} \gamma^5 C)_{\rho \, \tau } (\gamma^5 \hat{\Delta}  U )_\chi \nonumber \\ &&
+
 T_2^{\pi N} (x_1,x_2,x_3,\xi,\Delta^2) (\sigma_{P \mu} C)_{\rho \, \tau } (\gamma^\mu \hat{\Delta}  U )_\chi
+
\frac{1}{M}  T_3^{\pi N} (x_1,x_2,x_3,\xi,\Delta^2) (\sigma_{P \Delta } C)_{\rho \, \tau } (\hat{P} U)_\chi
 \nonumber \\ && +
\frac{1}{M}  T_4^{\pi N} (x_1,x_2,x_3,\xi,\Delta^2) (\sigma_{P \Delta } C)_{\rho \, \tau } (\hat{\Delta}   U )_\chi \big]\,,
\label{Decomposition_piN_TDAs_new}
\ee
where
$\mathcal{F}$
stands for the Fourier transform
\be
\mathcal{F} \equiv \mathcal{F}(x_1, \, x_2, \, x_3)(...)= (P \cdot n)^3 \int   \left[ \prod_{j=1}^3 \frac{d \lambda_j}{2 \pi}   \right]
e^{i \sum_{k=1}^3 x_k \lambda_k (P \cdot n)} (...)\,;
\label{Fourier_tr}
\ee
$f_\pi$ is the pion weak decay constant and $f_N$ is a constant with the dimension of energy squared;
$U$ is the usual Dirac spinor and $C$ is the charge conjugation matrix.

The price for avoiding the kinematical singularities in the invariant amplitudes in
(\ref{Decomposition_piN_TDAs_new})
is that apart from the leading twist contribution we have to keep some admixture
of subleading twist (see Appendix~\ref{app_c}).
The relationship between the parametrization involving pure twist-$3$ invariant amplitudes
employed in Refs.~\cite{Lansberg:2007ec,Pire:2010if} and that of
Eq.~(\ref{Decomposition_piN_TDAs_new})
is given by Eq.~(\ref{Old_to_new}).

We introduce the following notations for the leading twist Dirac structures occurring in
(\ref{Decomposition_piN_TDAs_new}):
\be
&&
(v_1^{\pi N})_{\rho \tau, \, \chi}= (\hat{P}C)_{\rho \tau} (\hat{P} U)_\chi\,;
\ \
(a_1^{\pi N})_{\rho \tau, \, \chi}=(\hat{P} \gamma^5 C)_{\rho \tau} (\gamma^5 \hat{P} U)_\chi\,;
\ \
(t_1^{\pi N})_{\rho \tau, \, \chi}=(\sigma_{P \mu} C)_{\rho \tau} (\gamma^\mu \hat{P} U)_\chi\,;
\nonumber \\ &&
(v_2^{\pi N})_{\rho \tau, \, \chi}=(\hat{P}C)_{\rho \tau}  (\hat{\Delta}  U)_\chi\,;
\ \
(a_2^{\pi N})_{\rho \tau, \, \chi}=(\hat{P} \gamma^5 C)_{\rho \tau} (\gamma^5 \hat{\Delta}  U )_\chi\,;
\ \
(t_2^{\pi N})_{\rho \tau, \, \chi}=(\sigma_{P \mu} C)_{\rho \tau} (\gamma^\mu \hat{\Delta}  U)_\chi\,;
\nonumber \\ &&
(t_3^{\pi N})_{\rho \tau, \, \chi}= \frac{1}{M} (\sigma_{P \Delta } C)_{\rho \tau} (\hat{P} U)_\chi\,;
\ \
(t_4^{\pi N})_{\rho \tau, \, \chi}=\frac{1}{M} (\sigma_{P \Delta } C)_{\rho \tau} (\hat{\Delta}   U)_\chi\,.
\label{Dirac_structures_PiN_TDA}
\ee
We also employ the shortened notation for the whole set of twist-$3$ Dirac structures:
\be
(s^{\pi N})_{\rho \tau, \, \chi}=
\big\{
(v_{1,2}^{\pi N})_{\rho \tau, \, \chi},\,
(a_{1,2}^{\pi N})_{\rho \tau, \, \chi},
(t_{1,2,3,4}^{\pi N})_{\rho \tau, \, \chi}
\big\}
\ee
and for the corresponding invariant amplitudes
\be
H_s^{\pi N}=\big\{V_{1,2}^{\pi N},\, A_{1,2}^{\pi N}, \, T_{1,2,3,4}^{\pi N} \big\}\,.
\ee

Each invariant amplitude $V_{1,2}^{\pi N}$, $A_{1,2}^{\pi N}$,  $T_{1,2,3,4}^{\pi N}$
of (\ref{Decomposition_piN_TDAs_new})
is a function of the longitudinal momentum fractions
$x_i$ ($i=\{1,\,2,\,3\}$),
skewness parameter
$\xi= -\frac{(\Delta \cdot n)}{2 (P \cdot n)}$
and the momentum transfer squared
$\Delta^2$. The support properties of $\pi N$ TDAs in the longitudinal momentum
fractions $x_i$ were established in
\cite{Pire:2010if}.

Now we are going to demonstrate that the $\pi N$ TDAs defined in
(\ref{Decomposition_piN_TDAs_new})
satisfy the polynomiality property.
Our demonstration generally repeats the usual way of arguing for the
case of GPDs (see {\it e.g.} \cite{Diehl}).

The
$(n_1,n_2,n_3)$-th ($n_1+n_2+n_3=N$)
Mellin moments of TDAs in $x_1$, $x_2$, $x_3$
lead to derivative operations acting on three quark fields:
\be
&&
4 (P \cdot n)^{n_1+n_2+n_3+3} \int d^3x   \, x_1^{n_1}  x_2^{n_2}  x_3^{n_3}  \nonumber \\ &&
% (p \cdot n)^3
 \int   \left[ \prod_{k=1}^3 \frac{d \lambda_k}{2 \pi}   \right]
e^{i \sum_{k=1}^3 x_k \lambda_k (P \cdot n)}
\langle \pi(P + \frac{\Delta}{2}) | %\Psi_\alpha ( \lambda_1 n) \Psi_\beta ( \lambda_2 n)  \Psi_\gamma ( \lambda_3 n)
\widehat{O}_{\rho \, \tau \, \chi}( \lambda_1 n, \,\lambda_2 n, \, \lambda_3 n )
| N(P - \frac{\Delta}{2}) \rangle
\nonumber \\ &&
=(P \cdot n)^{n_1+n_2+n_3 }
 \frac{i f_N}{f_\pi M}
 %
 %\times
\sum_{s
%{\rm Dirac \atop structures \; }
}
(s^{\pi N})_{\rho \tau, \, \chi}
%\int d^3x
\int_{-1+\xi}^{1+\xi}  dx_1 \int_{-1+\xi}^{1+\xi}  dx_2 \int_{-1+\xi}^{1+\xi}  dx_3
\, x_1^{n_1}  x_2^{n_2}  x_3^{n_3}
\nonumber \\ && \times
\delta(x_1+x_2+x_3-2 \xi) H_s^{\pi N}(x_1, x_2, x_3, \xi,\Delta^2)
\nonumber \\ &&
=
4(-1)^{n_1+n_2+n_3}
\langle \pi(P + \frac{\Delta}{2}) |
 \left[ (i \vec{\partial}^+)^{n_1} \Psi_\rho( 0)  \right]
\left[ (i \vec{\partial}^+)^{n_2}  \Psi_\tau( 0)  \right]
\left[ (i \vec{\partial}^+)^{n_3}  \Psi_\chi( 0)  \right]
 | N(P - \frac{\Delta}{2}) \rangle\,.
 \nonumber \\ &&
 \label{Moment_step_X}
\ee
Hence,  the Mellin moments of nucleon to meson TDAs  are expressed through the
form factors of the local twist-$3$ operators:
\be
\widehat{O}_{\rho \tau \chi }^{\; \mu_1...\mu_{n_1}, \, \nu_1...\nu_{n_2}, \, \lambda_1...\lambda_{ n_3}}(0)=
\left[ i\vec{D}^{\mu_1}...\,i\vec{D}^{\mu_{n_1}}  \Psi_\rho \right]
\left[ i\vec{D}^{\nu_1}...\,i\vec{D}^{\nu_{n_2}}  \Psi_\tau \right]
\left[ i\vec{D}^{\lambda_1}...\,i\vec{D}^{\lambda_{n_3}}  \Psi_\chi \right]\,,
\label{local_op_derivatives}
\ee
where $\vec{D}^\mu = \vec{\partial}^\mu -\frac{i g}{2} A^{l \, \mu} \lambda^l $
is the covariant derivative ($ \lambda^l$ stand  here for the Gell-Mann matrices).
Note that in (\ref{Moment_step_X}), (\ref{local_op_derivatives}) we omit color indices.

%We employ the following conventions:
%lettres from the middle of the Greek alphabet are reserved for the Lorentz indices $  \mu_i, \, \nu_i, \, \lambda_i  ={0,\,1,\,2,\,3}$;
%letters from the beginning of the Greek alphabet are reserved for the Dirac indices $ \alpha, \, \beta, \, \gamma=\{1,2,3,4\}$

Introducing the shortened notation
\be
(\Delta^\mu)^i (P^\mu)^{n_1-i} \equiv \Delta^{\mu_1}...\Delta^{\mu_i} P^{\mu_{i+1}}... P^{\mu_{n_1}}
\ee
we  write down the following parametrization for the $\pi N$ matrix element of the local
operator
(\ref{local_op_derivatives}):
\be
&&
4 \langle \pi |
\widehat{O}_{\rho \tau \chi}^{ \;\mu_1...\mu_{n_1}, \, \nu_1...\nu_{n_2}, \, \lambda_1...\lambda_{ n_3}}(0)
| N \rangle
=i \frac{f_N}{f_\pi M}
\nonumber \\ &&
\times \Big[\sum_{s }
(s^{\pi N})_{\rho \tau, \, \chi}
\sum_{i=0}^{n_1} \sum_{j=0}^{n_2}  \sum_{k=0}^{n_3} A^{s\; (n_1, n_2, n_3)}_{ijk}(\Delta^2)
(\Delta^\mu)^i (P^\mu)^{n_1-i}
(\Delta^\nu)^j (P^\nu)^{n_2-j}
(\Delta^\lambda)^i (P^\lambda)^{n_3-i}
\nonumber \\ &&
+\left\{
%(v_1)_{\rho \tau \chi}
(\hat{\Delta}C)_{\rho \tau} (\hat{P} U)_\chi
C^{V_1\; (n_1, n_2, n_3)}_{N+1}(\Delta^2)
+ %(v_2)_{\rho \tau \chi}
(\hat{\Delta}C)_{\rho \tau} (\hat{\Delta} U)_\chi
C^{V_2\; (n_1, n_2, n_3)}_{N+1}(\Delta^2)
\right.
\nonumber \\ &&
\left.
+ %(a_1)_{\rho \tau \chi}
(\hat{\Delta} \gamma^5 C)_{\rho \tau} (\gamma^5 \hat{P} U)_\chi
C^{A_1\; (n_1, n_2, n_3)}_{N+1}(\Delta^2)
+ %(a_2)_{\rho \tau \chi}
(\hat{\Delta} \gamma^5 C)_{\rho \tau} (\gamma^5 \hat{\Delta} U)_\chi
C^{A_2\; (n_1, n_2, n_3)}_{N+1}(\Delta^2)
\right.
\nonumber \\ &&
\left.
+ %(t_1)_{\rho \tau \chi}
(\sigma_{\Delta \mu} C)_{\rho \tau} (\gamma^\mu \hat{P} U)_\chi
C^{T_1\; (n_1, n_2, n_3)}_{N+1}(\Delta^2)
+ %(t_2)_{\rho \tau \chi}
(\sigma_{\Delta \mu} C)_{\rho \tau} (\gamma^\mu \hat{\Delta} U)_\chi
C^{T_2\; (n_1, n_2, n_3)}_{N+1}(\Delta^2)
\right\}
\nonumber \\ &&
\times (\Delta^\mu)^{n_1} (\Delta^\nu)^{n_2} (\Delta^\lambda)^{n_3}
\Big]\,,
\label{FF_decomposition_operator}
\ee
where
the sum in the first term
is over all independent Dirac structures (\ref{Dirac_structures_PiN_TDA});
$A^{s\; (n_1, n_2, n_3)}_{ijk}(\Delta^2)$
and
$C^{V_{1,2}, \,A_{1,2}, \, T_{1,2}\; (n_1, n_2, n_3)}_{N+1}(\Delta^2)$
denote the appropriate invariant form factors.

We introduce the compact notation for the Mellin moments of TDAs:
\be
&&
\langle x_1^{n_1} x_2^{n_2} x_3^{n_3} H_s^{\pi N} \rangle \nonumber \\ && =
\int_{-1+\xi}^{1+\xi}  dx_1 \int_{-1+\xi}^{1+\xi}  dx_2 \int_{-1+\xi}^{1+\xi}  dx_3
\delta(x_1+x_2+x_3-2\xi)
 x_1^{n_1} x_2^{n_2} x_3^{n_3}
H_s^{\pi N}(x_1,x_2,x_3,\xi,\Delta)\,.
\nonumber \\ &&
\ee
Now from
(\ref{FF_decomposition_operator})
we establish the following relations for
$(n_1,\,n_2,\,n_3)$-th ($n_1+n_2+n_3=N$) Mellin moments
of TDAs:
\be
&&
\langle x_1^{n_1} x_2^{n_2} x_3^{n_3} \{V_{1,2}, \, A_{1,2}, \, T_{1,2} \} \rangle
\nonumber \\ &&
 =
  %\big(
  \sum_{n=1}^{N} (-1)^{N-n}
(2\xi)^n
\sum_{i=0}^{n_1} \sum_{j=0}^{n_2}  \sum_{k=0}^{n_3}
\delta_{i+j+k,\,n} \;
A^{\{V_{1,2}, \, A_{1,2}, \, T_{1,2} \} \; (n_1, n_2, n_3)}_{ijk}(\Delta^2)
\nonumber \\ &&
-  (2 \xi)^{N+1} C^{\{V_{1,2}, \, A_{1,2}, \, T_{1,2} \} \; (n_1, n_2, n_3)}_{N+1}(\Delta^2) %\big)
\,;
\nonumber \\ &&
\nonumber \\ &&
\langle x_1^{n_1} x_2^{n_2} x_3^{n_3} \{  T_{3,4} \} \rangle
\nonumber \\ &&
 =
  %\big(
  \sum_{n=1}^{N} (-1)^{N-n}
(2\xi)^n
\sum_{i=0}^{n_1} \sum_{j=0}^{n_2}  \sum_{k=0}^{n_3}
\delta_{i+j+k,\,n} \;
A^{\{T_{3,4} \} \; (n_1, n_2, n_3)}_{ijk}(\Delta^2)\,.
\nonumber \\ &&
\ee
Thus, we conclude that the $\pi N$ TDAs defined in
(\ref{Decomposition_piN_TDAs_new}) indeed satisfy the polynomiality property.
For $n_1+n_2+n_3=N$ the highest power of $\xi$ occurring in $(n_1,n_2,n_3)$-th Mellin moment
of  $\{V_{1,2}^{\pi N}, \, A_{1,2}^{\pi N}, \, T_{1,2}^{\pi N}\}$ is $N+1$ while for
$T_{3,4}^{\pi N}$ it is $N$.
Consequently, the TDAs
$\{V_{1,2}^{\pi N}, \, A_{1,2}^{\pi N}, \, T_{1,2}^{\pi N}\}$
include an analogue of the $D$-term contribution
\cite{Polyakov:1999gs}
which generates the highest possible power of $\xi$.
Note that, exactly as in the case of GPDs, the spectral representation
\cite{Pire:2010if}
cannot produces the highest possible power of $\xi$ in the Mellin moments.
Therefore, the complete parametrization of $\pi N$ TDAs requires adding a separate $D$-term contribution
to the spectral representation or a singular modification of corresponding spectral densities
in the spirit of Ref.~\cite{Radyushkin:2011dh}.

\section{Isospin parametrization for leading twist baryon distribution amplitudes}
\label{Section_Isopin_par_DA}

\subsection{Notes on the operator in question}
\label{Notes_on_the_operator}

Below we review the group-theoretical properties
of the three-quark operator
(\ref{Def_operator}) under the
$SU(2)$
isospin symmetry group.
Throughout the rest of the paper we adopt the following conventions:
\begin{itemize}
\item Letters from the beginning of the Greek alphabet are reserved for the
$SU(2)$ isospin indices
$ \alpha,\,\beta,\,\gamma,\, \iota, \, \kappa  ={1,\,2 }$.
\item We have to distinguish between upper (contravariant) and lower (covariant) $SU(2)$ isospin indices.
We introduce the totally antisymmetric tensor $\varepsilon_{\alpha \beta}$ for lowering indices and $\varepsilon^{\alpha \beta}$
for rising indices ($\varepsilon_{1 \,2}=\varepsilon^{1 \,2}=1$):
$\Psi ^\alpha \varepsilon_{\alpha \beta} = \Psi_\beta$,
$\Psi_\alpha \varepsilon^{\alpha \beta} = \Psi^\beta$
and
$\delta^\alpha_{\; \beta}= -\varepsilon^\alpha_{\; \beta}= \varepsilon_\beta^{\; \, \alpha}$.
\item Letters from the middle of the Greek alphabet $\lambda$, $\mu$, $\nu$ denote the Lorentz indices.
\item Letters from the second half of the Greek alphabet $\rho, \, \tau,\, \chi$ are reserved for the Dirac indices.
\item Letters from the beginning of the Latin alphabet
$a,b,c\,...$
are reserved for indices of the adjoint representation of the $SU(2)$ isospin group.
\item Letters $c_1$, $c_2$, $c_3$ stand for $SU(3)$ color indices.
\end{itemize}
To simplify our formulas we will often skip the color and the Dirac indices when they are irrelevant for the discussion.
We will also often employ the shortened notation for the arguments of the operator (\ref{Def_operator}):
$\widehat{O}^{\alpha \beta \gamma}_{\rho   \tau \chi}(z_1,\,z_2,z_3) \equiv \widehat{O}^{\alpha \beta \gamma}_{\rho   \tau \chi}(1,2,3)$.

The operator (\ref{Def_operator}) transforms according to the
\be
2 \otimes 2 \otimes 2= 4\oplus 2 \oplus 2
\label{tensor_prod_decomp}
\ee
representation of the isospin
$SU(2)$.
To find out the operators transforming according to the isospin-$\frac{3}{2}$ and  isospin-$\frac{1}{2}$
representations we single out the totally symmetric and totally
antisymmetric  parts of (\ref{Def_operator}):
\be
\widehat{O}^{\alpha \beta \gamma}%=\Psi^\alpha \Psi^\beta \Psi^\gamma
=
\widehat{O}^{[\alpha \beta \gamma]}+
\widehat{O}^{\{\alpha \beta \gamma\}}+
\widehat{\widetilde{O}}^{ \alpha \beta \gamma}\,.
\ee
The totally symmetric part
\be
\widehat{O}^{\{\alpha \beta \gamma\}}=
\frac{1}{6}
\left(
\Psi^\alpha \Psi^\beta \Psi^\gamma+
\Psi^\beta \Psi^\alpha \Psi^\gamma+
\Psi^\alpha \Psi^\gamma \Psi^\beta+
\Psi^\beta \Psi^\gamma \Psi^\alpha+
\Psi^\gamma \Psi^\beta \Psi^\alpha+
\Psi^\gamma \Psi^\alpha \Psi^\beta
\right)
\ee
obviously transforms according to isospin-$\frac{3}{2}$ representation.
The totally antisymmetric part
\be
\widehat{O}^{[\alpha \beta \gamma]}=
\frac{1}{6}
\left(
\Psi^\alpha \Psi^\beta \Psi^\gamma-
\Psi^\beta \Psi^\alpha \Psi^\gamma-
\Psi^\alpha \Psi^\gamma \Psi^\beta+
\Psi^\beta \Psi^\gamma \Psi^\alpha-
\Psi^\gamma \Psi^\beta \Psi^\alpha+
\Psi^\gamma \Psi^\alpha \Psi^\beta
\right)
\ee
is zero in $SU(2)$.
The explicit expression for the remaining part
$\widehat{\widetilde{O}}^{\alpha \beta \gamma}$
reads:
\be
\widehat{\widetilde{O}}^{\alpha \beta \gamma}=\frac{1}{3} \left( 2 \Psi^\alpha \Psi^\beta \Psi^\gamma -
\Psi^\beta \Psi^\gamma \Psi^\alpha-
\Psi^\gamma \Psi^\alpha \Psi^\beta
\right)\,.
\ee
One can represent
$\widehat{\widetilde{O}}^{\alpha \beta \gamma}$
as a sum of three operators which are  antisymmetric in pairs of indices
$[\alpha, \beta]$, $[\alpha, \gamma]$ and  $[\beta, \gamma]$:
\be
\widehat{\widetilde{O}}^{\alpha \beta \gamma}= \widehat{O}_1^{[\alpha \beta] \gamma}+ \widehat{O}_2^{ [\alpha \check{\beta}  \gamma] }+
\widehat{O}_3^{\alpha [\beta \gamma]}\,.
\ee
%The notation for the   operator
%$\widehat{O}_2^{ [\alpha \check{\beta}  \gamma]}$
%may be misleading. It is meant that
%$\widehat{\widetilde{O}}^{\alpha \beta \gamma}$
%was antisymmetrized
%in $\alpha, \, \gamma$.
%
The explicit expressions for the operators $\widehat{O}_{1,2,3}^{.[..].}$
read
\be
\hspace*{-1.5cm}
&&
\widehat{O}_1^{[\alpha \beta] \gamma}=
\frac{1}{9}
(2 \Psi^\alpha \Psi^\beta \Psi^\gamma +\Psi^\alpha \Psi^\gamma \Psi^\beta -2 \Psi^\beta \Psi^\alpha \Psi^\gamma
-\Psi^\beta \Psi^\gamma \Psi^\alpha -\Psi^\gamma
   \Psi^\alpha \Psi^\beta +\Psi^\gamma \Psi^\beta \Psi^\alpha )\,; \nonumber \\ \hspace*{-1.5cm} &&
%%%%%%%%%%%%%%%%%%%%%%%%%%%%%%%
 \widehat{O}_2^{ [\alpha \check{\beta}  \gamma] }=
 \frac{1}{9} (2 \Psi^\alpha \Psi^\beta \Psi^\gamma +\Psi^\alpha \Psi^\gamma \Psi^\beta +\Psi^\beta \Psi^\alpha \Psi^\gamma
 -\Psi^\beta \Psi^\gamma \Psi^\alpha -\Psi^\gamma
   \Psi^\alpha \Psi^\beta -2 \Psi^\gamma \Psi^\beta \Psi^\alpha )\,;
   \nonumber \\ \hspace*{-1.5cm} &&
%%%%%%%%%%%%%%%%%%%%%%%%%%%%%%%%%%%%
\widehat{O}_3^{\alpha [\beta \gamma]}=
\frac{1}{9} (2 \Psi^\alpha \Psi^\beta \Psi^\gamma -2 \Psi^\alpha \Psi^\gamma \Psi^\beta +\Psi^\beta \Psi^\alpha \Psi^\gamma
-\Psi^\beta \Psi^\gamma \Psi^\alpha -\Psi^\gamma
   \Psi^\alpha \Psi^\beta +\Psi^\gamma \Psi^\beta \Psi^\alpha ) \,.
      %\nonumber \\ &&
   \label{Operators_O123}
\ee
Contracting operators (\ref{Operators_O123}) with the appropriate $\varepsilon$
tensor we get a spinor transforming according to the fundamental representation of $SU(2)$ .
Note that only two operators in (\ref{Operators_O123})  are independent due to the relation
\be
\varepsilon_{\alpha \beta}  \widehat{O}_1^{[\alpha \beta] \delta }-
\varepsilon_{\alpha \gamma}  \widehat{O}_2^{ [\alpha \check{\delta}  \gamma] }+
\varepsilon_{\beta \gamma} \widehat{O}_3^{\delta [\beta \gamma]}=0\,.
\ee
Thus, in   complete accordance with
(\ref{tensor_prod_decomp}),
the tensor decomposition of the  three-quark operator
(\ref{Def_operator})
%$\widehat{O}^{\alpha \beta \gamma}$
involves two copies of operators transforming according to
the isospin-$\frac{1}{2}$ representations of the isospin group.

\subsection{Case of nucleon DA}
In this subsection we suggest  convenient notations for the leading twist nucleon distribution amplitude.
We introduce  the isospin parametrization for the leading twist nucleon DA and rederive the familiar results
\cite{Korenblit,Chernyak_Nucleon_wave}
for the symmetry properties of the nucleon DA. This technique is applied in the
next subsection  to the analysis of the more involved cases of isospin
structure and symmetry properties of $\Delta(1232)$ DA and $\pi N$ TDAs and GDAs.

The leading twist nucleon DA
\cite{Chernyak_Nucleon_wave}
is defined through the matrix element of the three-quark operator
$\widehat{O}^{\alpha \beta \gamma}_{\rho \tau \chi}(1,2,3)$
between a nucleon state and the vacuum. Being guided by the principle of invariance under $SU(2)$ isospin we can put down
the following isospin decomposition for the matrix element in question:
\be
&&
4 \langle 0 | \widehat{O}^{\alpha \beta \gamma}_{\rho \tau \chi}
%(z_1,\,z_2,\,z_3)
(1,2,3)
| N_\iota(p_N) \rangle
\nonumber \\ &&
= \varepsilon^{\alpha \beta} \delta^\gamma_\iota {M_1^N}_{\rho \tau \chi}
%(z_1,\,z_2,\,z_3)
(1,2,3)
+
\varepsilon^{\alpha \gamma} \delta^\beta_\iota {M_2^N}_{\rho \tau \chi}
%(z_1,\,z_2,\,z_3)
(1,2,3)
%\nonumber \\ &&
+
\varepsilon^{\beta \gamma} \delta^\alpha_\iota {M_3^N}_{\rho \tau \chi}
%(z_1,\,z_2,\,z_3)
(1,2,3)
\,.
%\nonumber \\ &&
\label{nucleon_DA_isospin_dec_1}
\ee
The three  isospin invariant amplitudes are not independent %since each of the three isospin structures in
%(\ref{nucleon_DA_isospin_dec_1})
%can be traded in favor of two other structures with the help of
due to the identity
\be
\varepsilon^{\alpha \beta} \delta^\gamma_\iota+\varepsilon^{\beta \gamma} \delta^\alpha_\iota-\varepsilon^{\alpha \gamma} \delta^\beta_\iota=0\,.
\ee
It is worth   emphasizing that the fact we have only two independent invariant isospin amplitudes meets with the
consequences of the Wigner-Eckart theorem
\cite{Georgi}
for the matrix element of the three-quark operator $\widehat{O}$.
Indeed, as we checked in
Sec.~\ref{Notes_on_the_operator},
the tensor decomposition of the operator
$\widehat{O}$
involves two independent copies of operators transforming according to
the isospin-$\frac{1}{2}$ representations of the isospin group.
However, it should be properly taken into account that the nucleon DA also
possesses specific properties under group of permutations of three quark fields occurring in
the operator $\widehat{O}$.

To address this issue we introduce the following notations for the combinations of the isotopic amplitudes defined in
(\ref{nucleon_DA_isospin_dec_1}) which, as it is demonstrated below,
 are symmetric under permutation of the appropriate quark fields in
the operator
$\widehat{O}^{\alpha \beta \gamma}_{\rho \tau \chi}(z_1,\,z_2,z_3)$:
\be
&&
{M_2^N}_{\rho \tau \chi}(z_1,\,z_2,\,z_3)+{M_3^N}_{\rho \tau \chi}(z_1,\,z_2,\,z_3) \equiv M^{N\,\{12\}}_{\rho \tau \chi}(z_1,\,z_2,\,z_3)\,;
\nonumber \\ &&
{M_1^N}_{\rho \tau \chi}(z_1,\,z_2,\,z_3)-{M_3^N}_{\rho \tau \chi}(z_1,\,z_2,\,z_3) \equiv M^{N\,\{13\}}_{\rho \tau \chi}(z_1,\,z_2,\,z_3)\,;
\nonumber \\ &&
-{M_1^N}_{\rho \tau \chi}(z_1,\,z_2,\,z_3)-{M_2^N}_{\rho \tau \chi}(z_1,\,z_2,\,z_3) \equiv M^{N\,\{23\}}_{\rho \tau \chi}(z_1,\,z_2,\,z_3)\,.
\label{Def_combinations_NDA}
\ee
Note that these combinations satisfy the identity
\be
M^{N\,\{12\}}_{\rho \tau \chi}(z_1,\,z_2,\,z_3)+
M^{N\,\{13\}}_{\rho \tau \chi}(z_1,\,z_2,\,z_3)+
M^{N\,\{23\}}_{\rho \tau \chi}(z_1,\,z_2,\,z_3)=0\,.
\label{Isospin_Id_NDA}
\ee
In fact this is nothing but the familiar isospin identity for nucleon DA.
Indeed, one may check that
\be
&&
4 \langle 0| \widehat{O}^{uud}_{\rho \tau \chi}(z_1,z_2,z_3)|N_p(p_N) \rangle=
-4 \langle 0| \widehat{O}^{ddu}_{\rho \tau \chi}(z_1,z_2,z_3)|N_n(p_N) \rangle=
M^{N\,\{12\}}_{\rho \tau \chi}(z_1,z_2,z_3)\,;
\nonumber \\ &&
4 \langle 0| \widehat{O}^{udu}_{\rho \tau \chi}(z_1,z_2,z_3)|N_p(p_N) \rangle=
-4 \langle 0| \widehat{O}^{dud}_{\rho \tau \chi}(z_1,z_2,z_3)|N_n(p_N) \rangle=
M^{N\,\{13\}}_{\rho \tau \chi}(z_1,\,z_2,\,z_3)\,;
\nonumber \\ &&
4 \langle 0| \widehat{O}^{duu}_{\rho \tau \chi}(z_1,z_2,z_3)|N_p(p_N) \rangle=
-4 \langle 0| \widehat{O}^{udd}_{\rho \tau \chi}(z_1,z_2,z_3)|N_n(p_N) \rangle=
M^{N\,\{23\}}_{\rho \tau \chi}(z_1,\,z_2,\,z_3)\,
\ee
and recover the usual form of the isospin identity from
(\ref{Isospin_Id_NDA}).
Note that the neutron DA differs from that of the proton only by the overall sign, as it is well known.

To derive further symmetry properties of the nucleon DA under permutation of their arguments we employ
the fact that quarks field operators in
(\ref{Def_operator})
anticommute.
This allows in addition to
(\ref{Isospin_Id_NDA})
to establish the following relations for the isospin amplitudes:
\be
&&
M^{N\,\{12\}}_{\rho \tau \chi}(1,\,2,\,3)=M^{N\,\{12\}}_{ \tau \rho \chi }(2,\,1,\,3)\,; \ \ \
M^{N\,\{13\}}_{\rho \tau \chi}(1,\,2,\,3)=M^{N\,\{13\}}_{ \chi \tau \rho }(3,\,2,\,1)\,;
\nonumber \\ &&
M^{N\,\{23\}}_{\rho \tau \chi}(1,\,2,\,3)
=M^{N\,\{23\}}_{ \rho \chi \tau }(1,\,3,\,2)\,; %\ \ \
\nonumber \\ &&
M^{N\,\{23\}}_{\rho \tau \chi}(1,2,3)=M^{N\,\{12\}}_{\tau  \chi \rho}(2,3,1)\,; \ \ \
M^{N\,\{13\}}_{\rho \tau \chi}(1,2,3)=M^{N\,\{12\}}_{\rho  \chi \tau}( 1, 3, 2)\,; \ \ \
\label{Relations_NDA_permutations}
\ee
For example, the last identity in (\ref{Relations_NDA_permutations}) is the consequence of the relations
\be
&&
\langle 0| \varepsilon_{c_1 c_2 c_3 } \Psi_\rho^{c_1 \,\alpha} (1) \Psi_\tau^{c_2 \,\beta} (2) \Psi_\chi^{c_3 \,\gamma} (3) | N_\iota(p_N) \rangle
=-
\langle 0| \varepsilon_{c_1 c_2 c_3 } \Psi_\rho^{c_1 \,\alpha}(1)  \Psi_\chi^{c_3 \,\gamma} (3) \Psi_\tau^{c_2 \,\beta} (2)| N_\iota(p_N) \rangle
\nonumber \\ &&
=
\langle 0| \varepsilon_{c_1 c_2 c_3} \Psi_\rho^{c_1 \,\alpha}(1)  \Psi_\chi^{c_2 \,\gamma} (3) \Psi_\tau^{c_3 \,\beta} (2)| N_\iota(p_N) \rangle\,.
\ee
The first three identities in
(\ref{Relations_NDA_permutations})
justify our definitions
(\ref{Def_combinations_NDA}) while the two last ones further constrain isospin invariant amplitudes.

We choose to express all invariant isospin amplitudes through
$M^{N\,\{12\}}$.
This allows to write down the following invariant isospin parametrization for the nucleon DA:
\be
4 \langle 0 | \widehat{O}^{\alpha \beta \gamma}_{\rho \tau \chi}
%(z_1,\,z_2,\,z_3)
(1,2,3)
| N_\iota(p_N) \rangle=
\varepsilon^{\alpha \beta} \delta^\gamma_\iota M^{N\,\{12\}}_{\rho \chi \tau }(1,3,2)+
\varepsilon^{\alpha \gamma} \delta^\beta_\iota M^{N\,\{12\}}_{\rho  \tau \chi}(1,2,3)\,.
\label{Isospin_parmetrization_N_DA}
\ee

The next step is to consider the effect of the relations
(\ref{Isospin_Id_NDA})
and
(\ref{Relations_NDA_permutations})
for the nucleon DA.
To the leading twist accuracy we neglect mass effects ($p_N \to p$ , where $p$ is lightlike)
and employ the standard parametrization
for the  invariant amplitude symmetric under the exchange of the two first quark field operators:
\be
&&
M^{N\,\{12\}}_{\rho \tau \chi}(z_1,\,z_2,\,z_3)=
f_N
\Big[
V^p(z_1,z_2,z_3) v_{\rho \tau, \, \chi}^N+
A^p(z_1,z_2,z_3) a_{\rho \tau, \, \chi}^N+
T^p(z_1,z_2,z_3) t_{\rho \tau, \, \chi}^N
\Big]\,,
\nonumber \\ &&
\label{Decomposition_AVT_nucleon_DA}
\ee
where $\{v^N, a^N, t^N\}_{\rho \tau, \, \chi} $
are the conventional Dirac structures:
\be
&&
v_{\rho \tau, \, \chi}^N=(\hat{p} C)_{\rho \tau} (\gamma^5 U(p))_\chi\,;
\ \
a_{\rho \tau, \, \chi}^N=(\hat{p} \gamma^5 C)_{\rho \tau} ( U(p))_\chi
\,;
\ \
t_{\rho \tau, \, \chi}^N=(\sigma_{p \mu}  C)_{\rho \tau} ( \gamma^\mu \gamma^5 U(p))_\chi\,.
\nonumber \\ &&
\label{DA structures}
\ee
The symmetry relations
(\ref{symmetry_Dirac_Nucleon})
for the Dirac structures
(\ref{DA structures})
under the interchange of the two first Dirac indices together with
(\ref{Relations_NDA_permutations})
lead to the familiar symmetry properties:
\be
V^p(1,2,3)=V^p(2,1,3)\,; \ \ \ T^p(1,2,3)=T^p(2,1,3)\,; \ \ \ A^p(1,2,3)=-A^p(2,1,3)\,.
\ee
Next, using symmetry relations
(\ref{Relations_NDA_permutations})
and isospin identity
(\ref{Isospin_Id_NDA})
together with the Fierz transformation
(\ref{Fierz_nucleon_structures})
for the Dirac structures
(\ref{DA structures}),
one may establish the well known relation for twist-$3$ nucleon DAs \cite{Korenblit,Chernyak_Nucleon_wave,Chernyak:1983ej}:
\be
2T^p(1,2,3)= (V^p-A^p)(1,3,2)+(V^p-A^p)(2,3,1)\,.
\ee
This reflects the fact that at leading twist there is only one independent nucleon DA,
usually denoted as $\phi^N$:
\be
\phi^N \equiv V^p-A^p\,.
\label{def_phi_N}
\ee
 The DAs $V^p$, $A^p$ and $T^p$ are expressed through this latter function
according to
\be
&&
2V^p(1,2,3)=   \phi^N(1,2,3)+ \phi^N(2,1,3)\,; \ \ \  2A^p(1,2,3)=   -\phi^N(1,2,3)+ \phi^N(2,1,3)\,;
\nonumber \\ &&
2T^p(1,2,3)=\phi^N(1,3,2)+\phi^N(2,3,1)\,.
\ee

\subsection{Case of $\Delta(1232)$ DA}

In this subsection we introduce the invariant isospin notations for the leading twist DA of
$\Delta(1232)$
resonance \cite{Farrar,Chernyak_Delta}. With respect to $SU(2)$ isospin group $\Delta$ resonance state
%$|\Delta \rangle$
represents a spin tensor with one covariant spinor index and one vector index:
\be
I_a |\Delta_{b\,\iota} \rangle=
\left\{
i \varepsilon_{abc} \delta^\kappa_{\;\;\iota}+ \frac{1}{2} (\sigma_a)^\kappa_{\;\;\iota} \delta_{bc}
\right\}|\Delta_{c \, \kappa} \rangle\,.
\ee
It is natural to choose the isospin conventions for $\Delta$ resonance so that
the isospin classification of $\Delta$ states coincide  with that for
isospin-$\frac{3}{2}$ $\pi N$ states
(\ref{isospin-3/2_piN_states}).

With respect to the Lorentz group $\Delta$ resonance field is described with the help of the Rarita-Schwinger spin-tensor
$\mathcal{U}^\mu_{\iota}(p_\Delta,s_\Delta)$.
As usual,
$\bar{\mathcal{U}} ^\mu (p_\Delta,s_\Delta)= (\mathcal{U}^\mu(p_\Delta,s_\Delta))^\dag \gamma_0$.
For $p^2_\Delta=M_\Delta^2$
spin-tensor
$\mathcal{U}^\mu_{\iota}$ satisfies the following auxiliary conditions:
\be
&&
(\hat{p}_\Delta-M_\Delta) \mathcal{U}^\mu (p_\Delta,s_\Delta)|_{p^2_\Delta=M_\Delta^2}=0\,; \ \ \nonumber \\ &&
\bar{\mathcal{U}} ^\mu (p_\Delta,s_\Delta) \mathcal{U}_\mu (p_\Delta,s_\Delta)|_{p_\Delta^2=M_\Delta^2}=-2 M_\Delta\,; \nonumber \\ &&
\gamma^\mu \mathcal{U}_\mu (p_\Delta,s_\Delta)|_{p_\Delta^2=M_\Delta^2} =
p_\Delta^\mu \mathcal{U}_\mu (p_\Delta,s_\Delta)|_{p_\Delta^2=M_\Delta^2}=0\,.
\label{Conditions_for_Delta_spinor}
\ee

Being guided by the invariance under the isospin group we may write the following tensor decomposition
for the matrix element of the three-quark operator between $\Delta$ resonance state and the vacuum:
\be
4 \langle 0 | \widehat{O}^{\alpha \beta \gamma}_{\rho \tau \chi}(z_1,z_2,z_3)\,|\Delta_{a\,\iota}(p_\Delta) \rangle=
(f_a)^{\{\alpha \beta \gamma\}}_{\ \ \ \ \iota} \, M^\Delta_{\rho \tau \chi} ( 1, 2, 3)\,.
\label{isospin_parametrization_Delta_DA}
\ee
Here
 $(f_a)^{\{\alpha \beta \gamma\}}_{\ \ \ \ \iota}$
stands for the only  tensor totally symmetric in $\alpha$, $\beta$, $\gamma$ one can construct out of the existing structures:
\be
&&
(f_a)^{\{\alpha \beta \gamma\}}_{\ \ \ \ \iota}
\nonumber \\ &&
=
\frac{1}{6} \left(
(\sigma_a)^{\alpha}_{\; \delta} \varepsilon^{\delta \beta} \delta^{\gamma}_{\; \iota}+
(\sigma_a)^{\alpha}_{\; \delta} \varepsilon^{\delta \gamma} \delta^{\beta}_{\; \iota}+
(\sigma_a)^{\beta}_{\; \delta} \varepsilon^{\delta \alpha} \delta^{\gamma}_{\; \iota}+
(\sigma_a)^{\beta}_{\; \delta} \varepsilon^{\delta \gamma} \delta^{\alpha}_{\; \iota}+
(\sigma_a)^{\gamma}_{\; \delta} \varepsilon^{\delta \alpha} \delta^{\beta}_{\; \iota}+
(\sigma_a)^{\gamma}_{\; \delta} \varepsilon^{\delta \beta} \delta^{\alpha}_{\; \iota}
\right)
\nonumber \\ &&
\equiv \frac{1}{3}
\left(
(\sigma_a)^{\alpha}_{\; \delta} \varepsilon^{\delta \beta} \delta^{\gamma}_{\; \iota}
+
(\sigma_a)^{\alpha}_{\; \delta} \varepsilon^{\delta \gamma} \delta^{\beta}_{\; \iota}+
(\sigma_a)^{\beta}_{\; \delta} \varepsilon^{\delta \gamma} \delta^{\alpha}_{\; \iota}
\right)\,; \ \ \ {\rm  since} \ \ \
(\sigma_a)^{\alpha}_{\; \delta} \varepsilon^{\delta \beta}= (\sigma_a)^{\beta}_{\; \delta} \varepsilon^{\delta \alpha}\,.
\label{Def_ta_tensor}
\ee
One may check that the convolutions of the invariant tensor
$(f_a)^{\{\alpha \beta \gamma\}}_{\ \ \ \ \iota}$
with the isospin projecting operators
(\ref{isospin_projecting_oper})
respect the following properties:
\be
{P^{3/2}}^{\;\;  \kappa}_{b \ \  a \iota}
(f_b)^{\{\alpha \beta \gamma\}}_{\ \ \ \ \kappa}=
(f_a)^{\{\alpha \beta \gamma\}}_{\ \ \ \ \iota}\,;
\ \ \
{P^{1/2}}^{\;\;  \kappa}_{b \ \  a \iota}
(f_b)^{\{\alpha \beta \gamma\}}_{\ \ \ \ \kappa}=0\,.
\label{Projecting_ta_tensor}
\ee

We employ the following parametrization
for the leading twist invariant amplitude
$M^\Delta_{\rho \tau \chi} ( 1, 2, 3)$:
\be
&&
M^\Delta_{\rho \tau \chi} ( 1, 2, 3)
\nonumber \\ &&
=
-\frac{\lambda_\Delta^{\frac{1}{2}}}{\sqrt{2}}
\left\{
v^\Delta_{\rho \tau,\, \chi} V^\Delta(1,2,3)+ a^\Delta_{\rho \tau,\, \chi} A^\Delta(1,2,3)+
t^\Delta_{\rho \tau,\, \chi}T^\Delta(1,2,3)
\right\}-
\frac{f_\Delta^{\frac{3}{2}} }{\sqrt{2}} \varphi_{\rho \tau,\, \chi}^\Delta \, \phi^{ \Delta_{3/2} } (1,2,3)\,,
\nonumber \\ &&
\label{Parametrization_Delta_DA_FZ}
\ee
where
$\{v^\Delta, a^\Delta, t^\Delta, \varphi^\Delta \}_{\rho \tau,\, \chi} $
are the usual  Dirac structures
\be
&&
v^\Delta_{\rho \tau,\, \chi}= (\gamma_\mu C)_{\rho \tau} \, \mathcal{U}^\mu_\chi\,;
\ \ \
a^\Delta_{\rho \tau,\, \chi}= (\gamma_\mu \gamma_5 C)_{\rho \tau} \, (\gamma_5\mathcal{U}^\mu)_\chi\,;
\ \ \
t^\Delta_{\rho \tau,\, \chi}= \frac{1}{2} (\sigma_{\mu \nu} C)_{\rho \tau}(\gamma^\mu \mathcal{U}^\nu)_\chi\,;
\nonumber \\ &&
\varphi_{\rho \tau,\, \chi}^\Delta= (\sigma_{\mu \nu} C)_{\rho \tau} (p^\mu \mathcal{U}^\nu- \frac{1}{2} M_\Delta \gamma^\mu \mathcal{U}^\nu)_\chi\,
\label{Dirac_structures_Delta}
\ee
and the constants $\lambda_\Delta^{\frac{1}{2}}$,  $f_\Delta^{\frac{3}{2}}$ are defined in  Ref.~\cite{Farrar}.
The  factor $-\frac{1}{\sqrt{2}}$
in (\ref{Parametrization_Delta_DA_FZ}) ensures matching with
the parametrization of \cite{Farrar} for the $uuu$ DA of $|\Delta^{++} \rangle$
({\it c.f.}  Eq.~(\ref{Delta++for_isospin_Id})).
The DAs $V^\Delta$, $A^\Delta$, $T^\Delta$ and $\phi^{ \Delta_{3/2} }$ in
(\ref{Parametrization_Delta_DA_FZ})
thus coincide with those
of Refs.~\cite{Farrar,Braun:1999te}.

The isospin identities for the invariant amplitude
$M^\Delta_{\rho \tau \chi} (z_1,z_2,z_3)$
are derived analogously to how this was done for the nucleon case in the previous subsection.
Consider
\be
4 \langle 0 | \varepsilon_{c_1 c_2 c_3} u_\rho^{c_1}(z_1) u_\tau^{c_2}(z_2) u_\chi^{c_3}(z_3)| \Delta^{++} \rangle=
-\sqrt{2}
M^\Delta_{\rho \tau \chi}  ( 1, 2, 3)\,.
\label{Delta++for_isospin_Id}
\ee
The invariance under permutations of three $u$-quark fields in
(\ref{Delta++for_isospin_Id})
leads to the complete symmetry of the invariant matrix element under simultaneous permutations of
the arguments and of the Dirac indices:
\be
&&
M^\Delta_{\rho \tau \chi} (1,2,3)
=
M^\Delta_{ \rho \chi \tau} (1,3,2)
=
M^\Delta_{\tau \rho \chi} (2,1,3)
 \nonumber \\ &&
=
M^\Delta_{ \tau \chi \rho} (2,3,1)
=
M^\Delta_{  \chi \tau \rho} (3,2,1)
=
M^\Delta_{ \chi \rho  \tau} (3,1,2)
 \,.
 \label{symmetries_MI32}
\ee
Employing
(\ref{symmetries_MI32})
together with the well-known symmetry relations
(\ref{symmetry_Dirac_structures_Delta})
for the Dirac structures (\ref{Dirac_structures_Delta})
and the twist-$3$ Fierz transformations (\ref{Fierz_Delta_structures})
one establishes the familiar relations
\cite{Farrar}
for the invariant functions
$V^\Delta$, $A^\Delta$, $T^\Delta$
and
$\phi^{ \Delta_{3/2} }$
defined in
(\ref{Parametrization_Delta_DA_FZ}).
Introducing the notation
$\phi^{ \Delta_{1/2}  } (1,2,3)=V^\Delta(1,2,3)-A^\Delta(1,2,3)$
these relations can be written as
\be
&&
2V^\Delta(1,2,3)=   \phi^{\Delta_{1/2}}(1,2,3)+\phi^{\Delta_{1/2}}(2,1,3)\,; \ \ \  2A^\Delta(1,2,3)=  -\phi^{\Delta_{1/2}}(1,2,3)+ \phi^{\Delta_{1/2}}(2,1,3)
\nonumber \\ &&
T^\Delta(1,2,3)=\phi^{\Delta_{1/2}}(2,3,1)\,;
\label{Isospin_and_Sym_Relations_DeltaDA}
\ee
together with the consistency condition
\be
\phi^{\Delta_{1/2}}(1,2,3)=\phi^{\Delta_{1/2}}(3,2,1)\,.
\ee
Meanwhile, $\phi^{\Delta_{3/2}}(1,2,3)$ turns out to be totally symmetric.

\section{Isospin parametrization for $\pi N$ TDA and GDA}
\label{Section_isospin_TDAs}

Let us consider now the matrix element of three-quark operator
$\widehat{O}^{\alpha \beta \gamma}_{\rho \tau \chi}(z_1,\,z_2,\,z_3)$
between $\pi N$ states both in TDA and GDA regimes.
From the point of view
of the isospin symmetry the two regimes can be analyzed on the same footing
since the pion field $\pi_a$
transforms according to the adjoint representation of the isospin group.
So below we present the isospin decomposition of $\pi N$ TDA.
The expression for $\pi N$ GDA is exactly the same.

Isospin decomposition for $\pi N$ TDA should involve both the isospin-$\frac{3}{2}$ and
isospin-$\frac{1}{2}$
parts.
Thus, analogously to the cases of $\Delta$ and nucleon DAs, we can write the following isospin decomposition:
\be
&&
4\langle  \pi_a | \widehat{O}^{\alpha \beta \gamma}_{\rho \tau \chi}(z_1,\,z_2,\,z_3) | N_\iota \rangle
= (f_a)^{\{\alpha \beta \gamma \}}_{\ \ \ \ \iota}
M^{(\pi N)_{3/2}}_{\rho \tau \chi}
%(z_1,z_2,z_3)
(1,2,3)
 +
\varepsilon^{\alpha \beta} (\sigma_a)^\gamma_{\ \iota} {M^{(\pi N)_{1/2}}_1}_{\rho \tau \chi}
%(z_1,z_2,z_3)
(1,2,3) \nonumber \\ &&
+
\varepsilon^{\alpha \gamma} (\sigma_a)^\beta_{\ \iota} {M^{(\pi N)_{1/2}}_2}_{\rho \tau \chi}
%(z_1,z_2,z_3)
(1,2,3)
+\varepsilon^{\beta \gamma} (\sigma_a)^\alpha_{\ \iota} {M^{(\pi N)_{1/2}}_3}_{\rho \tau \chi}
%(z_1,z_2,z_3)
(1,2,3)
\nonumber \\ &&
=(f_a)^{\{\alpha \beta \gamma\}}_{\ \ \ \ \iota} M^{(\pi N)_{3/2}}_{\rho \tau \chi}
%(z_1,z_2,z_3)
(1,2,3)+
\varepsilon^{\alpha \beta} (\sigma_a)^\gamma_{\ \iota}  M^{(\pi N)_{1/2} \, \{1 3 \}}_{\rho \tau \chi}
%(z_1,z_2,z_3)
(1,2,3)
+\varepsilon^{\alpha \gamma} (\sigma_a)^\beta_{\ \iota} M^{(\pi N)_{1/2} \, \{1 2 \}}_{\rho \tau \chi}
%(z_1,z_2,z_3)
(1,2,3)\,,
\nonumber \\ &&
\label{GDA_isospin_dec}
\ee
where
$(f_a)^{\{\alpha \beta \gamma\}}_{\ \ \ \ \iota}$
is the symmetric tensor defined in
(\ref{Def_ta_tensor}).
To put down the last equality in (\ref{GDA_isospin_dec}) we employed the identity
\be
\varepsilon^{\beta \gamma}  (\sigma_a)^\alpha_{\; \; \iota}=
-\varepsilon^{\alpha \beta} (\sigma_a)^\gamma_{\; \; \iota}+\varepsilon^{\alpha \gamma} (\sigma_a)^\beta_{\; \; \iota}
\ee
to eliminate the third structure corresponding to the isospin-$\frac{1}{2}$ representation. Analogously to
(\ref{Def_combinations_NDA})
we introduce the notations
\be
&&
{M^{(\pi N)_{1/2}}_2}_{\rho \tau \chi}
%(z_1,z_2,z_3)
(1,2,3)
+{M^{(\pi N)_{1/2}}_3}_{\rho \tau \chi}
%(z_1,z_2,z_3)
(1,2,3)
\equiv M^{(\pi N)_{1/2} \; \{1 2 \}}_{\rho \tau \chi}
%(z_1,z_2,z_3)
(1,2,3)\,;
\nonumber \\ &&
{M^{(\pi N)_{1/2}}_1}_{\rho \tau \chi}
%(z_1,z_2,z_3)
(1,2,3)
 -{M^{(\pi N)_{1/2}}_3}_{\rho \tau \chi}
 %(z_1,z_2,z_3)
 (1,2,3)
\equiv M^{(\pi N)_{1/2}\, \{1 3 \}}_{\rho \tau \chi}
%(z_1,z_2,z_3)
(1,2,3)\,;
\nonumber \\ &&
-{M_1^{(\pi N)_{1/2}}}_{\rho \tau \chi}(1,2,3)-{M_2^{(\pi N)_{1/2}}}_{\rho \tau \chi}(1,2,3)
\equiv M^{{(\pi N)_{1/2}}\,\{23\}}_{\rho \tau \chi}(1,2,3)\,.
\label{Def_combinations_piNTDA}
\ee

Our present goal is to establish the isospin and permutation symmetry identities for $\pi N$ TDA invariant isotopic
amplitudes. We will show that isotopic and permutation symmetry reduces the number of independent $\pi N$ TDAs
from $16$ functions ($8$ both for the isospin-$\frac{1}{2}$ and isospin-$\frac{3}{2}$ parts as in Eq.~(\ref{Decomposition_piN_TDAs_new})) to just
$8$ independent functions.

\subsection{Isospin-$\frac{1}{2}$ case}

One may check that the isospin-$\frac{1}{2}$ invariant amplitudes satisfy the set of identities
analogous to the isospin invariant amplitudes for nucleon DA.
The isospin identity reads
\be
M^{(\pi N)_{1/2}\,\{12\}}_{\rho \tau \chi}(1,2,3)+
M^{(\pi N)_{1/2}\,\{13\}}_{\rho \tau \chi}(1,2,3)+
M^{(\pi N)_{1/2}\,\{23\}}_{\rho \tau \chi}(1,2,3)=0\,.
\label{Isospin_Id_piNTDA}
\ee
The permutation symmetry results in the set of identities analogous to
(\ref{Relations_NDA_permutations}):
\be
&&
M^{(\pi N)_{1/2}\,\{12\}}_{\rho \tau \chi}
(1,2,3)
=M^{(\pi N)_{1/2}\,\{12\}}_{ \tau \rho \chi }(2,1,3)\,; \ \ \
M^{(\pi N)_{1/2}\,\{13\}}_{\rho \tau \chi}(1,2,3)=M^{(\pi N)_{1/2}\,\{13\}}_{ \chi \tau \rho }(3,2,1)\,;
\nonumber \\ &&
M^{(\pi N)_{1/2}\,\{23\}}_{\rho \tau \chi}(1,2,3)
=M^{(\pi N)_{1/2}\,\{23\}}_{ \rho \chi \tau }(1,3,2)\,;
\nonumber \\ &&
M^{(\pi N)_{1/2}\,\{13\}}_{\rho \tau \chi}(1,2,3)=M^{(\pi N)_{1/2}\,\{12\}}_{\rho  \chi \tau}( 1, 3, 2)\,; \ \ \
M^{(\pi N)_{1/2}\,\{23\}}_{\rho \tau \chi}(1,2,3)=M^{(\pi N)_{1/2}\,\{12\}}_{\tau  \chi \rho}(2,3,1)\,.
\nonumber \\ &&
\label{Relations_piNTDA_permutations}
\ee

For
$M^{(\pi N)_{1/2}\,\{12\}}_{\rho \tau \chi}(1,2,3)$
we introduce the parametrization
(\ref{Decomposition_piN_TDAs_new})
and define $8$ leading twist isospin-$\frac{1}{2}$ $\pi N$ TDAs:
\be
&&
M^{(\pi N)_{1/2}\,\{12\}}_{\rho \tau \chi}(1,2,3)
\nonumber \\ &&
=i \frac{f_N}{f_\pi M}     \big[ V_1^{ (\pi N)_{1/2}} (1,2,3)
%(\hat{P}C)_{\rho \, \tau } (\hat{P} U)_\chi
(v_1^{\pi N})_{\rho \tau, \, \chi}
+
   A_1^{(\pi N)_{1/2}} (1,2,3)
   (a_1^{\pi N})_{\rho \tau, \, \chi}
  % (\hat{P} \gamma^5 C)_{\rho \, \tau } (\gamma^5 \hat{P} U)_\chi
 %  \nonumber \\ &&
+
  T_1^{(\pi N)_{1/2}} (1,2,3)
   (t_1^{\pi N})_{\rho \tau, \, \chi}
  %(\sigma_{P \mu} C)_{\rho \, \tau } (\gamma^\mu \hat{P} U)_\chi
\nonumber \\ &&
+    V_2^{(\pi N)_{1/2}} (1,2,3)
   (v_2^{\pi N})_{\rho \tau,, \, \chi}
% (\hat{P}C)_{\rho \, \tau }  (\hat{\Delta}  U )_\chi
+
  A_2^{(\pi N)_{1/2}} (1,2,3)
  (a_2^{\pi N})_{\rho \tau, \, \chi}
  %(\hat{P} \gamma^5 C)_{\rho \, \tau } (\gamma^5 \hat{\Delta}  U )_\chi \nonumber \\ &&
+
 T_2^{(\pi N)_{1/2}} (1,2,3)
 (t_2^{\pi N})_{\rho \tau, \, \chi}   %(\sigma_{P \mu} C)_{\rho \, \tau } (\gamma^\mu \hat{\Delta}  U )_\chi
 \nonumber \\ &&
+
\frac{1}{M}  T_3^{(\pi N)_{1/2}} (1,2,3)
(t_3^{\pi N})_{\rho \tau, \, \chi}
%(\sigma_{P \Delta } C)_{\rho \, \tau } (\hat{P} U)_\chi
% \nonumber \\ && 
+
\frac{1}{M}  T_4^{(\pi N)_{1/2}}  (1,2,3)(t_4^{\pi N})_{\rho \tau, \, \chi}   % (\sigma_{P \Delta } C)_{\rho \, \tau } (\hat{\Delta}   U )_\chi
\big].
\label{Decomposition_piN_TDAs_new_M12}
\ee

One may check that the permutation symmetry relations
(\ref{Relations_piNTDA_permutations})
result in the familiar symmetry
properties of the isospin-$\frac{1}{2}$ $\pi N$ TDAs:
\be
&&
V_{1,2}^{(\pi N)_{1/2}}(1,2,3)=V_{1,2}^{(\pi N)_{1/2}}(2,1,3)\,;  \ \ \
T_{1,2,3,4}^{(\pi N)_{1/2}}(1,2,3)=T_{1,2,3,4}^{(\pi N)_{1/2}}(2,1,3)\,;
\nonumber \\ &&
A_{1,2}^{(\pi N)_{1/2}}(1,2,3)=-A_{1,2}^{(\pi N)_{1/2}}(2,1,3)\,.
\,; \ \ \
\ee

We introduce two independent isospin-$\frac{1}{2}$ $\pi N$ TDAs:
\be
\phi^{(\pi N)_{1/2}}_{1,2}(1,2,3) \equiv V^{(\pi N)_{1/2}}_{1,2}(1,2,3)-A^{(\pi N)_{1/2}}_{1,2}(1,2,3)\,.
\label{independent1/2}
\ee
Employing the Fierz transformations
(\ref{Fierz_for_piNTDA_structures}),  (\ref{Fierz for t34}),
one establishes the consequences of the isospin symmetry relation (\ref{Isospin_Id_piNTDA}):
\be
T_{3,4}^{(\pi N)_{1/2}}(1,2,3)+ T_{3,4}^{(\pi N)_{1/2}}(1,3,2)+ T_{3,4}^{(\pi N)_{1/2}}(2,3,1)=0
\label{symmetry_T34}
\ee
and
\be
&&
2 T_{1,2}^{(\pi N)_{1/2}}(1,2,3)
= \phi^{(\pi N)_{1/2}}_{1,2}(1,3,2) + \phi^{(\pi N)_{1/2}}_{1,2}(2,3,1) \nonumber \\ && +
2g_{1,2}(\xi,\Delta^2) T_3^{(\pi N)_{1/2}}(1,2,3) + 2h_{1,2}(\xi,\Delta^2) T_4^{(\pi N)_{1/2}}(1,2,3)\,;
\nonumber \\ &&
2 V_{1,2}^{(\pi N)_{1/2}} (1,2,3)= \phi^{(\pi N)_{1/2}}_{1,2}(1,2,3)+\phi^{(\pi N)_{1/2}}_{1,2}(2,1,3)\,;
\nonumber \\ &&
2 A_{1,2}^{(\pi N)_{1/2}}(1,2,3)= -\phi^{(\pi N)_{1/2}}_{1,2}(1,2,3)+\phi^{(\pi N)_{1/2}}_{1,2}(2,1,3)\,,
\ee
where
$g_{1,2}(\xi,\Delta^2)$,
$h_{1,2}(\xi,\Delta^2)$
are defined in (\ref{Fierz for t34}).
We conclude that the parametrization of the isospin-$\frac{1}{2}$ $\pi N$ TDAs/GDAs require $4$ independent functions:
$\phi^{(\pi N)_{1/2}}_{1,2}$ and $T_{3,4}^{(\pi N)_{1/2}}$.
The latter should satisfy the symmetry relations (\ref{symmetry_T34}).

\subsection{Isospin-$\frac{3}{2}$ case}
The consequences of the isotopic and permutation symmetries for the isospin-$\frac{3}{2}$
invariant amplitude
$M^{(\pi N)_{3/2}}$
are analogous to that for $\Delta(1232)$ DA
(\ref{symmetries_MI32}).
It turns out to be   completely symmetric  under simultaneous permutations of
the arguments and the Dirac indices:
\be
&&
M^{(\pi N)_{3/2}}_{\rho \tau \chi} (1,2,3)
=
M^{(\pi N)_{3/2}}_{ \rho \chi \tau} (1,3,2)
=
M^{(\pi N)_{3/2}}_{\tau \rho \chi} (2,1,3)
 \nonumber \\ &&
=
M^{(\pi N)_{3/2}}_{ \tau \chi \rho} (2,3,1)
=
M^{(\pi N)_{3/2}}_{  \chi \tau \rho} (3,2,1)
=
M^{(\pi N)_{3/2}}_{ \chi \rho  \tau} (3,1,2)
 \,.
 \label{symmetries_MI32piNTDA}
\ee

Again, in accordance with (\ref{Decomposition_piN_TDAs_new}),
we introduce the following parametrization for
the leading twist isospin-$\frac{3}{2}$
$\pi N$
TDAs:
\be
&&
M^{(\pi N)_{3/2} }_{\rho \tau \chi}(1,2,3)
\nonumber \\ &&
=i \frac{f_N}{f_\pi M}     \big[ V_1^{ (\pi N)_{3/2}} (1,2,3)
%(\hat{P}C)_{\rho \, \tau } (\hat{P} U)_\chi
(v_1^{\pi N})_{\rho \tau, \, \chi}
+
   A_1^{(\pi N)_{3/2}} (1,2,3)
   (a_1^{\pi N})_{\rho \tau, \, \chi}
  % (\hat{P} \gamma^5 C)_{\rho \, \tau } (\gamma^5 \hat{P} U)_\chi
 %  \nonumber \\ &&
+
  T_1^{(\pi N)_{3/2}} (1,2,3)
   (t_1^{\pi N})_{\rho \tau, \, \chi}
  %(\sigma_{P \mu} C)_{\rho \, \tau } (\gamma^\mu \hat{P} U)_\chi
\nonumber \\ &&
+    V_2^{(\pi N)_{3/2}} (1,2,3)
   (v_2^{\pi N})_{\rho \tau, \, \chi}
% (\hat{P}C)_{\rho \, \tau }  (\hat{\Delta}  U )_\chi
+
  A_2^{(\pi N)_{3/2}} (1,2,3)
  (a_2^{\pi N})_{\rho \tau, \, \chi}
  %(\hat{P} \gamma^5 C)_{\rho \, \tau } (\gamma^5 \hat{\Delta}  U )_\chi \nonumber \\ &&
+
 T_2^{(\pi N)_{3/2}} (1,2,3)
 (t_2^{\pi N})_{\rho \tau, \, \chi}   %(\sigma_{P \mu} C)_{\rho \, \tau } (\gamma^\mu \hat{\Delta}  U )_\chi
 \nonumber \\ &&
+
\frac{1}{M}  T_3^{(\pi N)_{3/2}} (1,2,3)
(t_3^{\pi N})_{\rho \tau, \, \chi}
%(\sigma_{P \Delta } C)_{\rho \, \tau } (\hat{P} U)_\chi
% \nonumber \\ && 
+
\frac{1}{M}  T_4^{(\pi N)_{3/2}}  (1,2,3)(t_4^{\pi N})_{\rho \tau, \, \chi}   % (\sigma_{P \Delta } C)_{\rho \, \tau } (\hat{\Delta}   U )_\chi
\big].
\label{Decomposition_piN_TDAs_new_M32}
\ee
Analogously to the isospin-$\frac{1}{2}$ case, employing the Fierz identities of
Appendix~\ref{App_Fierz},
one may check that the permutation symmetry relations
(\ref{symmetries_MI32piNTDA})
result in the following symmetry properties of the isospin-$\frac{3}{2}$ $\pi N$ TDAs:
\be
&&
V_{1,2}^{(\pi N)_{3/2}}(1,2,3)=V_{1,2}^{(\pi N)_{3/2}}(2,1,3)\,;  \ \ \
T_{1,2 }^{(\pi N)_{3/2}}(1,2,3)=T_{1,2}^{(\pi N)_{3/2}}(2,1,3)\,;
\nonumber \\ &&
A_{1,2}^{(\pi N)_{3/2}}(1,2,3)=-A_{1,2}^{(\pi N)_{3/2}}(2,1,3)\,,
\ee
while $T_{3,4 }^{(\pi N)_{3/2}}$ is totally symmetric.

Introducing two independent isospin-$\frac{3}{2}$ $\pi N$ TDAs:
\be
\phi^{(\pi N)_{3/2}}_{1,2}(1,2,3) \equiv -V^{(\pi N)_{3/2}}_{1,2}(1,2,3)+A^{(\pi N)_{3/2}}_{1,2}(1,2,3)\,,
\label{independent3/2}
\ee
and employing further consequences of permutation and isotopic symmetry relations
(\ref{symmetries_MI32piNTDA})
one may express isospin-$\frac{3}{2}$ TDAs as
\be
&&
2T_{1,2 }^{(\pi N)_{3/2}}(1,2,3)
\nonumber \\ && =\phi^{(\pi N)_{3/2}}_{1,2}(1,3,2) +
2g_{1,2}(\xi,\Delta^2) T_3^{(\pi N)_{3/2}}(1,2,3) + 2h_{1,2}(\xi,\Delta^2) T_4^{(\pi N)_{3/2}}(1,2,3)\,;
\nonumber \\ &&
2 V^{(\pi N)_{3/2}}_{1,2}(1,2,3) = -\phi^{(\pi N)_{3/2}}_{1,2}(1,2,3)-\phi^{(\pi N)_{3/2}}_{1,2}(2,1,3)\,;
\nonumber \\ &&
2 A^{(\pi N)_{3/2}}_{1,2}(1,2,3) =  \phi^{(\pi N)_{3/2}}_{1,2}(1,2,3)-\phi^{(\pi N)_{3/2}}_{1,2}(2,1,3)\,,
\label{Isospin_3/2_TDA_final_symmetry}
\ee
where
$g_{1,2}(\xi,\Delta^2)$,
$h_{1,2}(\xi,\Delta^2)$
are defined in (\ref{Def_g12_h12}).
The consistency condition for
(\ref{Isospin_3/2_TDA_final_symmetry})
may be   established from  (\ref{symmetries_MI32piNTDA}):
\be
\phi^{(\pi N)_{3/2}}_{1,2}(1,2,3) = \phi^{(\pi N)_{3/2}}_{1,2}(3,2,1)\,.
\label{consistency_for_isospin3/2TDAsym}
\ee

Thus, we conclude that the parametrization of the isospin-$\frac{3}{2}$
$\pi N$
TDAs/GDAs  involves $4$ independent functions:
$T_{3,4 }^{(\pi N)_{3/2}}$
(completely symmetric under permutation of their variables)
and
$\phi^{(\pi N)_{3/2}}_{1,2}$
(symmetric under permutation
$1 \leftrightarrow 3$
{\it cf.} Eq.~(\ref{consistency_for_isospin3/2TDAsym})).

\section{Chiral constraints for $\pi N$ TDAs }
\label{Section_Soft_pion}
In this section we rederive for $\pi N$ GDAs the soft pion theorem \cite{Pobylitsa:2001cz} proposed in
\cite{BLP1}
to be valid  at a scale $Q^2 \gg \Lambda_{\rm QCD}^3/m$.
%Our derivation is in general analogous to that presented in \cite{BLP1}.
Our technique of handling isospin developed in
Sec.~\ref{Section_Isopin_par_DA}, \ref{Section_isospin_TDAs}
permits to distinguish between the isospin-$\frac{1}{2}$ and
isospin-$\frac{3}{2}$
$\pi N$ GDAs. This allows to fully take into account the consequences of the isotopic and permutation symmetries for
$\pi N$ GDAs. Using crossing between $\pi N$ GDAs and $\pi N$ TDAs discussed in
Sec.~\ref{Section_Intro}
we simultaneously argue that the soft pion theorem for  $\pi N$ GDAs  constrains $\pi N$ TDAs in the chiral limit ($m \rightarrow 0$).
The problem of validity of analytic continuation in $\Delta^2$ existing for $m \ne 0$ has the same
status as that for the case of pion GPDs v.s. $2\pi$ GDAs \cite{Polyakov:1998ze} (see also discussion in \cite{Kivel:2002ia}).
Assuming  smallness of nonanalytic corrections to the relevant matrix element
in the narrow domain in  $(\Delta^2,\,\xi)$-plane defined by
the inequalities
\be
(M-m)^2 < \Delta^2 < (M+m)^2\,; \ \ \  \frac{M-m}{M+m} < \xi < \frac{M+m}{M-m}
\ee
(see left panel of Fig.~\ref{Fig1})
one may argue that the soft pion limit
provides  us with the reference point for realistic modeling of $\pi N$ TDAs.

Let us consider the matrix element of the three-quark operator
$\widehat{O}^{\alpha \beta \gamma}_{\rho  \tau \chi}(z_1,\,z_2,\,z_3)$
in the regime of %generalized distribution amplitude
$\pi N$ GDA:
\be
\langle 0 | \widehat{O}^{\alpha \beta \gamma}_{\rho  \tau \chi}(z_1,\,z_2,\,z_3) | \pi_a(-p_\pi)  N_\iota(p_1) \rangle \,.
\label{input_soft_pion}
\ee
According to the to the partial conservation of axial current
(PCAC) hypothesis
(see {\it e.g.} \cite{Alfaro_red_book}),
a soft pion theorem
\cite{Pobylitsa:2001cz}
is valid for the matrix element (\ref{input_soft_pion}):
\be
\langle 0 | \widehat{O}^{\alpha \beta \gamma}_{\rho  \tau \chi}(z_1,\,z_2,\,z_3) | \pi_a N_\iota \rangle =
-\frac{i}{f_\pi}
\langle 0 |
\left[
\widehat{Q}_5^a, \, \widehat{O}^{\alpha \beta \gamma}_{\rho  \tau \chi}(z_1,\,z_2,\,z_3)\right]
 |N_\iota \rangle \,.
 \label{soft_pion_theorem_pobyl}
\ee
The commutator of the chiral charge operator
$\widehat{Q}_5^a$
with the quark field operators is given by
\be
\left[\widehat{Q}_5^a,\ \Psi^\alpha_\eta \right]= - \frac{1}{2}
 (\sigma_a)^\alpha_{\;\delta}  \gamma^5_{\eta \tau} \Psi^\delta_\tau\,,
\ee
where $\sigma_a$ are the Pauli matrices.

Computing the commutator of the chiral charge with the operator $\widehat{O}$ in
(\ref{soft_pion_theorem_pobyl})
with the help of the chain rule $[A,BCD]=[A,B]CD+B[A,C]D+BC[A,D]$
we get:
 \be
 &&
4\langle 0 | \widehat{O}^{\alpha \beta \gamma}_{\rho \tau \chi}(z_1,\,z_2,\,z_3) | \pi_a N_\iota \rangle =
 4\frac{i}{2 f_\pi}
 \Big\{
 (\sigma_a)^\alpha_{\; \delta}  \gamma^5_{\rho \eta}\langle 0| \widehat{O}^{\delta \beta \gamma}_{\eta \tau \chi}(z_1,\,z_2,\,z_3)|N_\iota \rangle
\nonumber \\ &&  +
  (\sigma_a)^\beta_{\; \delta}  \gamma^5_{\tau \eta}\langle 0| \widehat{O}^{\alpha \delta \gamma}_{\rho \eta \chi}(z_1,\,z_2,\,z_3)|N_\iota \rangle+
   (\sigma_a)^\gamma_{\; \delta}  \gamma^5_{\chi \eta}\langle 0| \widehat{O}^{\alpha \beta \delta}_{\rho \tau \eta}(z_1,\,z_2,\,z_3)|N_\iota \rangle
 \Big\}\nonumber \\ &&
  =\frac{i}{2 f_\pi}
 \Big\{
 (\sigma_a)^\alpha_{\; \delta}  \varepsilon^{\delta \beta} \delta^\gamma_\iota \gamma^5_{\rho \eta} M^{N \, \{13\}}_{\eta \tau \chi}( 1, 2, 3)+
 (\sigma_a)^\alpha_{\; \delta } \varepsilon^{\delta \gamma} \delta^\beta_\iota  \gamma^5_{\rho \eta} M^{N \,\{12\}}_{\eta \tau \chi}( 1, 2, 3)
 \nonumber \\ &&
+(\sigma_a)^\beta_{\; \delta}  \varepsilon^{\alpha \delta} \delta^\gamma_\iota \gamma^5_{\tau \eta} M^{N \,\{13\}}_{\rho \eta \chi}( 1, 2, 3)+
 (\sigma_a)^\beta_{\; \delta}  \varepsilon^{\alpha \gamma} \delta^\delta_\iota \gamma^5_{\tau \eta} M^{N \, \{12\}}_{\rho \eta \chi}( 1, 2, 3)
 \nonumber \\ &&
 +(\sigma_a)^\gamma_{\; \delta}  \varepsilon^{\alpha \beta} \delta^\delta_\iota \gamma^5_{\chi \eta} M^{N \, \{13\}}_{\rho \tau \eta}( 1, 2, 3)+
 (\sigma_a)^\gamma_{\; \delta}  \varepsilon^{\alpha \delta} \delta^\beta_\iota \gamma^5_{\chi \eta} M^{N \, \{12\}}_{\rho \tau \eta}( 1, 2, 3)
   \Big\}\,.
  \label{soft_pion_full_glory}
\ee
In the last equality we used the general isospin parametrization
(\ref{Isospin_parmetrization_N_DA})
for the nucleon DA.

Our present goal is to single out the contributions coming from
(\ref{soft_pion_full_glory})
into the invariant   isospin amplitudes
$  M^{(\pi N)_{3/2}}_{\rho \tau \chi}( 1, 2, 3)$,
$ M^{(\pi N)_{1/2} \, \{12\}}_{\rho \tau \chi}( 1, 2, 3)$
and
$ M^{(\pi N)_{1/2} \, \{13\}}_{\rho \tau \chi}( 1, 2, 3)$
% $M^{\{12  \}}_{\eta \xi \zeta}(z_1,z_2,z_3)$
in the isospin decomposition for $\pi N$ TDA/GDA (\ref{GDA_isospin_dec}).

Using   the parametrization
(\ref{Decomposition_AVT_nucleon_DA})
for
$M^{N \, \{12\}}_{\rho \tau \chi}( 1, 2, 3)$,
the  isospin decomposition
(\ref{nucleon_DA_isospin_dec_1})
and symmetry relations
(\ref{Relations_NDA_permutations})
for the nucleon DA  together with the Fierz identities from Appendix~\ref{App_Fierz}
one may check that
\be
&&
\nonumber M^{(\pi N)_{3/2}}_{\rho \tau \chi}(z_1,z_2,z_3)
 =\frac{i}{2 f_\pi}
 \Big\{
\gamma^5_{\rho \eta} M^{N \, \{12\}}_{\eta \tau \chi}(z_1,z_2,z_3)
-\gamma^5_{\chi \eta} M^{N \, \{12\}}_{\rho \tau \eta}(z_1,z_2,z_3)
\\ &&
+\gamma^5_{\rho \eta} M^{N \, \{13\}}_{\eta \tau \chi}(z_1,z_2,z_3)-
 \gamma^5_{\tau \eta} M^{N \, \{13\}}_{\rho \eta \chi}(z_1,z_2,z_3)  \Big\}
\nonumber \\ &&
=\frac{i f_N}{ f_\pi}
 \Big\{
- \left( \gamma^5_{\chi \eta} v^N_{\rho \tau, \, \eta} \right) \frac{1}{2} \left[
\phi_N(1,2,3)+\phi_N(2,1,3)+\phi_N(3,2,1)+\phi_N(3,1,2)
\right]
\nonumber \\ &&
- \left( \gamma^5_{\chi \eta} a^N_{\rho \tau, \, \eta} \right)
\frac{1}{2} \left[
-\phi_N(1,2,3)+\phi_N(2,1,3)-\phi_N(3,2,1)+\phi_N(3,1,2)
\right]
\nonumber \\ &&
- \left( \gamma^5_{\chi \eta} t^N_{\rho \tau, \, \eta} \right)
\frac{1}{2} \left[
\phi_N(1,3,2)+\phi_N(2,3,1)
\right]
  \Big\}\,,
\label{M_T_sym_soft_pion}
\ee
where $\phi_N$ is the leading twist nucleon DA (\ref{def_phi_N}).
The invariant  amplitude (\ref{M_T_sym_soft_pion}) satisfies the
isospin-$\frac{3}{2}$
symmetry relations (\ref{symmetries_MI32piNTDA}).
This provides an additional cross-check.

\be
&&
\nonumber M^{(\pi N)_{1/2} \, \{ 12\}}_{\rho \tau \chi}(z_1,z_2,z_3)
 =\frac{i}{2 f_\pi} \frac{1}{3}
 \Big\{
 \gamma^5_{\rho \eta} M^{N \, \{12\}}_{\eta \tau \chi}(z_1,z_2,z_3)+
 3 \gamma^5_{\tau \eta} M^{N \, \{12\}}_{\rho \eta \chi}(z_1,z_2,z_3)
 \nonumber \\ &&
 -\gamma^5_{\chi \eta} M^{N \, \{12\}}_{\rho \tau \eta}(z_1,z_2,z_3)
 -2\gamma^5_{\rho \eta} M^{N \, \{13\}}_{\eta \tau \chi}(z_1,z_2,z_3)
 +
 2 \gamma^5_{\tau \eta} M^{N \, \{13\}}_{\rho \eta \chi}(z_1,z_2,z_3)
  \Big\}
   \nonumber \\ &&
   =\frac{i}{  f_\pi}  \Big\{
- \left( \gamma^5_{\chi \eta} v^N_{\rho \tau, \, \eta} \right) \frac{1}{12} \left[ -\phi_N(1,2,3)-\phi_N(2,1,3)-4(\phi_N(3,1,2)+\phi_N(3,2,1))  \right]
\nonumber \\ &&
- \left( \gamma^5_{\chi \eta} a^N_{\rho \tau, \, \eta} \right) \frac{1}{12}
\left[\phi_N(1,2,3)-\phi_N(2,1,3) - 4(\phi_N(3,1,2)-\phi_N(3,2,1)) \right]
\nonumber \\ &&
- \left( \gamma^5_{\chi \eta} t^N_{\rho \tau, \, \eta} \right)
\frac{5}{12}
\left[ \phi_N(1,3,2)+\phi_N(2,3,1) \right]
  \Big\}\,.
  \label{M12_12_piN}
\ee
Analogously,
\be
&&
\nonumber M^{(\pi N)_{1/2} \, \{ 13\}}_{\rho \tau \chi}(z_1,z_2,z_3)
 =\frac{i}{2 f_\pi} \frac{1}{3}
 \Big\{
 \gamma^5_{\rho \eta} M^{N \, \{13\}}_{\eta \tau \chi}(z_1,z_2,z_3)+
 3 \gamma^5_{\chi \eta} M^{N \, \{13\}}_{\rho \tau \eta}(z_1,z_2,z_3)
 \nonumber \\ &&
 -\gamma^5_{\tau \eta} M^{N \, \{13\}}_{\rho \eta \chi}(z_1,z_2,z_3)
 -2\gamma^5_{\rho \eta} M^{N \, \{12\}}_{\eta \tau \chi}(z_1,z_2,z_3)
 +
 2 \gamma^5_{\chi \eta} M^{N \, \{12\}}_{\rho \tau \eta}(z_1,z_2,z_3)
 \Big\}\,.
 \label{M12_13_piN}
\ee
Again one may check that $ M^{(\pi N)_{1/2} \, \{ 12\}}$
$ M^{(\pi N)_{1/2} \, \{ 13\}}$
(\ref{M12_12_piN}), (\ref{M12_13_piN})
computed from the soft pion theorem satisfy the isospin-$\frac{1}{2}$
and permutation
symmetry relations
(\ref{Isospin_Id_piNTDA}), (\ref{Relations_piNTDA_permutations}).

In particular for $p \pi^0$ GDA we get
\be
&&
4 \langle 0 | u_\rho(1)  u_\tau(2) d_\chi(3)| p \pi^0 \rangle=
 M^{(\pi N)_{1/2} \, \{ 12\}}_{\rho \tau \chi}(1,2,3) + \frac{2}{3}  M^{(\pi N)_{3/2}  }_{\rho \tau \chi}(1,2,3)
\nonumber \\ &&
=
\frac{i f_N}{f_\pi} \Big\{
- \left( \gamma^5_{\chi \eta} v^N_{\rho \tau, \, \eta} \right) \frac{1}{2} V^p(1,2,3)
- \left( \gamma^5_{\chi \eta} a^N_{\rho \tau, \, \eta} \right) \frac{1}{2} A^p(1,2,3)
- \left( \gamma^5_{\chi \eta} t^N_{\rho \tau, \, \eta} \right) \frac{3}{2} T^p(1,2,3)
\Big\}\,;
\nonumber \\ &&
4\langle 0 | u_\rho(1)  u_\tau(2) d_\chi(3)| n \pi^+ \rangle
= -4\langle 0 | d_\rho(1)  d_\tau(2) u_\chi(3)| p \pi^- \rangle
\nonumber \\ &&
=
  \sqrt{2} M^{(\pi N)_{1/2} \, \{ 12\}}_{\rho \tau \chi}(1,2,3)
  - \frac{ \sqrt{2}}{3}  M^{(\pi N)_{3/2}  }_{\rho \tau \chi}(1,2,3)
  \nonumber \\ &&
  =\frac{i f_N}{f_\pi} \Big\{
  - \left( \gamma^5_{\chi \eta} v^N_{\rho \tau, \, \eta} \right)  \frac{1}{2 \sqrt{2}}
  \left[ -\phi_N(1,2,3)-\phi_N(2,1,3)-2(\phi_N(3,1,2)+\phi_N(3,2,1)\right]
    \nonumber \\ &&
 - \left( \gamma^5_{\chi \eta} a^N_{\rho \tau, \, \eta} \right)
 \frac{1}{2 \sqrt{2}}  \left[ \phi_N(1,2,3)-\phi_N(2,1,3)-2(\phi_N(3,1,2)-\phi_N(3,2,1))\right]
    \nonumber \\ &&
 - \left( \gamma^5_{\chi \eta} t^N_{\rho \tau, \, \eta} \right) \frac{1}{2 \sqrt{2}}
 \left[ \phi_N(1,3,2)+\phi_N(2,3,1) \right]
  \Big\}\,.
\ee
So we recover the result of \cite{BLP1}.

Now we can establish the consequences of the soft pion theorem (\ref{soft_pion_theorem_pobyl}) for
$\pi N$ TDAs.  Applying crossing to
the matrix element (\ref{input_soft_pion}) is trivial up to the problem of appropriate analytic continuation in $\Delta^2$.
The contributions to $\pi N$ TDAs occurring in the parametrization
(\ref{Decomposition_piN_TDAs_new})
can be established with the help of the relations between the Dirac structures
(\ref{Dirac_structures_PiN_TDA}) and those of
(\ref{M_T_sym_soft_pion}), (\ref{M12_12_piN}):
\be
&&
 \gamma^5_{\chi \eta} v^N_{\rho \tau, \, \eta}  = \frac{1}{M}
\left(
 {v_1}_{\rho \tau, \, \chi}^{(\pi N)}- \frac{1}{2}  {v_2 }_{\rho \tau, \, \chi}^{(\pi N)}
\right)\,;
\nonumber \\
&&
 \gamma^5_{\chi \eta} a^N_{\rho \tau, \, \eta}  = \frac{1}{M}
\left(
 {a_1}_{\rho \tau, \, \chi}^{(\pi N)}- \frac{1}{2}  {a_2 }_{\rho \tau, \, \chi}^{(\pi N)}
\right)\,;
\nonumber \\  &&
 \gamma^5_{\chi \eta} t^N_{\rho \tau, \, \eta}  = -\frac{1}{M}
\left(
 {t_1}_{\rho \tau, \, \chi}^{(\pi N)}- \frac{1}{2}  {t_2 }_{\rho \tau, \, \chi}^{(\pi N)}
\right)\,.
\ee

One may check that in the chiral limit this results in the following contributions to
the independent isospin-$\frac{1}{2}$ and isospin-$\frac{3}{2}$ $\pi N$ TDAs
(\ref{independent1/2}), (\ref{independent3/2}) regular at $\Delta^2=M^2$:
\be
&&
\left. \phi^{(\pi N)_{1/2}}_1(x_1,x_2,x_3, \xi=1, \Delta^2=M^2)\right|_{\rm soft \atop pion}= 
\frac{1}{24} \phi^N \left(\frac{x_1}{2},\frac{x_2}{2},\frac{x_3}{2} \right)+\frac{1}{6} \phi^N \left(\frac{x_3}{2},\frac{x_2}{2},\frac{x_1}{2} \right)\,;
\nonumber \\ &&
\left. \phi^{(\pi N)_{1/2}}_2(x_1,x_2,x_3, \xi=1, \Delta^2=M^2)\right|_{\rm soft \atop pion}=
\left.- \frac{1}{2} \phi^{(\pi N)_{1/2}}_1(x_1,x_2,x_3, \xi=1, \Delta^2=M^2)\right|_{\rm soft \atop pion}\,;
\nonumber \\ &&
\left. \phi^{(\pi N)_{3/2}}_1(x_1,x_2,x_3, \xi=1, \Delta^2=M^2)\right|_{\rm soft \atop pion}= \frac{1}{4}  \left( \phi^N\left(\frac{x_1}{2},\frac{x_2}{2},\frac{x_3}{2}\right)+ \phi^N\left(\frac{x_3}{2},\frac{x_2}{2},\frac{x_1}{2}\right) \right)\,;
\nonumber \\ &&
\left. \phi^{(\pi N)_{3/2}}_2(x_1,x_2,x_3, \xi=1, \Delta^2=M^2)\right|_{\rm soft \atop pion}=
\left.- \frac{1}{2} \phi^{(\pi N)_{3/2}}_1(x_1,x_2,x_3, \xi=1, \Delta^2=M^2)\right|_{\rm soft \atop pion}\,.
\nonumber \\ &&
\ee
The singular at
$\Delta^2=M^2$
contribution from the
$u$-channel nucleon exchange pole is considered in the next Section.

\section{$u$-channel $N$ and $\Delta$ exchange  contribution into $\pi N$ TDAs}
\label{Section_u-channel_resonance}

In this section, by employing the results of Sec.~\ref{Section_Isopin_par_DA}, \ref{Section_isospin_TDAs},
 we construct a simple resonance exchange model for the isospin-$\frac{1}{2}$
and isospin-$\frac{3}{2}$ $\pi N$ TDAs. It represents a consistent model for $\pi N$
TDAs in the Efremov-Radyushkin-Brodsky-Lepage (ERBL)-like region and satisfies the appropriate symmetry relations established in
Sec.~\ref{Section_isospin_TDAs} as well as the polynomiality conditions of Sec.~\ref{Section_polynomiality}.
It turns out, in particular, that the nucleon exchange results in a pure $D$-term contribution supplementary to
the spectral representation of
\cite{Pire:2010if}. Let us also mention that the nucleon pole contribution may become  dominant in the
near to threshold kinematics of the reaction
(\ref{Direct_channel_reaction}),
where
$\Delta^2-M^2$
is small enough.

\subsection{Nucleon exchange contribution}
The effective Hamiltonian for $\pi \bar{N} N$ interaction can be written as (see {\it e.g.} \cite{EricsonWeise}):
\be
\mathcal{H}_{\rm eff}(\pi N N)=  i g_{\pi N N} \bar{N}_\alpha (\sigma_a)^\alpha_{\;\beta} \gamma_5 N^\beta \pi_a\,.
\ee

After the reduction the matrix element in question reads:
\be
&&
\langle \pi_a(p_\pi) | \widehat{O}^{\alpha \, \beta \, \gamma}_{\rho \tau \chi}( \lambda_1 n, \,\lambda_2 n, \, \lambda_3 n )| N_{\iota}(p_1,s_1) \rangle
\nonumber \\ &&
= \sum_{s_p}\langle 0 |
\widehat{O}^{\alpha  \beta   \gamma}_{\rho \tau \chi}( \lambda_1 n, \,\lambda_2 n, \, \lambda_3 n )
| N_\kappa(-\Delta, s_p) \rangle (\sigma_a)^\kappa_{\ \  \iota}
\frac{i g_{\pi NN} \, \bar{U}_\varrho (-\Delta, s_p) }{\Delta^2-M^2} \left( \gamma^5 U(p_1, s_1) \right)_\varrho\,.
\nonumber \\ &&
\label{matrix_el}
\ee
$\pi N$ TDAs are computed form the matrix element
(\ref{matrix_el})
with the help of the Fourier transform (\ref{Fourier_tr}).
%:
%\be
%\mathcal{F}(x_1, \, x_2, \, x_3)(...)= (p \cdot n)^3 \int   \left[ \prod_{k=1}^3 \frac{d \lambda_k}{2 \pi}   \right]
%e^{i \sum_{k=1}^3 x_k \lambda_k (p \cdot n)} (...)
%\ee

Let us first consider isospin structure of
(\ref{matrix_el}).
Employing the isospin decomposition of the nucleon DA
(\ref{Isospin_parmetrization_N_DA})
one may check that
(\ref{matrix_el})
contributes only into the invariant amplitudes
$M^{(\pi N)_{1/2}\,\{12\}}_{\rho   \tau \chi}$
and
$M^{(\pi N)_{1/2}\,\{13\}}_{\rho  \tau  \chi}$.

The inverse Fourier transform  allowing to express the matrix element in
the second line of (\ref{matrix_el})
through the nucleon DA
$M^{N\,\{12\}}_{\rho   \tau \chi}(y_1, \, y_2,\, y_3)$
reads:
\be
\mathcal{F}^{-1} (\lambda_k (-\Delta \cdot n) )(...)
=   \int d^3y \delta(1-y_1-y_2-y_3) e^{i (p \cdot n) 2 \xi \sum_{k=1}^3 y_k \lambda_k} (...)\,.
\ee

Calculation of the Fourier transform (\ref{Fourier_tr}) of (\ref{matrix_el}) gives:
\be
&&
4 \mathcal{F}(x_1,x_2,x_3)
\frac{1}{4}
\big[ \mathcal{F}^{-1} (\lambda_k (-\Delta \cdot n) ) \big[
M^{N\,\{12\}}_{\rho   \tau \chi}(y_1, \, y_2,\, y_3) \big] \big] \nonumber \\ &&
= (p \cdot n)^3 \int_0^1 dy_1 dy_2 dy_3 \delta(1-y_1-y_2-y_3)
\big[ \prod_{k=1}^3 \frac{1}{2 \pi} \int d \lambda_k e^{i \lambda_k (x_k-2 \xi y_k) (p \cdot n)}  \big]
M^{N\,\{12\}}_{\rho   \tau \chi}(y_1, \, y_2,\, y_3)
 \nonumber \\ &&
 =\frac{1}{(2 \xi)^2} \delta(x_1+x_2+x_3-2\xi) \left[ \prod_{k=1}^3 \theta(0 \le x_k \le 2 \xi)  \right]
 M^{N\,\{12\}}_{\rho   \tau \chi} \left( \frac{x_1}{2 \xi}, \frac{x_2}{2 \xi}, \frac{x_3}{2 \xi} \right)\,.
\ee

Thus, we obtain the following result for the contribution of the matrix element  (\ref{matrix_el}) into $\pi N$ TDA:
\be
&&
M^{(\pi N)_{1/2}\,\{12\}}_{\rho   \tau \chi}(x_1, x_2, x_3)
=\frac{1}{(2 \xi)^2} \delta(x_1+x_2+x_3-2\xi) \left[ \prod_{k=1}^3 \theta(0 \le x_k \le 2 \xi)  \right] \nonumber \\ &&
\times
 f_N
 \sum_{s_p}
 \big\{
V^p(\frac{x_1}{2 \xi},\frac{x_2}{2 \xi},\frac{x_3}{2 \xi})  (-\hat{\Delta} C  )_{\rho \, \tau} (\gamma^5 {U}(-\Delta,s_p))_{\chi}
\nonumber \\ &&
+A^p(\frac{x_1}{2 \xi},\frac{x_2}{2 \xi},\frac{x_3}{2 \xi})  \left( -\hat{\Delta}  \gamma^5 C\right)_{\rho \, \tau}U(-\Delta,s_p)_\chi
\nonumber \\ &&
+ T^p(\frac{x_1}{2 \xi},\frac{x_2}{2 \xi},\frac{x_3}{2 \xi}) \left(\sigma_{-\Delta \, \mu} C \right)_{\rho \, \tau} \left(\gamma^\mu \gamma^5 U (-\Delta,s_p) \right)_\chi
\big\}
\frac{i g_{\pi NN}\, \bar{U}_\varrho (-\Delta, s_p) }{\Delta^2-M^2} \left( \gamma^5 U(p_1, s_1) \right)_\varrho\,.
\nonumber \\ &&
\label{Nucleon_exchang_contr_into_TDA}
\ee
Now it is straightforward to trace the contribution of the nucleon exchange matrix element into the
particular invariant functions occurring in the parametrization of $\pi N$ TDA. For this issue, employing
formulas given in the  Appendix~\ref{app_c}, one has to express the Dirac structures in
(\ref{Nucleon_exchang_contr_into_TDA})
in terms of standard ones.
For example, let us consider the first term in
(\ref{Nucleon_exchang_contr_into_TDA}).
To the leading twist accuracy
\be
&&
 \sum_{s_p}  (-\hat{\Delta} C  )_{\rho \, \tau} (\gamma^5 {U}(-\Delta,s_p))_{\chi} \left( \bar{U} (-\Delta, s_p) \right)_\varrho
 \left( \gamma^5 U(p_1, s_1) \right)_\varrho
 \nonumber \\ &&
 =
 2 \xi (\hat{P} C)_{\rho \tau} ((\hat{\Delta} U(p_1,s_1))_\chi+ M( U(p_1,s_1))_\chi
  \nonumber \\ &&
 =2 \xi (\hat{P} C)_{\rho \tau} ((\hat{P} U(p_1,s_1))_\chi+  \xi (\hat{P} C)_{\rho \tau} ((\hat{\Delta} U(p_1,s_1))_\chi\,.
\ee
Finally, one establishes the expressions for the contribution of the nucleon exchange into the isospin-$\frac{1}{2}$ $\pi N$ TDAs
\be
&&
%\big\{ V_1^{(\pi N)_{1/2}}, \, A_1^{(\pi N)_{1/2}}, \, T_1^{(\pi N)_{1/2}} \big\}(x_1,x_2,x_3) \ \ \Leftarrow \ \   %\delta(x_1+x_2+x_3-2\xi)
\big\{ V_1, \, A_1 , \, T_1  \big\}^{(\pi N)_{1/2}} (x_1,x_2,x_3,\xi,\Delta^2)\Big|_{N(940)}
%
 %\ \ \Leftarrow
%\left[ \prod_{k=1}^3 \theta(0 \le x_k \le 2 \xi)  \right]
\nonumber \\ &&
 =\Theta_{\rm ERBL}(x_1,x_2,x_3) \times  (g_{\pi NN}) \frac{M f_\pi}{\Delta^2-M^2} 2 \xi  \frac{1}{(2 \xi)^2}
 \big\{ V^p,\,A^p, \,T^p  \big\}\left( \frac{x_1}{2 \xi}, \frac{x_2}{2 \xi}, \frac{x_3}{2 \xi} \right)\,;
  \nonumber \\ &&
%\big\{ V_2^{(\pi N)_{1/2}}, \, A_2^{(\pi N)_{1/2}}, \, T_2^{(\pi N)_{1/2}} \big\}(x_1,x_2,x_3) \ \ \Leftarrow \ \   %\delta(x_1+x_2+x_3-2\xi)
\big\{ V_2, \, A_2 , \, T_2  \big\}^{(\pi N)_{1/2}} (x_1,x_2,x_3,\xi,\Delta^2)\Big|_{N(940)} %\ \ \Leftarrow
%\left[ \prod_{k=1}^3 \theta(0 \le x_k \le 2 \xi)  \right]
  \nonumber \\ &&
  =\Theta_{\rm ERBL}(x_1,x_2,x_3) \times  (g_{\pi NN}) \frac{M f_\pi}{\Delta^2-M^2}   \xi  \frac{1}{(2 \xi)^2}
 \big\{ V^p,\,A^p, \,T^p  \big\} \left( \frac{x_1}{2 \xi}, \frac{x_2}{2 \xi}, \frac{x_3}{2 \xi} \right)\,,
\label{Nucleon_exchange_contr_VAT}
\ee
where we introduced the notation
\be
\Theta_{\rm ERBL}(x_1,x_2,x_3)  \equiv  \prod_{k=1}^3 \theta(0 \le x_k \le 2 \xi)\,.
\label{theta_ERBL}
\ee
Notice that
(\ref{Nucleon_exchange_contr_VAT})
is a pure $D$- term contribution. It is nonzero only in the ERBL-like region and its $(n_1,n_2,n_3)$-th ( $n_1+n_2+n_3=N$)
Mellin moments
give  rise to   monomials of $\xi$  of the maximal allowed power $N+1$.

\subsection{$\Delta(1232)$ exchange contribution}

The effective Hamiltonian for $\Delta N \pi$ interaction reads (see {\it e.g.} \cite{SemenovTianShansky:2007hv}):
\be
\mathcal{H}_{\rm eff}(\pi N  {\Delta}) =
g_{\pi N  {\Delta}}
\overline{N}_\kappa \,
%P_{  \frac{3}{2} }
{P^{ 3/2}}_{b \ \ a \, \iota}^{\; \kappa}
 R_{\mu \, a}^{\,  \iota} \,
\partial^{\mu}   \pi_b
+ h.c.,
\label{Effective_Lag_Delta}
\ee
where $P^{  (3/2)}$
denotes the isospin-$\frac{3}{2}$ projecting operator (\ref{isospin_projecting_oper}).
$g_{\pi N  {\Delta}}$ is a dimensional coupling constant.
%Computation from average PDG \cite{PDG}
%${\Delta(1232)} \rightarrow \pi N$  decay width gives:
%$
%|g_{\pi N  {\Delta}}| \approx 7.6 \ \ {\rm GeV}^{-1}\,.
%$
As usual, the $\Delta$ resonance is described with the help of the Rarita-Schwinger spin-tensor
$\mathcal{U}^\mu_{\rho}
%(-\Delta,s_\Delta)
$
which satisfies the auxiliary conditions (\ref{Conditions_for_Delta_spinor}).

After the reduction the matrix element in question reads:
\be
&&
\langle \pi_a(p_\pi) | \widehat{O}^{\alpha  \beta   \gamma}_{\rho \tau \chi}( \lambda_1 n, \,\lambda_2 n, \, \lambda_3 n )| N_{\iota}(p_1,s_1) \rangle
\nonumber \\ &&
= \sum_{s_\Delta}\langle 0 |
\widehat{O}^{\alpha  \beta   \gamma}_{\rho \tau \chi}( \lambda_1 n, \,\lambda_2 n, \, \lambda_3 n )
| \Delta_{b \kappa}(-\Delta, s_\Delta) \rangle
{P^{3/2}}^{\;\;  \kappa}_{b \ \  a \iota}
\frac{   g_{\pi N \Delta} \, \bar{\mathcal U}_\varrho^\nu  (-\Delta, s_\Delta) }{\Delta^2-M^2} ( i P_\nu) \left(    U(p_1, s_1) \right)_\varrho\,.
\nonumber \\ &&
\label{matrix_el_Delta}
\ee
For the matrix element involving $\Delta$ we employ the parametrization
(\ref{isospin_parametrization_Delta_DA})
with
$M^\Delta_{\rho \tau \chi}$
given by (\ref{Parametrization_Delta_DA_FZ}).
As a consequence of the identity (\ref{Projecting_ta_tensor})
$\Delta$ exchange populates only the isospin-$\frac{3}{2}$ $\pi N$ TDAs.

To compute the    on-shell  numerator of graph (\ref{matrix_el_Delta}) we employ the method of
contracted projectors \cite{Weinberg:1964cn} (see also Appendix I of Chapter~I of \cite{Alfaro_red_book}).
We introduce the corresponding on-shell spin sum:
\begin{eqnarray}
&& \Pi^{\mu }_{\nu  \rho \tau}(-\Delta)|_{(-\Delta)^2=M_\Delta^2}
\equiv
\sum\limits_{s_\Delta= -3/2}^{3/2}
{\mathcal U}_{\rho \; \nu}(-\Delta, s_{\Delta}) \,
\overline{\mathcal U}^{\mu}_{\tau\, }(-\Delta, s_{\Delta})\,,
\label{spin-sum-fermion}
\end{eqnarray}
which carries two Dirac indices as well as two Lorentz indices.
The contracted projector is defined as
\begin{eqnarray}
&&{\cal P}^{(\frac{3}{2})}_{\rho\tau}( \Sigma, P, -\Delta) \equiv
\Sigma^{\nu }
\Pi^{\mu }_{\nu  \rho \tau}(-\Delta)
{P}_{\mu }\,,
\label{cpr5}
\end{eqnarray}
where $P_\mu=\frac{1}{2}(p_1+p_\pi)_\mu$ and $\Sigma$ in principle may be an arbitrary vector.
In order to keep with the $u$-channel baryon resonance exchange picture of the
$\gamma^* N \rightarrow N \pi$
reaction (\ref{Direct_channel_reaction}), $\Sigma$ should be chosen as:
\be
\Sigma_\mu= \frac{1}{2} (q+p_2)_\mu\,.
\ee
We also introduce the components of $P_\mu$ and $\Sigma_\mu$ transverse with respect to
$\Delta_\mu$   denoted as $\s{P}_\mu$ and $\s{\Sigma}_\mu$%
\footnote{Not to be confused with the contraction with $\gamma$ matrices.
We rather adopt  Dirac's ``hat'' notation for this issue.}:
\be
\s{P}_\mu \equiv P_\mu- \frac{(P \cdot \Delta)}{\Delta^2} \Delta_\mu\,;  \ \ \
\s{\Sigma}_\mu \equiv \Sigma_\mu- \frac{(\Sigma \cdot \Delta)}{\Delta^2} \Delta_\mu\,;
\ee
Then the explicit expression for the    on-shell  contracted projector reads \cite{Alfaro_red_book}:
\be
&&
\left. {\cal P}^{(\frac{3}{2})}_{\rho\tau}( \Sigma, P, -\Delta) \right|_{(-\Delta)^2=M_\Delta^2}
\nonumber \\ &&
=
-\frac{1}{3}
|\s{P}| |\s{\Sigma}|
\big\{
P_{2}' \left(
 \frac{(\s{\Sigma} \cdot \s{P})}{|\s{\Sigma}| |\s{P}|}
\right)
- \frac{\hat{\s{\Sigma}} \hat{\s{P}}  }{|\s{\Sigma}| |\s{P}|}
P_{1}' \left(
 \frac{(\s{\Sigma} \cdot \s{P})}{|\s{\Sigma}| |\s{P}|}
\right)
\big\}
( -\hat{\Delta}+M_\Delta)
%%%%%%%%%%%%%%%%%%%%%%%%%
\nonumber \\ && =
-\frac{1}{3} |\s{\Sigma}| |\s{P}|
\big\{
 3 \frac{(\s{\Sigma} \cdot \s{P})}{|\s{\Sigma}| |\s{P}|} - \frac{1}{|\s{\Sigma}| |\s{P}|}   \hat{\s{\Sigma}} \hat{\s{P}}  \big\}
 ( -\hat{\Delta}+M_\Delta)\,,
\ee
where $P_{k}'(...)$ stands for the derivative of the $k$-th Legendre polynomial. Note that the argument of the polynomials
is the cosine of the $u$-channel center-of-mass frame scattering angle at
$(-\Delta)^2=M_\Delta^2$:
\be
\cos \theta_u= \frac{(\s{\Sigma} \cdot \s{P})}{|\s{\Sigma}| |\s{P}|}=
\frac{1- \xi \frac{M^2 -m^2}{\Delta^2}}{\xi \sqrt{1+ \frac{(M^2-m^2)^2}{(\Delta^2)^2}- \frac{2(M^2+m^2)}{\Delta^2}}}+ O \left(
\frac{1}{Q^2}
\right)\,,
\ee

For our purpose we also need the derivative of the contracted projector:
\be
&&
\frac{\partial}{\partial \Sigma^\mu} {\cal P}^{(\frac{3}{2})} ( \Sigma, P, -\Delta)=
\big\{
%\left( -P^\mu+ \frac{(P \cdot \Delta)}{\Delta^2} \Delta^\mu \right)
- {\s{P}}^\mu
+
\frac{1}{3}  \left(\gamma^\mu- \frac{\Delta^\mu}{\Delta^2} \hat{\Delta} \right) \hat{\s{P}}
\big\} (-\hat{\Delta}+M_\Delta)\,. \nonumber \\ &&
\ee
The calculation of contributions of graph
(\ref{matrix_el_Delta})
into the appropriate invariant form factors
is then straightforward and analogous to that for the case of nucleon exchange.
For example to trace the contributions into $V_{1,2}^{(\pi N)_{3/2}}$
one has to decompose
\be
&&
(\gamma_\mu C)_{\rho \tau}
\left(
\sum_{s_\Delta} \mathcal{U}^\mu(-\Delta, s_\Delta)  \bar{\mathcal{U}}_\nu(-\Delta, s_\Delta) P^\nu U(p_1,s_1)
\right)_\chi \nonumber \\ &&
=(\gamma_\mu C)_{\rho \tau}
\left(
\frac{\partial}{\partial \Sigma^\mu} {\cal P}^{(\frac{3}{2})} (  \Sigma, P, -\Delta) U(p_1,s_1)
\right)_\chi
\ee
over the basis of the Dirac structures of (\ref{Decomposition_piN_TDAs_new_M32}).

After some algebra one may work out the following contributions
of
(\ref{matrix_el_Delta})
into the invariant form factors  (\ref{Decomposition_piN_TDAs_new_M32}) to the leading twist-$3$:
\be
&&
 \left\{ V_{1,2}^{(  \pi N)_{3/2}}, \,  A_{1,2}^{(  \pi N)_{3/2}} \right\} (x_1,x_2,x_3,\xi,\Delta^2)\Big|_{\Delta(1232)}
%\; \Leftarrow \;   %\delta(x_1+x_2+x_3-2\xi)
%\left[ \prod_{k=1}^3 \theta(0 \le x_k \le 2 \xi)  \right]
  \nonumber \\ &&
=-\Theta_{\rm ERBL}(x_1,x_2,x_3)
 \frac{1}{(2 \xi)^2}\left\{ V^\Delta, \, A^\Delta \right\} \left( \frac{x_1}{2 \xi}, \frac{x_2}{2 \xi}, \frac{x_3}{2 \xi} \right)
%  \nonumber \\ &&
%
% \times
 \frac{ g_{\pi N \Delta} \lambda_{\Delta}^{\frac{1}{2}} M f_\pi}{ \sqrt{2}(\Delta^2-M_\Delta^2) f_N} \;
R_{1,2}( \xi,  M_\Delta)\,;
  \nonumber \\
&&
T_{1,2}^{(\pi N)_{3/2}}(x_1,x_2,x_3, \xi,\Delta^2)\Big|_{\Delta(1232)}
%\ \ \Leftarrow \ \   %\delta(x_1+x_2+x_3-2\xi)
\nonumber \\ && %\times
=-\Theta_{\rm ERBL}(x_1,x_2,x_3)
\Big\{
%\left[ \prod_{k=1}^3 \theta(0 \le x_k \le 2 \xi)  \right]
\frac{1}{(2 \xi)^2} T^\Delta \left( \frac{x_1}{2 \xi}, \frac{x_2}{2 \xi}, \frac{x_3}{2 \xi} \right)
% \times
\frac{ g_{\pi N \Delta} \lambda_{\Delta}^{\frac{1}{2}} M f_\pi}{\sqrt{2}(\Delta^2-M_\Delta^2) f_N}
 R_{1,2}( \xi,  M_\Delta)
  \nonumber \\ &&
 +
 \frac{1}{(2 \xi)^2} \phi^\Delta \left( \frac{x_1}{2 \xi}, \frac{x_2}{2 \xi}, \frac{x_3}{2 \xi} \right)
  %\times
 \frac{  g_{\pi N \Delta} f_{\Delta}^{\frac{3}{2}} M^2 f_\pi}{\sqrt{2}(\Delta^2-M_\Delta^2) f_N}
 %\left( - \frac{M_\Delta}{M} \right)
\widetilde{ R}_{1,2}(\xi, M_\Delta ) \Big\}\,;
  \nonumber \\ &&
T_{3,4}^{(\pi N)_{3/2}}(x_1,x_2,x_3, \xi, \Delta^2)\Big|_{\Delta(1232)}
%\ \ \Leftarrow \ \   %\delta(x_1+x_2+x_3-2\xi)
%\left[ \prod_{k=1}^3 \theta(0 \le x_k \le 2 \xi)  \right]
\nonumber \\ &&
=-\Theta_{\rm ERBL}(x_1,x_2,x_3)
\frac{1}{(2 \xi)^2} \phi^\Delta \left( \frac{x_1}{2 \xi}, \frac{x_2}{2 \xi}, \frac{x_3}{2 \xi} \right)
%  \nonumber \\ &&
 \times
 \frac{  g_{\pi N \Delta} f_{\Delta}^{\frac{3}{2}} M^2 f_\pi}{\sqrt{2}(\Delta^2-M_\Delta^2) f_N}
 %\left( - \frac{M_\Delta}{M} \right)
 R_{3,4}(M_\Delta)\,.
   \nonumber \\ &&
   \label{Contribution_of_Delta_exchange_final}
\ee
Here $\lambda_{\Delta}^{\frac{1}{2}}$ is a dimensional constant  with the dimension $[{\rm GeV}]^3$
and $f_{\Delta}^{\frac{3}{2}}$ is a dimensional constant  with the dimension $[{\rm GeV}]^2$.
In Ref.~\cite{Farrar} the following numerical values are quoted:
\be
&&
|\lambda_{\Delta}^{\frac{1}{2}}| \equiv \sqrt{\frac{3}{2}} M_\Delta |f_{\Delta}^{\frac{1}{2}}|=(1.8 \pm 0.3) \times 10^{-2} \ \ {\rm GeV}^3\,;
\nonumber \\ &&
|f_{\Delta}^{\frac{3}{2}}|=1.4 \times 10^{-2} \ \ {\rm GeV}^2\,.
\ee

The functions $R_{1,2}$, $\widetilde{R}_{1,2}$ are determined by  residue at the
pole $\Delta^2=M_\Delta^2$.
They read as:
\be
&&
R_1(\xi,   M_\Delta )=
\frac{(\xi -3) M_\Delta^2+2 M   M_\Delta \xi +4 \left(M^2-m^2 \right) \xi }{3 M M_\Delta}\,;
\nonumber \\ &&
R_2(\xi,  M_\Delta )=
\frac{-4 \xi  M^3+\left(4 \xi  m^2+6 M_\Delta^2\right) M-M_\Delta^3 (\xi -3)+4 m^2 M_\Delta \xi }{6 M M_\Delta^2}\,;
\nonumber \\ &&
R_3(M_\Delta )=- \frac{M_\Delta}{M}\,; \ \ \ R_4(M_\Delta )=1+ \frac{M_\Delta}{2M}\,;
\nonumber \\ &&
\widetilde{R_1}(\xi,   M_\Delta )= \frac{\left(M (M+M_\Delta)-m^2\right) \xi }{M^2}
-M_\Delta^2 \frac{(1-\xi)}{2 M^2}\,;
\nonumber \\ &&
\widetilde{R_2}(\xi,   M_\Delta )=
\frac{    M M_\Delta+(m^2+M^2) \xi }{2 M^2}+ M_\Delta^2 \frac{(1-\xi)}{4 M^2}\,.
\label{R_on_shell}
\ee

The crucial point is that the $\Delta$ exchange contribution into $\pi N$ TDAs
should satisfy symmetry relations for the isospin-$\frac{3}{2}$ TDAs
established in Sec.~\ref{Section_isospin_TDAs}.
Employing the set of  the Fierz identities, one may check that the part of
(\ref{Contribution_of_Delta_exchange_final})
involving contributions of
$V^\Delta$, $A^\Delta$, $T^\Delta$
decouples and satisfies the symmetry relations
(\ref{symmetries_MI32piNTDA})
as a consequence of symmetry relations
(\ref{Isospin_and_Sym_Relations_DeltaDA})
for
$\Delta$ DAs.

The situation with the contribution involving $\phi^\Delta$ is more complicated.
This turns out to be due to the fact that the Fierz identities
(\ref{Fierz for t34})
for the tensor structures
$t_3^{ \pi N }$,
$t_4^{ \pi N }$
involve coefficient functions with explicit dependence on
$\xi$
and
$\Delta^2$.
In order to satisfy symmetry relation
(\ref{symmetries_MI32piNTDA})
we have to add a polynomial background to
$\pi N$ TDAs $T_{1,2}$.
Symmetry relation
(\ref{symmetries_MI32piNTDA})
fixes the background uniquely. This provides the following final result
for   $\Delta(1232)$ exchange contributions into
$T_{1,2}^{(\pi N)_{3/2}}$:
\be
&&
\left. T_{1}^{(\pi N)_{3/2}}(x_1,x_2,x_3, \xi, \Delta^2) \right|_{\Delta(1232)}
\nonumber \\ &&
=-\Theta_{\rm ERBL}(x_1,x_2,x_3)
\Big\{
\frac{1}{(2 \xi)^2} T^\Delta \left( \frac{x_1}{2 \xi}, \frac{x_2}{2 \xi}, \frac{x_3}{2 \xi} \right)
\frac{ g_{\pi N \Delta} \lambda_{\Delta}^{\frac{1}{2}} M f_\pi}{\sqrt{2} f_N (\Delta^2-M_\Delta^2) }
 R_{1}( \xi,  M_\Delta)
  \nonumber \\ &&
 +
 \frac{1}{(2 \xi)^2} \phi^\Delta \left( \frac{x_1}{2 \xi}, \frac{x_2}{2 \xi}, \frac{x_3}{2 \xi} \right)
 \frac{  g_{\pi N \Delta} f_{\Delta}^{\frac{3}{2}} M^2 f_\pi}{ \sqrt{2} f_N  }
 \left(
\frac{\widetilde{ R}_{1 }(\xi, M_\Delta )}{ \Delta^2-M_\Delta^2 } - \frac{1-\xi}{2 M^2} \right)\Big\}\,;
\ee
\be
&&
\left. T_{2}^{(\pi N)_{3/2}}(x_1,x_2,x_3, \xi, \Delta^2) \right|_{\Delta(1232)}
\nonumber \\ &&
=
-\Theta_{\rm ERBL}(x_1,x_2,x_3)
\Big\{
\frac{1}{(2 \xi)^2} T^\Delta \left( \frac{x_1}{2 \xi}, \frac{x_2}{2 \xi}, \frac{x_3}{2 \xi} \right)
\frac{ g_{\pi N \Delta} \lambda_{\Delta}^{\frac{1}{2}} M f_\pi}{\sqrt{2}f_N (\Delta^2-M_\Delta^2) }
 R_{ 2}( \xi,  M_\Delta)
  \nonumber \\ &&
 +
 \frac{1}{(2 \xi)^2} \phi^\Delta \left( \frac{x_1}{2 \xi}, \frac{x_2}{2 \xi}, \frac{x_3}{2 \xi} \right)
 \frac{  g_{\pi N \Delta} f_{\Delta}^{\frac{3}{2}} M^2 f_\pi}{ \sqrt{2} f_N }
 \left(
\frac{\widetilde{ R}_{2 }(\xi, M_\Delta )}{ \Delta^2-M_\Delta^2 } + \frac{1-\xi}{4 M^2} \right)\Big\}\,.
\ee

\section{Conclusions}
\label{Section_conclusions}

We considered  general symmetry properties of $\pi N$ transition distribution amplitudes.
We showed that the Lorentz invariance results in the polynomiality property of the
Mellin moments of TDAs in the longitudinal momentum fractions.
Analogously to the GPD case, we revealed the presence of a $D$-term contribution for
the $\pi N$ TDAs $V_{1,2}$, $A_{1,2}$ and $T_{1,2}$ generating
the highest power monomials of the Mellin moments.

The detailed account of the isospin and permutation symmetries allowed us to
provide a unified description of all isotopic channels in terms of eight
independent  $\pi N$ TDAs.
The general constraints derived here should be satisfied by
any realistic model  of TDAs.

The crossing relation between $\pi N$ TDAs and GDAs lead us to establish
a soft pion theorem for the isospin-$\frac{1}{2}$ and isospin-$\frac{3}{2}$ $\pi N$ TDAs.
This yields normalization conditions for $\pi N$ TDAs.

We also presented a simple resonance exchange model for $\pi N$ TDAs considering nucleon
and $\Delta(1232)$ exchanges in the
isospin-$\frac{1}{2}$ and isospin-$\frac{3}{2}$ channels, respectively. Nucleon exchange
may be considered as a pure $D$-term contribution complementary to the spectral representation
for TDAs in terms of quadruple distributions.

This work opens the way to various consistent models of baryon to meson TDAs to be confronted with experimental data.

\section*{Acknowledgements}
%We would especially like to thank
We are   thankful to Vladimir Braun,  Jean-Philippe Lansberg, Alexandr Manashov,
Vladimir Vereshagin,
Samuel~Wallon and Christian Weiss
for many discussions and helpful comments.
This work is supported in part by  Polish Grant No 202249235
and by the French-Polish Collaboration Agreement Polonium.
K.S. also acknowledges  partial support by Consortium Physique des Deux Infinis (P2I).
%This work was supported by the Polish Grant N202 249235.

\setcounter{section}{0}
\setcounter{equation}{0}
\renewcommand{\thesection}{\Alph{section}}
\renewcommand{\theequation}{\thesection\arabic{equation}}

\section{Isotopic invariance and isospin classification of $\pi N$ states}
Generators of the
$SU(2)$
isospin group satisfy the familiar commutation relation:
\be
[I_a,\,I_b]= i \varepsilon_{abc} I_c\,.
\label{Ispin}
\ee
When constructing   spinor representations of
$SU(2)$
one has to distinguish between the covariant and the contravariant representations.
We choose to transform
$\bar{N}_\alpha$
field  according to the covariant representation and to transform
$N^\alpha$
field  according to the contravariant representation
\be
\left[
I_a, \bar{N}_\alpha
\right]
= \frac{1}{2} \left( \sigma_a \right)^\beta_{\; \alpha} \bar{N}_\beta\,;
\ \ \
\left[
I_a, N^\alpha\right]=- \frac{1}{2} \left( \sigma_a \right)^\beta_\alpha N^\alpha\,,
\label{isospin_commutators_with_N}
\ee
where
$\sigma_a$
are the Pauli matrices.

We adopt the following standard convention upon the nucleon field
\cite{Itzykson}:
\be
&&
N^\alpha(x)=
\int \frac{d^3 k}{(2 \pi)^3}  \frac{M}{k_0}
\sum_{s=1,\, 2}
\big\{
e^{i k x} d^{\dag \, \alpha}(k,s) V(k,s)+e^{-i k x} b^{ \alpha}(k,s) U(k,s)
\big\} \,; \nonumber \\
&&
\bar{N}_\alpha(x)=
\int \frac{d^3 k}{(2 \pi)^3}  \frac{M}{k_0}
\sum_{s=1,\, 2}
\big\{
e^{i k x} {b^\dag}_\alpha (k,s) \bar{U}(k,s)+e^{-i k x} d_\alpha(k,s) \bar{V}(k,s)
\big\}\,.
\label{def_N_fiels}
\ee
Here spinors
$U(k,s)$
and
$\bar{U}(k,s) \equiv U^\dag (k,s) \gamma_0 $
describe a nucleon respectively in the initial and final states,
while spinors
$\bar{V}(k,s)\equiv V^\dag (k,s) \gamma_0$
and
$V(k,s)$
describe an antinucleon in the initial and final states.

The creation and annihilation operators in
(\ref{def_N_fiels})
satisfy the usual anticommutation relations for fermions \cite{Itzykson}:
\be
&&
\left\{
b^{ \alpha}(p,s), \, {b^\dag}_\beta(p',s')\right\}=(2 \pi)^3 \frac{p_0}{M} \delta^3(p-p') \delta_{ss'} \delta^{\alpha}_\beta\,;
\nonumber \\ &&
\left\{
d_\alpha(p,s), \,
{d}^{\dag \, \beta} (p',s')
\right\}
=(2 \pi)^3 \frac{p_0}{M} \delta^3(p-p') \delta_{ss'} \delta^\beta_\alpha \,.
\ee
The ``in'' nucleon state $|N_\alpha \rangle$ is defined according to:
\be
|N_1 \rangle \equiv | N_p(p,s) \rangle= {b}^{\dag}_1 (p,s) |0 \rangle \, ;  \ \ \  |N_2 \rangle \equiv  | N_n(p,s) \rangle= {b}^{\dag}_2 (p,s) |0 \rangle\,.
\label{nucleon_in_states}
\ee
Analogously,
the ``in'' antiparticle state  $|\bar{N}^\alpha \rangle$ is defined as:
\be
|\bar{N}^1 \rangle \equiv | N^{\bar{p}}(p,s) \rangle={d}^{\dag \,1 }  (p,s) |0 \rangle
 \, ;  \ \ \
|\bar{N}^2 \rangle \equiv | N^{\bar{n}}(p,s) \rangle={d}^{\dag \,2 }  (p,s) |0 \rangle\,.
\label{antinucleon_states}
\ee

In order to check the consistency of our conventions
(\ref{isospin_commutators_with_N})
and
(\ref{def_N_fiels})
we should explicitly construct
the isospin and hypercharge operators and make sure that the  nucleon and antinucleon
states
(\ref{nucleon_in_states}), (\ref{antinucleon_states})
have the proper quantum numbers.

With the help of Noether's  theorem from the free nucleon Lagrangian
\be
\mathcal{L}= \frac{i}{2}
\left[
\bar{N}_\alpha \gamma^\mu (\partial_\mu N^\alpha)-
(\partial_\mu \bar{N}_\alpha)  \gamma^\mu N^\alpha
\right] -m \bar{N}_\alpha N^\alpha
\ee
employing
(\ref{isospin_commutators_with_N})
we construct the explicit expression for the nucleon isospin operator:
\be
&&
I_a^{(N)}= \int d^3x :\!N^\dag_\alpha (x)  \frac{{(\sigma_a)}^\alpha_{\; \beta}}{2} N^\beta(x)\!:
\nonumber \\ &&
=
\int \frac{d^3 k}{(2 \pi)^3}  \frac{M}{k_0} \sum_s
\left[
{b}^{\dag}_\alpha (k,s)
\frac{{(\sigma_a)}^\alpha_{\; \beta}}{2}
b^{\beta}(k,s)-
%%%
d^{\dag \, \beta}(k,s)
\frac{{(\sigma_a)}^\alpha_{\; \beta}}{2}
{d}_\alpha(k,s)
%%%%%%%%%%%%%%%%%%%%%%%%%%%%%%
%
\right]\,.
\ee
Thus, the isospin operator acts on the incoming nucleon state according to
\be
I_a^{(N)} |N_\alpha \rangle= \frac{1}{2} (\sigma_a)^\beta_{\, \alpha} |N_\beta \rangle\,;
\ \ \
I_a^{(N)} |\bar{N}^\alpha \rangle= -\frac{1}{2} (\sigma_a)^\alpha_{\, \beta} |\bar{N}^\beta \rangle\,.
\ee

We also introduce the hypercharge operator $Y$ according to
\be
[Y, \, \bar{N}_\alpha]=\bar{N}_\alpha\,; \ \ \ [Y,N^\alpha]=-N^\alpha\,.
\ee
The explicit expression for the nucleon hypercharge operator reads
\be
&&
Y^{(N)}=
\int \frac{d^3 k}{2 \pi^3}  \frac{M}{k_0} \sum_s
\left[
{b}^{\dag}_\alpha (k,s)
%\frac{{\sigma_a}^\alpha_{\; \beta}}{2}
b^{\alpha}(k,s)-
%%%
d^{\dag \, \alpha}(k,s)
%\frac{{\sigma_a}^\alpha_{\; \beta}}{2}
{d}_\alpha(k,s)
%%%%%%%%%%%%%%%%%%%%%%%%%%%%%%
%
\right]\,.
\ee
It acts on the nucleon states according to
\be
Y_a^{(N)} |N_\alpha \rangle= |N_\alpha \rangle\,;
\ \ \
Y_a^{(N)} |\bar{N}^\alpha \rangle= -|\bar{N}^\alpha \rangle\,.
\ee

Now we construct the nucleon charge operator employing the Gell-Mann-Nishijima formula
\be
Q^{(N)}=I_3^{(N)}+\frac{Y^{(N)}}{2}
\ee
and perform the classification of states
(\ref{nucleon_in_states}), (\ref{antinucleon_states})
to check the consistency of our conventions.

The case of pion field is simpler since for the adjoint representation of
$SU(2)$
there is no difference between covariant and contravariant representation.
Indeed, using
\be
[I_a, \pi_b]=i \varepsilon_{abc} \pi_c \equiv (t_a)_{cb} \pi_c\,
\ee
one may  check that %the generator in the contravariant representation
$(t_a)_{bc}= -(t_a)_{cb}\,$.

We describe pions with the help of real pseudoscalar field $\pi_a$:
\be
\pi_a(x)= \int \frac{d^3 k}{(2 \pi)^3 2k_0}
\left(
e^{ikx} a^+_a(k)+e^{-ikx} a^-_a(k)
\right)
\ee
and adopt the usual conventions of \cite{Itzykson} for the
commutation relations of the corresponding creation/anihilation operators
$a^\pm_a$.
Pion states are defined as
$|\pi_a \rangle= a^+_a |0 \rangle$.

The expression for the pion isospin operator reads:
\be
I_a^{(\pi)}=i \int d^3 x  :\! \pi_b(x) (t_a)_{bc} \, \partial_0 \pi_c(x)\!:= \int \frac{d^3 k}{(2 \pi)^3 2k_0}
(t_a)_{bc} \, a^+_b(k) \, a^-_c(k)\,.
\ee
Pion isospin operator acts on the pion state according to
\be
I_a^{(\pi)} |\pi_b \rangle=(t_a)_{cb} |\pi_c \rangle\,.
\ee
We may construct the usual charged combinations
\be
|\pi^\pm \rangle= | \frac{\pi_1 \pm i \pi_2}{\sqrt{2}} \rangle\,; \ \ \ |\pi^0 \rangle=|\pi_3 \rangle\,.
\ee

Now we perform the isospin classification of pion-nucleon states.
Let us consider the action of the isospin operator $I_a$ on the pion-nucleon state
\be
I_a | \pi_b N_\iota \rangle=
\left\{
I_a^{(N)} \cdot   \mathbf{1}^{(\pi)}+ I_a^{(\pi)} \cdot   \mathbf{1}^{(N)}
\right\}| \pi_b N_\iota \rangle=
\left\{
i \varepsilon_{abc} \delta^\kappa_{\;\;\iota}+ \frac{1}{2} (\sigma_a)^\kappa_{\;\;\iota} \delta_{bc}
\right\}| \pi_c N_\kappa \rangle\,.
\ee

The action of the operator of the total isospin $I^2$ on the pion-nucleon state then
reads:
\be
I^2| \pi_a N_\iota \rangle=
\left\{
\frac{11}{4} \delta_{ab} \delta^\kappa_{\; \iota}-i \varepsilon_{bac} (\sigma_c)^\kappa_{\; \iota}
\right\}
|\pi_b N_\kappa \rangle\,.
\ee

This allows to classify the pion-nucleon states with respects to total isospin
$I^2$
and its third projection
$I_3$
and compute the Clebsch-Gordan coefficients:
\be
|I,\,I_3 \rangle= \sum_{a, \iota} \mathcal{C}_{a}^{\; \iota}(I,I_3) |\pi_a N_\iota \rangle\,.
\ee
Let us emphasize that the calculation of the  Clebsch-Gordan coefficients is subject of
adopting a particular phase convention. There is much controversy on this point in the literature
(see discussion in \cite{Georgi}).
To be consistent we prefer to fix our own phase convention which turns out to be different from the so-called
Condon-Shortley and Wigner phase convention adopted {\it e.g.} in the tables in \cite{PDG}.

The calculation within our phase convention gives the following result for the isospin-$\frac{3}{2}$
$\pi N$
states:
\be
&&
\left|{  \frac{3}{2}}; \, \frac{3}{2} \right\rangle=\left| \frac{\pi_1 N_1+ i \pi_2 N_1 }{\sqrt{2}} \right\rangle\,;
\ \ \
\left|\frac{3}{2}; \, \frac{1}{2} \right\rangle=- \sqrt{\frac{2}{3}}  \left|
\pi_3 N_1
\right\rangle\
+
\sqrt{\frac{1}{3}}
\left| \frac{\pi_1 N_2+ i \pi_2 N_2 }{\sqrt{2}} \right\rangle\,;
\nonumber \\ &&
\left|\frac{3}{2}; \, -\frac{1}{2} \right\rangle= \sqrt{\frac{2}{3}}  \left|
\pi_3 N_2
\right\rangle\
+
\sqrt{\frac{1}{3}}
\left| \frac{\pi_1 N_1- i \pi_2 N_1 }{\sqrt{2}} \right\rangle\,;
\ \ \
\left|\frac{3}{2}; \, -\frac{3}{2} \right\rangle=\left| \frac{\pi_1 N_2- i \pi_2 N_2 }{\sqrt{2}} \right\rangle\,.
\nonumber \\ &&
\label{isospin-3/2_piN_states}
\ee
The expansion of the isospin-$\frac{1}{2}$
$\pi N$
states reads:
\be
&&
\left|\frac{1}{2}; \, \frac{1}{2} \right\rangle=  \sqrt{\frac{1}{3}}  \left|
\pi_3 N_1
\right\rangle\
+
\sqrt{\frac{2}{3}}
\left| \frac{\pi_1 N_2+ i \pi_2 N_2 }{\sqrt{2}} \right\rangle\,;
\nonumber \\ &&
\left|\frac{1}{2}; \, -\frac{1}{2} \right\rangle= -\sqrt{\frac{1}{3}}  \left|
\pi_3 N_2
\right\rangle\
+
\sqrt{\frac{2}{3}}
\left| \frac{\pi_1 N_1- i \pi_2 N_1 }{\sqrt{2}} \right\rangle\,.
\ee

The inverse expansion reads
\be
|\pi_a N_\iota \rangle=  \sum_{I, I_3}\mathcal{D}_{a \iota}(I,I_3)|I,I_3 \rangle\,, \ \ \
{\rm where} \ \ \
\mathcal{D}_{a \iota}(I,I_3) = (\mathcal{C}_{a}^{\; \iota}(I,I_3))^\dag\,.
\label{Inverse_Exp_Clebsh}
\ee
Note that the last equality in
(\ref{Inverse_Exp_Clebsh})
should be understood as the equality of the corresponding numerical values
(and not as that of $SU(2)$ spin-tensors).

We also compute the isospin projecting operators:
\be
{P^I}_{b \ \ a \, \iota}^{\; \kappa}= \sum_{I_3, I'_3}  \mathcal{C}_{b}^{\; \kappa}(I,I_3)  \mathcal{D}_{a \iota}(I,I'_3)
|I, I_3' \rangle  \langle I, I_3|\,.
\ee
The explicit expressions for the isospin projecting operators read  \cite{SemenovTianShansky:2007hv}:
\be
%&&
{P^{ 3/2}}_{b \ \ a \, \iota}^{\; \kappa}=
\frac{2}{3}
\big(
\delta_{ba} \delta^\kappa_{\; \iota}- \frac{i}{2} \varepsilon_{bac} (\sigma_c)^\kappa_{\; \iota}
\big)\,;
\ \ \
%\nonumber \\ &&
{P^{ 1/2}}_{b \ \ a \, \iota}^{\; \kappa}=
\frac{1}{3}
\big(
\delta_{ba} \delta^\kappa_{\; \iota} + i \varepsilon_{bac} (\sigma_c)^\kappa_{\; \iota}
\big)\,.
\label{isospin_projecting_oper}
\ee

\setcounter{equation}{0}
\section{Fierz identities}
\label{App_Fierz}
Employing the Fierz identity for
$\gamma$
matrices
(see {\it e.g.}
\cite{Borodulin:1995xd})
one may establish the following useful identity for arbitrary
Dirac structures $\Gamma$, $\Gamma'$:
\be
&&
(\Gamma C)_{\rho \tau} (\Gamma' U)_\chi \nonumber \\ &&
=\frac{1}{4}
\big\{
C_{\chi \tau} (\Gamma \Gamma' U)_\rho+
(\gamma^\mu C)_{\chi \tau} (\Gamma \gamma^\mu \Gamma' U)_\rho+
(\gamma^5 C)_{\chi \tau} (\Gamma \gamma^5 \Gamma' U)_\rho-
(\gamma^5 \gamma^\mu C)_{\chi \tau} (\Gamma \gamma^5 \gamma^\mu \Gamma' U)_\rho
\nonumber \\ &&
-\frac{1}{2} (\sigma^{\mu \nu} C)_{\chi \tau}  (\Gamma \sigma_{\mu \nu} \Gamma' U)_\rho
\big\}\,.
\label{Master_Fierz}
\ee
Here $U$ stands for an arbitrary spin-tensor with one Dirac index and $C$ is the charge conjugation matrix.

\subsection{Nucleon DA}
To the leasing twist-$3$ the parametrization of the nucleon DA involves
the following Dirac structures
\be
v_{\rho \tau, \, \chi}^N=(\hat{p} C)_{\rho \tau} (\gamma^5 U )_\chi\,;
\ \
a_{\rho \tau, \, \chi}^N=(\hat{p} \gamma^5 C)_{\rho \tau} ( U )_\chi
\,;
\ \
t_{\rho \tau, \, \chi}^N=(\sigma_{p \mu} C)_{\rho \tau} ( \gamma^\mu \gamma^5 U )_\chi\,.
\label{DA structures_App}
\ee
The Dirac structures
(\ref{DA structures_App})
satisfy symmetry relations:
\be
v_{\rho \tau, \, \chi}^N=v_{ \tau \rho, \, \chi}^N\,; \ \ \
a_{\rho \tau, \, \chi}^N=-a_{\tau \rho , \, \chi}^N\,; \ \ \
t_{\rho \tau, \, \chi}^N=t_{ \tau \rho, \, \chi}^N\,.
\label{symmetry_Dirac_Nucleon}
\ee

With the help of
(\ref{Master_Fierz})
one may establish the following set of the Fierz identities valid to the leading twist-$3$ accuracy:
\be
&&
v_{\rho \tau, \, \chi}^N= \frac{1}{2} \left(v^N-  a^N-  t^N \right)_{\chi \tau, \, \rho};
\;
a_{\rho \tau, \, \chi}^N= \frac{1}{2} \left(-v^N+  a^N-  t^N \right)_{\chi \tau, \, \rho};
\nonumber \\ &&
t_{\rho \tau, \, \chi}^N=   \left(-v^N-  a^N \right)_{\chi \tau, \, \rho}.
\label{Fierz_nucleon_structures}
\ee

\subsection{$\Delta(1232)$ DA}
Leading twist Dirac structures employed in the parametrization
(\ref{Parametrization_Delta_DA_FZ})
of
$\Delta(1232)$
resonance DA read:
\be
&&
v^\Delta_{\rho \tau,\, \chi}= (\gamma_\mu C)_{\rho \tau} \, \mathcal{U}^\mu_\chi\,;
\ \ \
a^\Delta_{\rho \tau,\, \chi}= (\gamma_\mu \gamma_5 C)_{\rho \tau} \, (\gamma_5\mathcal{U}^\mu)_\chi\,;
\ \ \
t^\Delta_{\rho \tau,\, \chi}= \frac{1}{2} (\sigma_{\mu \nu} C)_{\rho \tau}(\gamma^\mu \mathcal{U}^\nu)_\chi\,;
\nonumber \\ &&
\varphi_{\rho \tau,\, \chi}^\Delta= (\sigma_{\mu \nu} C)_{\rho \tau} (p^\mu \mathcal{U}^\nu- \frac{1}{2} M_\Delta \gamma^\mu \mathcal{U}^\nu)_\chi\,.
\label{Dirac_structures_Delta_App}
\ee
The Dirac structures
(\ref{Dirac_structures_Delta_App})
satisfy symmetry relations:
\be
v_{\rho \tau, \, \chi}^\Delta=v_{ \tau \rho, \, \chi}^\Delta\,; \ \ \
a_{\rho \tau, \, \chi}^\Delta=-a_{\tau \rho , \, \chi}^\Delta\,; \ \ \
t_{\rho \tau, \, \chi}^\Delta=t_{ \tau \rho, \, \chi}^\Delta\,; \ \ \
\varphi_{\rho \tau, \, \chi}^\Delta=\varphi_{ \tau \rho, \, \chi}^\Delta\,;
\label{symmetry_Dirac_structures_Delta}
\ee
The set of the corresponding Fierz identities valid to the leading twist-$3$ accuracy reads:
\be
&&
v^\Delta_{\rho \tau, \, \chi}= \big( \frac{1}{2} v^\Delta -\frac{1}{2} a^\Delta +t^\Delta \big)_{\chi \tau, \rho}\,;
\ \ \
a^\Delta_{\rho \tau, \, \chi}=  \big( -\frac{1}{2} v^\Delta +\frac{1}{2} a^\Delta +t^\Delta \big)_{\chi \tau, \rho}\,;
\nonumber \\ &&
t^\Delta_{\rho \tau, \, \chi}=  \big( \frac{1}{2} v^\Delta +\frac{1}{2} a^\Delta  )_{\chi \tau, \rho}\,;
\ \ \
\varphi_{\rho \tau, \, \chi}^\Delta=\varphi_{\chi \tau, \rho}^\Delta\,.
\label{Fierz_Delta_structures}
\ee

\subsection{$\pi N$ TDA}
Below we consider the properties of the Dirac structures (\ref{Dirac_structures_PiN_TDA})
occurring in the parametrization of the $\pi N$ TDA. %and GDA.
Firstly, one may check that the Dirac structures
(\ref{Dirac_structures_PiN_TDA})
satisfy symmetry relations:
\be
(v_{1,2}^{\pi N})_{\rho \tau, \, \chi}=(v_{1,2}^{\pi N})_{ \tau \rho, \, \chi} \,; \ \ \
(a_{1,2}^{\pi N})_{\rho \tau, \, \chi}=-(a_{1,2}^{\pi N})_{ \tau \rho, \, \chi} \,; \ \ \
(t_{1,2,3,4}^{\pi N})_{\rho \tau, \, \chi}=(t_{1,2,3,4}^{\pi N})_{ \tau \rho, \, \chi} \,.
\label{symmetry_Dirac_structures_PiN_TDA}
\ee

The set of the corresponding Fierz identities for the structures
$s_{1,2}^{\pi N}$
is similar to that for the case of the nucleon DA (\ref{Fierz_nucleon_structures}):
\be
&&
(v_{1,2}^{\pi N})_{\rho \tau, \, \chi}=
\frac{1}{2}  (v_{1,2}^{\pi N})_{\chi \tau, \, \rho}
-\frac{1}{2} (a_{1,2}^{\pi N})_{\chi \tau, \, \rho}
-\frac{1}{2} (t_{1,2}^{\pi N})_{\chi \tau, \, \rho}\,;
\nonumber \\ &&
(a_{1,2}^{\pi N})_{\rho \tau, \, \chi}=
-\frac{1}{2}  (v_{1,2}^{\pi N})_{\chi \tau, \, \rho}
+\frac{1}{2} (a_{1,2}^{\pi N})_{\chi \tau, \; \rho}
-\frac{1}{2} (t_{1,2}^{\pi N})_{\chi \tau, \; \rho}\,;
\nonumber \\ &&
(t_{1,2}^{\pi N})_{\rho \tau, \, \chi}=
-   (v_{1,2}^{\pi N})_{\chi \tau, \; \rho}
- (a_{1,2}^{\pi N})_{\chi \tau, \; \rho}\,.
\label{Fierz_for_piNTDA_structures}
\ee
The result for
$(t_{3,4}^{\pi N})$
is a  bit more involved:
\be
&&
(t_{3}^{\pi N})_{\rho \tau, \, \chi}
\nonumber \\ &&
=
(t_{3}^{\pi N})_{\chi \tau, \, \rho}+
g_1(\xi, \Delta^2)
\left( v_{1}^{\pi N}+a_{1}^{\pi N}+t_{1}^{\pi N} \right)_{\chi \tau, \, \rho} +
g_2(\xi, \Delta^2)
\left(
v_{2}^{\pi N}+a_{2}^{\pi N}+t_{2}^{\pi N}
\right)_{\chi \tau, \, \rho}\,;
 \nonumber \\ &&
 (t_{4}^{\pi N})_{\rho \tau, \, \chi}
\nonumber \\ &&
=
(t_{4}^{\pi N})_{\chi \tau, \, \rho}+
h_1(\xi, \Delta^2)
\left( v_{1}^{\pi N}+a_{1}^{\pi N}+t_{1}^{\pi N} \right)_{\chi \tau, \, \rho} +
h_2(\xi, \Delta^2)
\left(
v_{2}^{\pi N}+a_{2}^{\pi N}+t_{2}^{\pi N}
\right)_{\chi \tau, \, \rho} \,,
 \nonumber \\ &&
 \label{Fierz for t34}
\ee
where
\be
&&
g_1(\xi, \Delta^2)=\frac{-\Delta ^2 (1-\xi)-2 \left(m^2+M^2\right) \xi }{4 M^2} \,;
\nonumber \\ &&
g_2(\xi, \Delta^2)=\frac{2 \xi  m^2+2 M^2 (\xi -2)+\Delta ^2 (1-\xi)}{8 M^2}\,;
\nonumber \\ &&
h_1(\xi, \Delta^2)=\frac{-\Delta ^2(1-\xi) -2 \left(m^2-M^2\right) \xi }{2 M^2}\,;
\nonumber \\ &&
h_2(\xi, \Delta^2)=\frac{2 \left(m^2+M^2\right) \xi +\Delta ^2 (1-\xi )}{4 M^2}\,.
 \label{Def_g12_h12}
\ee

\setcounter{equation}{0}
\section{ Choice of independent Dirac structures}
\label{app_c}

Keeping the first-order corrections in the masses and $\Delta_T^2$ one can
establish the following Sudakov decomposition for the momenta of reaction (\ref{Direct_channel_reaction})
\cite{Lansberg:2007ec}:
\be
&&
\nonumber
p_1 = (1+\xi) p + \frac{M^2}{1+\xi}n \,; \\ &&
\nonumber
q \simeq- 2 \xi \Big(1+ \frac{(\Delta_T^2-M^2)}{Q^2}\Big)  p + \frac{Q^2}{2\xi\Big(1+   \frac{(\Delta_T^2-M^2)}{Q^2}\Big)} n\,; \\ &&
\nonumber
p_\pi = (1-\xi) p +\frac{m^2-\Delta_T^2}{1-\xi}n+ \Delta_T \, ;
\\ &&
\Delta = - 2 \xi p +\Big[\frac{m-\Delta_T^2}{1-\xi}- \frac{M^2}{1+\xi}\Big]n
+ \Delta_T \,.
\ee

Because of the Dirac equation
\be
\hat{p}_1 U(p_1,s_1)= M U(p_1,s_1)
\ee
one has two following relations  for the large ($U^+=  \hat{p} \hat{n}   U $)
and small ($U^-=  \hat{n} \hat{p}  U $) components of the nucleon Dirac spinor:
\be
&&
\hat{p} \, U(p_1,s_1)= \frac{M}{1+\xi} U^+(p_1,s_1)  \ \ \  \ (p= \frac{1}{1+\xi} p_1  -\frac{M^2}{(1+\xi)^2} n)\,;\nonumber  \\  &&
\hat{n}  \, U(p_1,s_1)=\frac{1+\xi}{M} U^-(p_1,s_1) \ \ \  \ (n= \frac{1+\xi}{M^2} p_1-\frac{(1+\xi)^2}{M^2} p )\,.
\ee

As the consequence of the Dirac equation we establish the following identities:
\be
&&
(\hat{P} \, U(p_1,s_1))_{\gamma} \nonumber \\ && = \frac{M}{1+\xi} U^+(p_1,s_1)_{\gamma} + \frac{1}{2} (\hat{\Delta}_T U(p_1,s_1))_{\gamma}
+\frac{(1+\xi)}{2M} \left[  \frac{M^2}{1+\xi } + \frac{m^2-\Delta_T^2}{1-\xi} \right] U^-(p_1,s_1)_{\gamma}\,; \nonumber \\ &&
\label{Hat_P_U}
\\  &&
(\hat{\Delta}\, U(p_1,s_1))_\gamma \nonumber \\ && = -2 \xi\frac{M}{1+\xi}  U^+(p_1,s_1)+ (\hat{\Delta}_T U(p_1,s_1))_{\gamma}
-\frac{(1+\xi)}{M} \left[  \frac{M^2}{1+\xi } - \frac{m^2-\Delta_T^2}{1-\xi} \right] U^-(p_1,s_1)_{\gamma}\,. \nonumber  \\  &&
%=-2 \xi (\hat{P} \, U(p_1,s_1))_{\gamma} +(1+\xi)(\hat{\Delta}_T U(p_1,s_1))_{\gamma} \nonumber  \\  &&-
%%%%%%
%\frac{M^2 (1-\xi )^2-m^2 (1+\xi )^2+ {\Delta_T^2} (1+\xi)^2}{M (1-\xi )} (U^-(p_1,s_1))_{\gamma}
%%%%%%
%\nonumber  \\  &&
\label{Hat_Delta_U}
\ee
The last term in (\ref{Hat_P_U}) and (\ref{Hat_Delta_U})
is of sub-leading twist while the two first terms are of the leading twist.
Thus, among the four  structures containing the leading twist contribution which  can be written as
\be
(\hat{P} \, U(p_1,s_1))_{\gamma}\,; \ \ (\hat{\Delta} \, U(p_1,s_1))_{\gamma}\,;
\ \ \
U(p_1,s_1)_{\gamma}\,; \ \ \ {\rm and} \ \ (\hat{\Delta}_T U(p_1,s_1))_{\gamma}
\ee
only two are independent.
In order to keep the traditional formulation of the polynomiality condition for
$\pi N$ TDAs and avoid the appearing of singular $\frac{1}{1+\xi}$
we choose $(\hat{P} \, U(p_1,s_1))_{\gamma}$ and $(\hat{\Delta} \, U(p_1,s_1))_{\gamma}$
to be the independent Dirac structures.

We establish the following useful relations:
\be
2 \xi (\hat{P} U)_\gamma+ (\hat{\Delta} U)_\gamma = (1+\xi) (\hat{\Delta}_T  U)_\gamma+
\left(\frac{4 \xi  M^2+\left(m^2-M^2-\Delta^2\right) (1+\xi )}{M} \right) U^-_\gamma\,.
\ee
From (\ref{Hat_P_U}) we also establish the relation:
\be
&&
M U^+(p_1,s_1)_{\gamma} \nonumber  = (1+\xi) (\hat{P} \, U(p_1,s_1))_{\gamma}
- \frac{1+\xi}{2} (\hat{\Delta}_T U(p_1,s_1))_{\gamma} \\ &&
-\frac{(1+\xi)^2}{2M} \left[  \frac{M^2}{1+\xi } + \frac{m^2-\Delta_T^2}{1-\xi} \right] U^-(p_1,s_1)_{\gamma}\,.
%\nonumber \\ &&
\ee
This results in
\be
%&&&
M U (p_1,s_1)_{\gamma} %\nonumber \\ &&
=   (\hat{P} \, U(p_1,s_1))_{\gamma}
-   \frac{1}{2} (\hat{\Delta}  U(p_1,s_1))_{\gamma}
%-\frac{(1+\xi)^2}{2M} \left[  \frac{M^2}{1+\xi } + \frac{m^2-\Delta_T^2}{1-\xi} \right] U^-(p_1,s_1)_{\gamma} \nonumber
+ \big\{ {\rm Twist-}4 \ \ {\rm terms}\big\}\,.
\ee

One may also check that
\be
&&
(\hat{P} \hat{\Delta} U(p_1,s_1))_\gamma= \frac{2(P^2-M^2)}{M}(\hat{P} \, U(p_1,s_1))_{\gamma}-
\frac{P^2}{M} (\hat{\Delta} \, U(p_1,s_1))_{\gamma}
+ \big\{ {\rm Twist-}4 \ \ {\rm terms}\big\}\,;
\nonumber \\ &&
(\hat{\Delta} \hat{P} U(p_1,s_1))_\gamma= 2(P \cdot \Delta) ( U(p_1,s_1))_\gamma -(\hat{P} \hat{\Delta} U(p_1,s_1))_\gamma
\nonumber \\ &&
=
\frac{\Delta^2}{2M} (\hat{P} \, U(p_1,s_1))_{\gamma}+ (M-\frac{\Delta^2}{4M})(\hat{\Delta} \, U(p_1,s_1))_{\gamma}
+ \big\{ {\rm Twist-}4 \ \ {\rm terms}\big\}\,.
\ee

The relation of new definition (\ref{Decomposition_piN_TDAs_new}) of
$\pi N$ TDAs
to that of
\cite{Lansberg:2007ec,Pire:2010if}
is given by
\be
&&
\left.{ \{V_{1},A_{1},T_{1}\}^{\pi N}} \right|_{\text{\cite{Lansberg:2007ec,Pire:2010if}}}=
\left. \left(
\frac{1}{1+\xi} \{V_{1},A_{1},T_{1}\}^{\pi N}  -
\frac{2 \xi}{1+\xi} \{V_{2},A_{2},T_{2}\}^{\pi N}
\right)
\right|_{\text{This work}} \,;  \nonumber \\  &&
%%%
%%%
\left.{\{V_{2},A_{2}\}^{\pi N}} \right|_{\text{\cite{Lansberg:2007ec,Pire:2010if}}}=
\left. \big(  {\{V_{2},A_{2}\}^{\pi N}}  +
\frac{1}{2}   {\{V_{1},A_{1}\}^{\pi N}} \big) \right|_{\text{This work}}\,;
 \nonumber \\  &&
\left.{T_{3}^{\pi N}} \right|_{\text{\cite{Lansberg:2007ec,Pire:2010if}}}=   \left.  {T_{2}^{\pi N}} \right|_{\text{This work}}+
\frac{1}{2}  \left.  {T_{1}^{\pi N}} \right|_{\text{This work}}
\nonumber \\ &&
%%%%%%%%%%%%%%%%%%%%%%%%%%%%%%%%%%%%%
\left.{T_{2}^{\pi N}} \right|_{\text{\cite{Lansberg:2007ec,Pire:2010if}}}=   \left.
\left(
\frac{1}{2} T_1^{\pi N}+T_2^{\pi N}+T_3^{\pi N}-2\xi T_4^{\pi N}
\right)
\right|_{\text{This work}}\,;
\nonumber \\ &&
\left.{T_{4}^{\pi N}} \right|_{\text{\cite{Lansberg:2007ec,Pire:2010if}}}=   \left. \left(  \frac{1+\xi}{2}  T_{3}^{\pi N}    +
 (1+\xi)  {T_{4}^{\pi N}} \right) \right|_{\text{This work}}\,.
 \label{Old_to_new}
\ee
These relations can be easily established employing Eq.~(\ref{Hat_P_U}),
(\ref{Hat_Delta_U})
and the identity
$\gamma^\mu \hat{\Delta}_T= \Delta_T^\mu + \sigma^{\mu \Delta_T}$.
Note the  appearance of $\frac{1}{1+ \xi}$
factors that are of pure kinematical origin and,  in particular, lead  to violation
of polynomiality property of TDAs.


\begin{thebibliography}{99}
%%%%%%%%%%%%%%%%%%%%%%%%%%%%%%%%%%%%%%%%%%%%%%%%%%%%%%%%%%%%%%%%%%%%
%            Pioneers TDA                                          %
%%%%%%%%%%%%%%%%%%%%%%%%%%%%%%%%%%%%%%%%%%%%%%%%%%%%%%%%%%%%%%%%%%%%
%\cite{Frankfurt:1999fp}
\bibitem{Frankfurt:1999fp}
  L.~L.~Frankfurt, P.~V.~Pobylitsa, M.~V.~Polyakov and M.~Strikman,
  %``Hard exclusive pseudoscalar meson electroproduction and spin structure  of
  %a nucleon,''
  Phys.\ Rev.\  D {\bf 60}, 014010 (1999)
  [arXiv:hep-ph/9901429].
  %%CITATION = PHRVA,D60,014010;%%

%\cite{Frankfurt:2002kz}
\bibitem{Frankfurt:2002kz}
  L.~Frankfurt, M.~V.~Polyakov, M.~Strikman, D.~Zhalov and M.~Zhalov,
  %``Novel hard semiexclusive processes and color singlet clusters in hadrons,''
  arXiv:hep-ph/0211263.
  %%CITATION = HEP-PH/0211263;%%

%\bibitem{Pire:2004ie}
%  B.~Pire, L.~Szymanowski,
  %``Hadron annihilation into two photons and backward VCS in the scaling regime of
%QCD,''
 % Phys.\ Rev.\  {\bf D71}, 111501 (2005).
  %[hep-ph/0411387].

%%%%%%%%%%%%%%%%%%%%%%%%%%%%%%%%%%%%%%%%%%%%%%%%%%%%%%%%%%%%%%%%%%%%

\bibitem{LPS}
  J.~P.~Lansberg, B.~Pire and L.~Szymanowski,
  %``Production of a pion in association with a high-Q2 dilepton pair in
  %antiproton-proton annihilation at GSI-FAIR,''
  Phys.\ Rev.\  D {\bf 76}, 111502 (2007)
  [arXiv:0710.1267 [hep-ph]].
  %%CITATION = PHRVA,D76,111502;%%
%%%%%%%%%%%%%%%%%%%%%%%%%%%%%%%%%%%%%%%%%%%%%%%%%%%%%%%%%%%%%%%%%%%%%%%%%%%%%%%%%%%%

%%%%%%%%%%%%%%%%%%%%%%%%%%%%%%%%%%%%%%%%%%%%%%%%%%%%%%%%%%%%%%%%%%%
%               Three-local quark operators                       %
%%%%%%%%%%%%%%%%%%%%%%%%%%%%%%%%%%%%%%%%%%%%%%%%%%%%%%%%%%%%%%%%%%%
%\cite{Radyushkin:1977gp}
\bibitem{Radyushkin:1977gp}
  A.~V.~Radyushkin,
  %``Deep elastic processes of composite particles in field theory and
  %asymptotic freedom,''
  arXiv:hep-ph/0410276.
  %%CITATION = HEP-PH/0410276;%%

%\cite{Efremov:1978rn}
\bibitem{Efremov:1978rn}
  A.~V.~Efremov and A.~V.~Radyushkin,
  %``Asymptotical Behavior Of Pion Electromagnetic Form-Factor In QCD,''
  Theor.\ Math.\ Phys.\  {\bf 42}, 97 (1980)
  [Teor.\ Mat.\ Fiz.\  {\bf 42}, 147 (1980)].
  %%CITATION = TMFZA,42,147;%%

%\cite{Lepage:1979zb}
\bibitem{Lepage:1980}
  G.~P.~Lepage and S.~J.~Brodsky, Phys. Rev. D {\bf 22}, 2157 (1980).

\bibitem{Chernyak:1983ej}
  V.~L.~Chernyak and A.~R.~Zhitnitsky,
  %``Asymptotic Behavior Of Exclusive Processes In QCD,''
  Phys.\ Rept.\  {\bf 112}, 173 (1984).
  %%CITATION = PRPLC,112,173;%%

%%%%%%%%%%%%%%%%%%%%%%%%%%%%%%%%%%%%%%%%%%%%%%%%%%%%%%%%%%%%%%%%%%%
%\cite{Stefanis:1999wy}
\bibitem{Stefanis:1999wy}
  N.~G.~Stefanis,
  %``The Physics of exclusive reactions in QCD: Theory and phenomenology,''
  Eur.\ Phys.\ J.\ direct C {\bf 7}, 1 (1999)
  [arXiv:hep-ph/9911375].
  %%CITATION = EPHJD,C7,1;%%

%\cite{Braun:1999te}
\bibitem{Braun:1999te}
  V.~M.~Braun, S.~E.~Derkachov, G.~P.~Korchemsky and A.~N.~Manashov,
  %``Baryon distribution amplitudes in QCD,''
  Nucl.\ Phys.\  B {\bf 553}, 355 (1999)
  [arXiv:hep-ph/9902375].
  %%CITATION = NUPHA,B553,355;%%



%%%%%%%%%%%%%%%%%%%%%%%%%%%%%%%%%%%%%%%%%%%%%%%%%%%%%%%%%%%%%%%%%%%%
%            Pioneers TDA II: Backward regime                      %
%%%%%%%%%%%%%%%%%%%%%%%%%%%%%%%%%%%%%%%%%%%%%%%%%%%%%%%%%%%%%%%%%%%%
%\cite{Pire:2005ax,Pire:2005mt,Lansberg:2007ec}
%\cite{Pire:2005ax}
\bibitem{Pire:2005ax}
  B.~Pire and L.~Szymanowski,
  %``QCD analysis of anti-p N --> gamma* pi in the scaling limit,''
  Phys.\ Lett.\  B {\bf 622}, 83 (2005)
  [arXiv:hep-ph/0504255].
  %%CITATION = PHLTA,B622,83;%%


\bibitem{Pasquini:2009ki}
  B.~Pasquini, M.~Pincetti and S.~Boffi,
  %``Parton content of the nucleon from distribution amplitudes and transition
  %distribution amplitudes,''
  Phys.\ Rev.\  D {\bf 80}, 014017 (2009)
  [arXiv:0905.4018 [hep-ph]].
  %%CITATION = PHRVA,D80,014017;%%

%\cite{Strikman:2009bd}
\bibitem{Strikman:2009bd}
  M.~Strikman and C.~Weiss,
  %``Chiral dynamics and partonic structure at large transverse distances,''
  Phys.\ Rev.\  D {\bf 80}, 114029 (2009)
  [arXiv:0906.3267 [hep-ph]].
  %%CITATION = PHRVA,D80,114029;%%

 %\cite{Pire:2005mt}
\bibitem{Pire:2005mt}
  B.~Pire and L.~Szymanowski,
  %``A QCD analysis of \bar p N -> gamma^* pi and \bar p N -> gamma^* gamma.
  %Where is the pion in the proton ?,''
  PoS {\bf HEP2005}, 103 (2006)
  [arXiv:hep-ph/0509368].
  %%CITATION = POSCI,HEP2005,103;%%


%\cite{Lansberg:2007ec}
\bibitem{Lansberg:2007ec}
  J.~P.~Lansberg, B.~Pire and L.~Szymanowski,
  %``Hard exclusive electroproduction of a pion in the backward region,''
  Phys.\ Rev.\  D {\bf 75}, 074004 (2007)
  [Erratum-ibid.\  D {\bf 77}, 019902 (2008)]
  [arXiv:hep-ph/0701125].
  %%CITATION = PHRVA,D75,074004;%%

%\cite{Collins:1996fb}
\bibitem{Collins:1996fb}
  J.~C.~Collins, L.~Frankfurt and M.~Strikman,
  %``Factorization for hard exclusive electroproduction of mesons in QCD,''
  Phys.\ Rev.\  D {\bf 56}, 2982 (1997)
  [arXiv:hep-ph/9611433].
  %%CITATION = PHRVA,D56,2982;%%







\bibitem{BLP1}
 %\cite{Braun:2006td}
%\bibitem{Braun:2006td}
  V.~M.~Braun, D.~Y.~Ivanov, A.~Lenz and A.~Peters,
  %``Deep inelastic pion electroproduction at threshold,''
  Phys.\ Rev.\  D {\bf 75}, 014021 (2007)
  [arXiv:hep-ph/0611386].
  %%CITATION = PHRVA,D75,014021;%%
\bibitem{BLP2}
V.~M.~Braun, D.~Y.~Ivanov and A.~Peters,
  %``Threshold Pion Electroproduction at Large Momentum Transfers,''
  Phys.\ Rev.\  D {\bf 77}, 034016 (2008).
  %%CITATION = PHRVA,D77,034016;%%
 %%CITATION = PHRVA,D75,014021;%%



%%%%%%%%%%%%%%%%%%%%%%%%%%%%%%%%%%%%%%%%%%%%%%%%%%%%%%%%%%%%%%%%%%%

%Crossing


%\cite{Polyakov:1998ze}
\bibitem{Polyakov:1998ze}
  M.~V.~Polyakov,
  %``Hard exclusive electroproduction of two pions and their resonances,''
  Nucl.\ Phys.\  B {\bf 555}, 231 (1999)
  [arXiv:hep-ph/9809483].
  %%CITATION = NUPHA,B555,231;%%


%\cite{Polyakov:1999gs}
\bibitem{Polyakov:1999gs}
  M.~V.~Polyakov and C.~Weiss,
  %``Skewed and double distributions in pion and nucleon,''
  Phys.\ Rev.\  D {\bf 60}, 114017 (1999)
  [arXiv:hep-ph/9902451].
  %%CITATION = PHRVA,D60,114017;%%


%%%%%%%%%%%%%%%%%%%%%%%%%%%%%%%%%%%%%%%%%%%%%%%%%%%%%%%%%%%%%%%%%%%
%                                                                 %
%                  Soft pion theorem                              %
%%%%%%%%%%%%%%%%%%%%%%%%%%%%%%%%%%%%%%%%%%%%%%%%%%%%%%%%%%%%%%%%%%%

%\cite{Pobylitsa:2001cz}
\bibitem{Pobylitsa:2001cz}
  P.~V.~Pobylitsa, M.~V.~Polyakov and M.~Strikman,
  %``Soft pion theorems for hard near threshold pion production,''
  Phys.\ Rev.\ Lett.\  {\bf 87}, 022001 (2001)
  [arXiv:hep-ph/0101279].
  %%CITATION = PRLTA,87,022001;%%




%\cite{Pire:2010if}
\bibitem{Pire:2010if}
   B.~Pire, K.~Semenov-Tian-Shansky and L.~Szymanowski,
  %``A Spectral representation for baryon to meson transition distribution
  %amplitudes,''
  Phys.\ Rev.\  D {\bf 82}, 094030 (2010)
  [arXiv:1008.0721 [hep-ph]].



\bibitem{Cross-sec-paper}
  B.~Pire, K.~Semenov-Tian-Shansky and L.~Szymanowski,
  %``A spectral representation for baryon to meson and baryon to photon
  %transition distribution amplitudes,''
  {\it in preparation}.


\bibitem{RDDA4}
  I.~V.~Musatov and A.~V.~Radyushkin,
  %``Evolution and models for skewed parton distributions,''
  Phys.\ Rev.\  D {\bf 61}, 074027 (2000)
  [arXiv:hep-ph/9905376].

%%%%%%%%%%%%%%%%%%%%%%%%%%%%%%%%%%%%%%%%%%%%%%%%%%%%%%%%%%%%%

\bibitem{pigamTDA1}
  W.~Broniowski, E.~R.~Arriola,
  %``Pion-photon Transition Distribution Amplitudes in the Spectral Quark Model,''
  Phys.\ Lett.\  {\bf B649}, 49-56. (2007).

\bibitem{pigamTDA2}
  B.~C.~Tiburzi,
  %``Estimates for pion-photon transition distributions,''
  Phys.\ Rev.\  {\bf D72}, 094001 (2005).

\bibitem{pigamTDA3}
  A.~Courtoy, S.~Noguera,
  %``The Pion-photon transition distribution amplitudes in the Nambu-Jona Lasinio
%model,
  Phys.\ Rev.\  {\bf D76},   094026 (2007).

\bibitem{pigamTDA4}
  P.~Kotko, M.~Praszalowicz,
  %``Covariant Non-local Chiral Quark Model and Pion-photon Transition Distribution
%Amplitudes,
  Phys.\ Rev.\  {\bf D80}, 074002 (2009).

\bibitem{pigamTDA5}
  J.~P.~Lansberg, B.~Pire, L.~Szymanowski,
  %``Exclusive meson pair production in gamma* gamma scattering at small momentum
%transfer,''
  Phys.\ Rev.\  {\bf D73},  074014 (2006)
  and Nucl.Phys.Proc. Suppl {\bf 184}, 239  (2008).


%%%%%%%%%%%%%%%%%%%%%%%%%%%%%%%%%%%%%%%%%%%%%%%%%%%%%%%%%%%%%

\bibitem{J-lab}
K.~Park, P.~Stoler, V.~Kubarovsky, {\it private communications}.

\bibitem{PANDA}
  M.~F.~M.~Lutz {\it et al.} [ PANDA Collaboration ],
  %``Physics Performance Report for PANDA: Strong Interaction Studies with Antiprotons,''
    [arXiv:0903.3905 [hep-ex]].

\bibitem{Diehl}
M. Diehl, Phys. Rept. {\bf 388}, 41 (2003) [hep-ph/0307382].




%\cite{Radyushkin:2011dh}
\bibitem{Radyushkin:2011dh}
  A.~V.~Radyushkin,
  %``Generalized Parton Distributions and Their Singularities,''
  Phys.\ Rev.\  D {\bf 83}, 076006 (2011)
  [arXiv:1101.2165 [hep-ph]].
  %%CITATION = PHRVA,D83,076006;%%















%%Nucleon DA pioneers %%%%%%%%%%%%%%%%%%%%%%%%%%%%%%%%%%%%%%%%%%%%%%%%%%%
\bibitem{Korenblit}
V.A. Avdeenko, V.L.Chernyak and S.E.~Korenblit; Yad. Fiz. 33, 481 (1981)
[Sov. J. Nucl. Phys. 33, 252 (1981)].



\bibitem{Chernyak_Nucleon_wave}
V.L. Chernyak and I.R. Zhitnitsky, Nucl. Phys. {\bf B 246}, 52 (1984).
%%Nucleon DA pioneers %%%%%%%%%%%%%%%%%%%%%%%%%%%%%%%%%%%%%%%%%%%%%%%%%%%

\bibitem{Georgi}
H.~Georgi,  {\em Lie Algebras in Particle Physics}
(Westview
Press, Boulder, CO, 1999).
%ISBN 978-0738202334

\bibitem{Farrar}
G.R. Farrar, H. Zhang, A.A. Ogloblin and I.R. Zhitnitsky, Nucl. Phys. B {\bf 311}
585 (1989).

\bibitem{Chernyak_Delta}
V.L. Chernyak, A.A. Ogloblin and I.R. Zhitnitsky, Z. Phys. C {\bf 42} (1989) 569.


%\cite{Kivel:2002ia}
\bibitem{Kivel:2002ia}
  N.~Kivel and M.~V.~Polyakov,
  %``One loop chiral corrections to hard exclusive processes: 1. Pion case,''
  arXiv:hep-ph/0203264.
  %%CITATION = HEP-PH/0203264;%%


\bibitem{Alfaro_red_book}
 V.~de Alfaro, S.~Fubini, G.~Furlan, C.~Rossetti,
 {\em Currents in Hadron Physics}
 (North-Holland, Amsterdam, 1973).

\bibitem{EricsonWeise}
T.~Ericson, W.~Weise, {\it Pions and Nuclei}
(Clarendon Press, Oxford, 1988).

%\cite{SemenovTianShansky:2007hv}
\bibitem{SemenovTianShansky:2007hv}
  K.~M.~Semenov-Tian-Shansky, A.~V.~Vereshagin and V.~V.~Vereshagin,
  %``Bootstrap and the physical values of $\pi N$ resonance parameters,''
  Phys.\ Rev.\  D {\bf 77}, 025028 (2008)
  [arXiv:0706.3672 [hep-ph]].
  %%CITATION = PHRVA,D77,025028;%%



%\cite{Weinberg:1964cn}
\bibitem{Weinberg:1964cn}
  S.~Weinberg,
  %``Feynman Rules for Any Spin,''
  Phys.\ Rev.\  {\bf 133}, B1318 (1964).
  %%CITATION = PHRVA,133,B1318;%%









\bibitem{Itzykson}
C.~Itzykson, J.~B.~Zuber, {\em Quantum Field Theory}  (McGraw-Hill,  New York, 1980).




\bibitem{PDG}
K. Nakamura et al. (Particle Data Group), J. Phys. G 37, 075021 (2010)

%\cite{Borodulin:1995xd}
\bibitem{Borodulin:1995xd}
  V.~I.~Borodulin, R.~N.~Rogalev, S.~R.~Slabospitsky,
  %``CORE: COmpendium of RElations: Version 2.1,''
  hep-ph/9507456.
\end{thebibliography}
\end{document}